\documentclass[twocolumn,aps,prd,amssymb,amsmath,eqsecnum,nofootinbib]{revtex4}
\usepackage{epsfig}
\begin{document}

\title{Two timescale analysis of extreme mass ratio inspirals in Kerr.  I. Orbital Motion}

%
%
\newcount\hh
\newcount\mm
\mm=\time
\hh=\time
\divide\hh by 60
\divide\mm by 60
\multiply\mm by 60
\mm=-\mm
\advance\mm by \time
\def\hhmm{\number\hh:\ifnum\mm<10{}0\fi\number\mm}


\date{draft of June 12, 2008; printed \today{} at \hhmm}
\author{Tanja Hinderer$^1$}
\author{ \'Eanna \'E. Flanagan$^{1,2}$}
\affiliation{$^1$ Center for Radiophysics and Space
Research, Cornell University, Ithaca, NY 14853, USA\\
$^2$ Laboratory for
Elementary Particle Physics, Cornell University, Ithaca, NY 14853,
USA}

\begin{abstract}

Inspirals of stellar mass compact objects into massive black holes
are an important source for future gravitational
wave detectors such as Advanced LIGO and LISA.
Detection of these
sources and extracting information from the
signal relies on accurate theoretical models of the binary
dynamics.
We cast the equations describing binary inspiral in the extreme mass
ratio limit in terms of action angle variables, and derive properties
of general solutions using a two-timescale expansion.
This provides a rigorous derivation of the prescription
for computing the leading order orbital motion.
As shown by Mino, this leading order or adiabatic motion requires
only knowledge of the orbit-averaged, dissipative piece of the self force.
The two timescale method also gives a framework for
calculating the post-adiabatic corrections.
For circular and for equatorial orbits, the leading order corrections
are suppressed by one power of the mass ratio, and give rise to phase
errors of order unity over a complete inspiral through the relativistic
regime.  These
post-1-adiabatic
corrections are generated by the fluctuating, dissipative piece of the
first order self force, by the conservative piece of the
first order self force, and by the orbit-averaged, dissipative piece of
the second order self force.
We also sketch a two-timescale expansion of the Einstein equation,
and deduce an analytic formula for the leading order, adiabatic
gravitational waveforms generated by an inspiral.

\end{abstract}

\maketitle

\def\be{\begin{equation}}
\def\ee{\end{equation}}
\def\bfx{{\bf x}}
\def\bfq{{\bf q}}
\def\bfJ{{\bf J}}
\def\bfk{{\bf k}}
\def\bfv{{\bf v}}
\def\bfzero{{\bf 0}}
\def\bfPsi{\mbox{\boldmath $\Psi$}}
\def\bfpsi{\mbox{\boldmath $\psi$}}
\def\bfnabla{\mbox{\boldmath $\nabla$}}
\def\bfsigma{\mbox{\boldmath $\sigma$}}
\def\bfomega{\mbox{\boldmath $\omega$}}
\def\bfOmega{\mbox{\boldmath $\Omega$}}
\def\bfTheta{\mbox{\boldmath $\Theta$}}
\def\bfcalJ{\mbox{\boldmath ${\cal J}$}}
\def\bea{\begin{eqnarray}}
\def\eea{\end{eqnarray}}
\def\nn{\nonumber}
\def\tt{{\tilde t}}
\def\ttau{{\tilde \tau}}
\newcommand{\bes}{\begin{subequations}}
\newcommand{\ees}{\end{subequations}}



\section{introduction and Summary}

\subsection{Background and Motivation}

\begin{figure}
\begin{center}
\epsfig{file=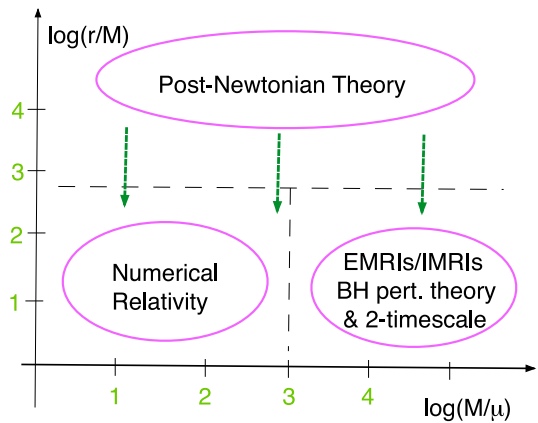,width=7.0cm}
\caption{
The parameter space of inspiralling compact binaries in general relativity, in
terms of the inverse mass ratio $M/\mu = 1/\varepsilon$ and the orbital radius $r$,
showing the different regimes and the computational techniques necessary in each
regime.  Individual binaries evolve downwards in the diagram (green
dashed arrows). }
\label{fig:paramspace}
\end{center}
\end{figure}

Recent years have seen great progress in our understanding of the two
body problem in general relativity.  Binary systems of compact bodies
undergo an inspiral driven by gravitational radiation reaction until
they merge.  As illustrated in Fig.\ \ref{fig:paramspace}, there are three
different regimes in the dynamics of
these systems, depending on the values of the total and reduced masses
$M$ and $\mu$ of the system and the orbital
separation $r$ : (i) The early, weak field regime at $r \gg M$, which
can be accurately modeled using post-Newtonian theory, see, for
example, the review \cite{PN}. (ii) The relativistic, equal mass
regime $r \sim M$, $\mu \sim M$, which must be treated using numerical
relativity.  Over the last few years, numerical relativists have
succeeded for the first time in simulating the merger of black hole
binaries, see, for example, the review \cite{Pretorius:2007nq} and
references therein.  (iii) The relativistic, extreme mass ratio regime
$r \sim M$, $\mu \ll M$.  Over timescales short compared to the
dephasing time $\sim M \sqrt{M/\mu}$, systems in this regime can be accurately
modeled using black hole perturbation theory\cite{Teukolsky:1973ha},
with the mass ratio
$\varepsilon \equiv \mu/M$ serving as the expansion parameter.
The subject of this paper is the approximation methods that are
necessary to treat such systems over the longer inspiral timescale $\sim M^2/\mu$
necessary for computation of complete inspirals.

This extreme mass ratio regime has direct
observational relevance: Compact objects spiraling into
much larger black holes are expected to be a key source for both LIGO
and LISA.
{\it Intermediate-mass-ratio inspirals} (IMRIs) are
inspirals of black holes or neutron stars into intermediate
mass ($50 \leq M \leq 1000 M_\odot)$ black holes; these would be visible to
Advanced LIGO out to distances of several hundred Mpc \cite{Brown2006},
where the event rate could be about $ 3 - 30$ per year
\cite{Brown2006,imris}.
{\it Extreme-mass-ratio inspirals} (EMRIs) are inspirals of stellar-mass compact
objects (black holes, neutron stars, or possibly white dwarfs) into massive
($10^4 \leq M \leq 10^7 M_\odot$) black holes in galactic nuclei; these will
be visible to LISA
out to redshifts $z \approx 1$ \cite{Finn:2000sy,Cutler:2002me,pau}.
It has been estimated \cite{emri-research,Gair:2004iv} that LISA should
see about $ 50$ such events per year, based on calculations of
stellar dynamics in galaxies' central cusps\cite{2006ApJ...645L.133H}.
Because of an IMRI's
or EMRI's small mass ratio $\varepsilon = \mu/M$, the small body lingers
in the large black hole's strong-curvature region for many wave cycles before
merger: hundreds of cycles for LIGO's IMRIs; hundreds of thousands
for LISA's EMRIs \cite{Finn:2000sy}.
In this relativistic regime the post-Newtonian approximation has
completely broken down, and full numerical relativity simulations become prohibitively
difficult as $\varepsilon$ is decreased.  Modeling of these sources
therefore requires a specialized approximation method.

Gravitational waves from these sources will be rich with
information \cite{Cutler:2002me,pau}:

\begin{itemize}

\item The waves
carry
not only the
details of the evolving orbit, but also 
a map of the large body's spacetime geometry, or
equivalently the values of all its multipole
moments, as well as details of the response of the horizon to tidal
forces~\cite{Ryan:1995wh,Li:2007qu}.
Extracting the map ({\it bothrodesy})
is a high priority for LISA,
which can achieve ultrahigh accuracy, and a moderate priority for
LIGO, which will have a lower (but still interesting) accuracy
~\cite{Brown2006}.
Measurements of the black hole's quadrupole (fractional accuracy about $10^{-3}$ for LISA
\cite{Ryan:1997hg,Barack:2006pq}, about $1$ for Advanced LIGO \cite{Brown2006})
will enable tests of the black hole's {\it no hair property}, that all of the
mass and current multipole moments are uniquely determined in terms of the first two,
the mass and spin.  Potentially, these measurements could lead to the
discovery of non-black-hole central objects such as boson stars
\cite{PhysRevLett.57.2485,1997PhRvD..55.6081R}
or naked singularities.

\item One can measure the mass
and spin of the central black hole with fractional accuracies of order
$10^{-4}$ for LISA \cite{Poisson:1996tc,sources} and  about $
10^{-2}$--$10^{-1}$ for Advanced LIGO \cite{Brown2006}.
Observing many
events will therefore provide a census of the masses and spins of the
massive central black holes in non-active galactic nuclei like M31 and
M32.
The spin can provide useful information about the hole's growth
history (mergers versus accretion) \cite{Hughes:2002ei}.

\item For LISA, one can measure the inspiralling
objects' masses with precision about $10^{-4}$, teaching us about the
stellar population
in the central parsec of galactic nuclei.

\item If the LISA event rate is large enough,
one can measure the Hubble constant $H_0$ to about $1\%$
\cite{2007arXiv0712.0618M}, which would indirectly aid
dark energy studies \cite{Hu:2004kn}.
The idea is to combine
the measured luminosity distance of cosmological ($z \sim 1/2$) EMRIs
with a statistical analysis of the redshifts of candidate
host galaxies located within the error box on the sky.

\end{itemize}

To realize the science goals for these sources requires accurate
theoretical models of the waveforms for matched filtering.
The accuracy requirement is roughly that the theoretical template's phase must
remain accurate to $\sim 1$ cycle over the $\sim \varepsilon^{-1}$
cycles of waveform in the highly relativistic regime
($\sim 10^2$ cycles for LIGO, $\sim 10^5$ for LISA).
For signal detection, the requirement is slightly less stringent than this,
while for parameter extraction the requirement is slightly more stringent:
The waveforms are characterized by 14 parameters, which makes a fully
coherent search of the entire data train computationally impossible.
Therefore, detection templates for LISA will use short segments of
the signal and require phase coherence for $\sim 10^4$ cycles
{\cite{Gair:2004iv}}. Once the presence of a signal has been
established, the source parameters will be extracted using
measurement templates that require a fractional phase accuracy of
order the reciprocal of the signal to noise ratio
{\cite{Gair:2004iv}}, in order to keep systematic errors
as small as the statistical errors.

\subsection{Methods of computing orbital motion and waveforms}
\label{ssec:methods}

A variety of approaches to computing waveforms have been pursued in
the community.  We now review these approaches in order to place the
present paper in context.
The foundation for all approaches is the fact that, since
$\varepsilon = \mu/M \ll 1$, the field of the compact object can be treated as a
small perturbation to the large black hole's gravitational field.  On
short timescales $\sim M$, the compact object moves on a geodesic of the Kerr geometry,
characterized by its conserved energy $E$, $z$-component of angular
momentum $L_z$, and Carter constant $Q$.  Over longer timescales $\sim
M/\varepsilon$,
radiation reaction causes the parameters $E$, $L_z$ and $Q$ to evolve
adiabatically
and the orbit to shrink.  The effect of the internal structure of the
object is negligible\footnote{There are two exceptions, where
corrections to the point-particle model can be important: (i) White
dwarf EMRIs, where tidal interactions can play a role
\protect{\cite{2005MNRAS.357..834R}}.
(ii) The effect due to the spin, if any, of the inspiralling object,
whose importance has been emphasized by Burko
\protect{\cite{2004PhRvD..69d4011B,2006CQGra..23.4281B}}.  While this effect is
at most marginally relevant for signal detection \protect{\cite{Barack:2003fp}}, it is likely
quite important for information extraction.  We neglect the spin
effect in the present paper, since it can be computed and included in
the waveforms relatively easily.}, so it can
be treated as a point particle.
At the end of the inspiral, the particle passes through an innermost
stable orbit where adiabaticity breaks down, and it transitions onto a
geodesic plunge orbit
\cite{Ori:2000zn,Buonanno:2000ef,O'Shaughnessy:2002ez,Sundararajan:2008bw}.  In this
paper we restrict
attention to the adiabatic portion of the motion.

\bigskip
\noindent
{\it Numerical Relativity:}
Numerical relativity has not yet been applied to the extreme mass
ratio regime.  However, given the recent successful simulations in the
equal mass regime $\varepsilon \sim 1$, one could contemplate trying
to perform simulations with smaller mass ratios.
There are a number of difficulties that arise as $\varepsilon$ gets
small: (i) The orbital timescale and the radiation
reaction timescale are separated by the large factor $\sim 1/\varepsilon$.  The huge number of
wave cycles implies an impractically large computation time.
(ii) There is a separation
of lengthscales: the compact object is smaller than the central black
hole by a factor $\varepsilon$.  (iii) Most importantly, in the strong field region
near the small object, the piece of the metric perturbation
responsible for radiation reaction is of order $\varepsilon$, and
since one requires errors in
the radiation reaction to be of order $\varepsilon$, the accuracy
requirement on the metric perturbation is of order $\varepsilon^2$.
These difficulties imply that numerical simulations will likely not be
possible in the extreme mass ratio regime in the foreseeable future,
unless major new techniques are devised to speed up computations.

\bigskip
\noindent
{\it Use of post-Newtonian methods:}
Approximate waveforms which are qualitatively similar to real waveforms
can be obtained using post-Newtonian methods
or using hybrid schemes containing some post-Newtonian elements
\cite{Glampedakis:2002cb,Barack:2003fp,Babak:2006uv}.
Although these waveforms are insufficiently accurate for the eventual
detection and data analysis of real signals,
they have been very useful for approximately scoping out the detectability of
inspiral events and the accuracy of parameter measurement, both for
LIGO \cite{Brown2006} and LISA \cite{Gair:2004iv,Barack:2003fp}.
They have the advantage that they can be computed relatively quickly.

\bigskip
\noindent
{\it Black hole perturbation theory -- first order:}
There is a long history of using first order perturbation theory
\cite{Teukolsky:1973ha} to
compute gravitational waveforms from particles in geodesic orbits
around black holes \cite{Cutler:1994pb,Shibata:1994xk,Glampedakis:2002ya,Hughes:1999bq}.
These computations have recently been extended to fully generic orbits
\cite{Hughes:2005qb,Drasco:2005kz,Drasco:2007gn}.
However first order perturbation theory is limited to producing ``snapshot'' waveforms
that neglect radiation reaction.\footnote{The source for the linearized
Einstein equation must be a conserved stress energy tensor, which
for a point particle requires a geodesic orbit.}
Such waveforms fall out of phase with true waveforms after a dephasing
time $\sim M/\sqrt{\varepsilon}$, the geometric mean of the orbital
and radiation reaction timescales, and so are of limited utility.\footnote{Drasco has argued that
snapshot waveforms may still be useful for signal detections in certain limited parts
of the IMRI/EMRI parameter space, since the phase coherence time is
actually $\sim 100 M / \sqrt{\varepsilon}$
\protect{\cite{Drasco:2006ws}}.}

\bigskip
\noindent
{\it Black hole perturbation theory -- second order:}
One can in principle go to second order in perturbation theory
\cite{1999PhRvD..59l4022C,1998gr.qc.....7077G,Lousto:2008vw}.
At this order, the particle's geodesic motion
must be corrected by self-force effects describing
its interaction with its own spacetime distortion.
This gravitational self force is analogous to the electromagnetic
Abraham-Lorentz-Dirac force.
Although a formal expression for the self force is known
{\cite{Mino:1996nk,Quinn:1996am}}, it has proved difficult to
translate this expression into a practical computational scheme for
Kerr black holes because of the mathematical complexity of the
self-field regularization which is required.  Research into this topic is ongoing; see,
for example the review \cite{Poisson:2003nc}
and Refs.\
\cite{Ori,Barack,Barack:2002mh,Gralla:2005et,Keidl:2006wk,Barack:2007tm,Vega:2007mc,Barack:2007we,Lousto:2008vw}
for various approaches and recent progress.

When the self-force is finally successfully computed, second order
perturbation theory will provide a self-consistent framework for
computing the orbital motion and the waveform, but only over short
timescales.  The second order waveforms will
fall out of phase with the true waveforms after only a dephasing time $\sim
M/\sqrt{\varepsilon}$ \footnote{The reason is as follows.
Geodesic orbits and true orbits become out of
phase by $\sim 1$ cycle after a dephasing time.
Therefore, since the linear metric perturbation is sourced by a geodesic
orbit, fractional errors in the linear metric perturbation must be of order unity.
Therefore the second order metric perturbation must become comparable to the first order
term after a dephasing time.}
\cite{Eannatalk,2006PThPh.115...43M}.
Computing accurate waveforms describing a full inspiral
therefore requires going beyond black hole perturbation theory.

\bigskip
\noindent
{\it Use of conservation laws:}
This well-explored method allows tracking an entire inspiral for
certain special classes of orbits.  Perturbation theory
is used to compute
the fluxes
of $E$ and $L_z$ to infinity and down the horizon for
geodesic orbits, and
imposing global conservation laws, one infers the rates of change of the orbital energy
and angular momentum.  For circular orbits
and equatorial orbits these determine the rate of change
of the Carter constant $Q$, and thus the inspiralling trajectory.
The computation can either be done in the frequency domain
\cite{Cutler:1994pb,Shibata:1994xk,Glampedakis:2002ya,Hughes:1999bq}, or in the time domain by
numerically integrating the Teukolsky equation as a 2+1 PDE with a
suitable numerical model of the point particle source
\cite{Krivan:1997hc,Burko:2002bt,
Martel:2003,Lopez-Aleman:2003ik,Khanna:2003qv,Pazos-Avalos:2004rp,Sopuerta:2005rd,Sopuerta:2005gz,Sundararajan:2007jg,Barton:2008eb}.
However, this method fails for generic orbits since there is no known
global conservation law associated with the Carter constant $Q$.

\bigskip
\noindent {\it Adiabatic approximation -- leading order:}
Over the last few years, it has been discovered how to compute
inspirals to leading order for generic orbits.  The method is based on the
Mino's realization \cite{Mino2003} that, in the adiabatic limit,
one needs only the time
averaged, dissipative piece of the first order self force, which can be
straightforwardly computed from the half retarded minus half advanced
prescription.  This sidesteps the difficulties associated with
regularization that impede computations of the full, first order self force.
From the half advanced minus half retarded prescription, one can derive an explicit formula for
a time-average of ${\dot Q}$ in terms of mode expansion
\cite{Hughes:2005qb,scalar,2005PThPh.114..509S,2006PThPh.115..873S,Drasco:2007a}.
Using this formula it will be straightforward to compute inspirals to
the leading order.

\bigskip
We now recap and assess the status of these various approaches.
All of the approaches described above have shortcomings and limitations \cite{2006PThPh.115...43M}.
Suppose that one computes the inspiral motion, either from
conservation laws, or from the
time-averaged dissipative piece of the first order self-force, or from
the exact first order self-force when it becomes available.
It is then necessary to compute the radiation generated by this inspiral.
One might be tempted to use linearized perturbation theory for this
purpose.  However, two problems then arise:

\begin{itemize}

\item As noted above, the use of linearized perturbation theory with
nongeodesic sources is mathematically inconsistent.  This inconsistency
has often been remarked upon, and various ad hoc methods of modifying
the linearized theory to get around the difficulty have been
suggested or implemented \cite{Quinn:1996am,2005PThPh.113..733M,Sundararajan:2008zm}.

\item A related problem is that the resulting waveforms will depend on
  the gauge chosen for the linearized metric perturbation, whereas the
  exact waveforms must be gauge invariant.

\end{itemize}
It has often been suggested that these problems can be resolved by
going to second order in perturbation theory \cite{Poisson:2003nc,Lousto:2008vw}.
However, as discussed above, a second order computation will be valid
only for a dephasing time, and not for a full inspiral.

Of course, the above problems are not fatal, since the motion is locally
very nearly geodesic, and so the inconsistencies and ambiguities
are suppressed by a factor $\sim \varepsilon$ relative to the leading
order waveforms.\footnote{This is true both for the instantaneous
amplitude and for the accumulated phase of the waveform.}
Nevertheless, it is clearly desirable to have a well
defined approximation method that gives a unique, consistent result
for the leading order waveform.
Also, for parameter extraction, it will be necessary to compute the
phase of the waveform beyond the leading order.  For this purpose it
will clearly be necessary to have a more fundamental computational
framework.

\subsection{The two timescale expansion method}

In this paper we describe an approximation scheme which addresses and
resolves all of the theoretical difficulties discussed above.
It is based on the fact that the systems evolve adiabatically: the
radiation reaction timescale $\sim M/\varepsilon$ is much longer
than the orbital timescale $\sim M$ \cite{Mino2003}.
It uses two-timescale expansions, which are
a systematic method for studying the cumulative effect of a small disturbance
on a dynamical system that is active over a long time \cite{Kevorkian}.

The essence of the method is simply an ansatz for the dependence of
the metric $g_{ab}(\varepsilon)$ on $\varepsilon$, and an ansatz for
the dependence of the orbital motion on $\varepsilon$, that are
justified a posteriori order by order via substitution into Einstein's equation.
The ansatz for the metric is more complex than the Taylor series
ansatz which underlies standard perturbation theory.
The two timescale method has roughly the same
relationship to black hole perturbation theory as post-Newtonian
theory has to perturbation theory of Minkowski spacetime.
The method is consistent with standard black hole perturbation theory locally
in time, at each instant, but extends the domain of validity to an entire inspiral.
The method provides a systematic procedure for computing the leading
order waveforms, which we call the adiabatic waveforms, as well as
higher order corrections.  We call the $O(\varepsilon)$ corrections
the post-1-adiabatic corrections, the $O(\varepsilon^2)$ corrections
post-2-adiabatic, etc., paralleling the standard terminology in
post-Newtonian theory.

The use of two timescale expansions in the extreme mass ratio regime
was first suggested in Refs.\ \cite{Eannatalk,2006APS..APRS11005H}.
The method has already been applied to some simplified model problems:
a computation of the inspiral of a point
particle in Schwarzschild subject to electromagnetic radiation
reaction forces by Pound and Poisson \cite{Pound:2007ti}, and a computation
of the scalar radiation generated by a inspiralling particle coupled to
a scalar field by Mino and Price \cite{Minoscalar}.
We will extend and generalize these analyses, and develop a complete
approximation scheme.

There are two, independent, parts to the the approximation scheme.
The first is a two timescale analysis of the inspiralling orbital
motion, which is the focus of the present paper.
Our formalism enables
us to give a rigorous derivation and clarification of the
prescription for computing the leading order motion that is valid
for all orbits, and resolves some controversies in
the literature \cite{Pound:2007ti}.
It also allows us to systematically calculate the higher order
corrections.  For these corrections, we restrict
attention to inspirals in Schwarzschild, and to circular and
equatorial inspirals in Kerr.  Fully generic inspirals in Kerr involve
a qualitatively new feature -- the occurrence of transient resonances
-- which we will discuss in the forthcoming papers \cite{FH08a,FH08b}.

The second part to the approximation scheme is the application of the
two timescale method to the Einstein equation, and a meshing of that
expansion to the analysis of the orbital motion.  This allows
computation of the observable gravitational waveforms, and is described
in detail in the forthcoming paper \cite{FH08c}.  We briefly sketch
this formalism in Sec.\ \ref{sec:intro:waves} below, and give an
analytic result for the leading order waveforms.

We note that alternative methods of attempting to overcome the problems with
standard perturbation theory, and of going beyond adiabatic order,
have been developed by Mino
\cite{2005PThPh.113..733M,2006PThPh.115...43M,2005CQGra..22S.717M,CQG22b,2005CQGra..22S.375M}.
These methods have some overlap with the method discussed here, but
differ in some crucial aspects.  We do not believe that these methods
provide a systematic framework for going to higher orders, unlike
the two-timescale method.

\subsection{Orbital Motion}

We now turn to a description of our two timescale analysis of the
orbital motion.  The first step is to exploit the Hamiltonian
structure of the unperturbed, geodesic motion to rewrite the governing
equations in terms of generalized action angle variables.
We start from the forced geodesic equation
\be
\frac{d^2x^\nu}{d\tau^2}+\Gamma^\nu_{\sigma\rho}\frac{dx^\sigma}{d\tau}
\frac{dx^\rho}{d\tau}=\varepsilon a^{(1)\,\nu}
+ \varepsilon^2 a^{(2)\,\nu}
+ O(\varepsilon^3).
\label{aselfdefinition0}
\ee
Here $\tau$ is proper time and $a^{(1)\,\nu}$ and $a^{(2)\,\nu}$ are
the first order and second order self-accelerations.
In Sec.\ \ref{sec:kerrapplication} we
augment these equations to describe the leading order backreaction
of the inspiral on the mass $M$ and spin $a$ of the black hole,
and show they can be rewritten as
[cf.\ Eqs.\ (\ref{eq:eomsimplified}) below]
\bes
\bea
\frac{dq_\alpha}{d\tau}&=&\omega_\alpha(J_\sigma)+\varepsilon
g^{(1)}_\alpha(q_r,q_\theta,J_\sigma)
+\varepsilon^2
g^{(2)}_\alpha(q_r,q_\theta,J_\sigma)
\nonumber \\
&&
+O(\varepsilon^3),\\
\frac{dJ_\lambda}{d\tau}&=&\varepsilon
G^{(1)}_\lambda(q_r,q_\theta,J_\sigma)+\varepsilon^2
G^{(2)}_\lambda(q_r,q_\theta,J_\sigma)
\nonumber \\
&&
+O(\varepsilon^3).
\eea
\label{orbital}\ees
Here the variables $J_\lambda$ are the three conserved quantities of
geodesic motion, with the dependence on the particle mass scaled out,
together with the black hole mass and spin parameters:
\be
J_\lambda = (E/\mu, L_z/\mu, Q/\mu^2, M, a).
\ee
The variables $q_\alpha = (q_r, q_\theta, q_\phi, q_t)$ are a set of
generalized angle variables associated with the $r$, $\theta$, $\phi$
and $t$ motions in Boyer-Lindquist coordinates, and are defined
more precisely in Sec.\ \ref{sec:kerrexplicit} below.
The variables $q_r$, $q_\theta$, and $q_\phi$ each increase by $2
\pi$ after one cycle of motion of the corresponding variable $r$,
$\theta$ or $\phi$.
The functions $\omega_\alpha(J_\sigma)$ are the fundamental frequencies of
geodesic motion in the Kerr metric.
The functions $g^{(1)}_\alpha$, $G^{(1)}_\lambda$ are currently not
known explicitly, but their exact form will not be
needed for the analysis of this paper.  They are
determined by
the first order self acceleration \cite{Mino:1996nk,Quinn:1996am}.
Similarly, the functions $g^{(2)}_\alpha$
and $G^{(2)}_\lambda$ are currently not known explicitly, and are
determined in part by the second order self acceleration
\cite{Rosenthal:2006nh,Rosenthal:2006iy,Chadthesis,Galley:2008ih,Chadtalk};
see Sec.\ \ref{sec:backreaction} for more details.

In Secs.\ \ref{sec:derivation_single} -- \ref{sec2:manyvariables}
below we analyze the differential equations (\ref{orbital}) using two
timescale expansions. In the non-resonant case, and up to
post-1-adiabatic order, the results can be summarized as follows.
Approximate solutions of the equations can be constructed via a series of
steps:
\begin{itemize}
\item We define the slow time variable $\ttau = \varepsilon
  \tau$.

\item We construct a set of functions $\psi_\alpha^{(0)}(\ttau)$, ${\cal
    J}_\lambda^{(0)}(\ttau)$, $\psi_\alpha^{(1)}(\ttau)$ and ${\cal
    J}_\lambda^{(1)}(\ttau)$ of the slow time.  These functions are
  defined by a set of differential equations that involve
  the functions $\omega_\alpha$, $g^{(1)}_\alpha$, $G^{(1)}_\lambda$,
  $g^{(2)}_\alpha$ and $G^{(2)}_\lambda$ and which are independent of
  $\varepsilon$ [Eqs.\ (\ref{jks}), (\ref{sec2:eq:Omega0ans}), (\ref{sec2:eq:calJ0ans}),
(\ref{sec2:eq:Omega1ans}), (\ref{sec2:eq:calJ1ans}) below].

\item We define a set of auxiliary phase variables $\psi_\alpha$ by
\be
\psi_\alpha(\tau,\varepsilon) = \frac{1}{\varepsilon}
\psi^{(0)}_\alpha(\varepsilon \tau) + \psi_\alpha^{(1)}(\varepsilon
\tau) + O(\varepsilon),
\label{aux}
\ee
where the $O(\varepsilon)$ symbol refers to the limit $\varepsilon \to
0$ at fixed $\ttau = \varepsilon \tau$.

\item Finally, the solution to post-1-adiabatic order is given by
\bes
\bea
\label{qalpha0}
q_\alpha(\tau,\varepsilon) &=& \psi_\alpha + O(\varepsilon), \\
J_\lambda(\tau,\varepsilon) &=& {\cal J}_\lambda^{(0)}(\varepsilon
\tau) + \varepsilon {\cal J}^{(1)} (\varepsilon \tau) \nonumber \\
&& + H_\lambda[ \psi_r, \psi_\theta, {\cal J}_\sigma^{(0)}(\varepsilon \tau) ]
+ O(\varepsilon^2),
\eea
\label{finalans}\ees
where the $O(\varepsilon)$ and $O(\varepsilon^2)$ symbols refer to
$\varepsilon \to 0$ at fixed $\ttau$ and $\psi_\alpha$.
Here $H_\lambda$ is a function which is periodic in its first two
arguments and which can computed from the function $G_\lambda^{(1)}$
[Eq.\ (\ref{Hlambdadef}) below].

\end{itemize}

We now turn to a discussion of the implications of the final result (\ref{finalans}).
First, we emphasize that the purpose of the analysis is {\it not} to give a
convenient, practical scheme to integrate the orbital equations of
motion.  Such a scheme is not needed, since once the
self-acceleration is known, it is straightforward
%
%
to numerically
integrate the forced geodesic equations (\ref{aselfdefinition0}).
Rather, the main benefit of the analysis is to give an analytic
understanding of the dependence of the motion on $\varepsilon$ in the
limit $\varepsilon \to 0$.
This serves two purposes.  First, it acts as a foundation for the
two timescale expansion of the Einstein equation and the computation
of waveforms (Sec.\ \ref{sec:intro:waves} below and Ref.\
\cite{FH08c}).  Second, it clarifies the utility of different
approximations to the self-force that have been proposed, by
determining which pieces of the self-force contribute to the adiabatic
order and post-1-adiabatic order motions \cite{Hughes:2005qb,scalar}.
This second issue is discussed in detail in Sec.\ \ref{sec:discussion}
below.  Here we give a brief summary.


Consider first the motion to adiabatic order, given by the functions
$\psi_\alpha^{(0)}$ and ${\cal J}_\lambda^{(0)}$.  These functions are
obtained from the differential equations [Eqs.\ (\ref{jks}),
(\ref{sec2:eq:Omega0ans}) and (\ref{sec2:eq:calJ0ans}) below]
\bes
\bea
\frac{d \psi_\alpha^{(0)}}{d \ttau}(\ttau) &=& \omega_\alpha[{\cal
  J}^{(0)}_\sigma(\ttau)] , \\
\frac{d {\cal J}_\lambda^{(0)}}{d \ttau}(\ttau) &=& \left<
  G_\lambda^{(1)} \right>[{\cal
  J}^{(0)}_\sigma(\ttau)] ,
\label{minoeqn}
\eea
\label{motionans}\ees
where $\langle \ldots \rangle$ denotes the average\footnote{This phase
  space average is uniquely determined by the
dynamics of the system, and resolves concerns in the literature about
inherent ambiguities in the choice of averaging \protect{\cite{Pound:2007ti}}.}
 over the 2-torus
\be
\left<
  G_\lambda^{(1)} \right>(J_\sigma) \equiv \frac{1}{(2 \pi)^2}
\int_0^{2 \pi} dq_r \int_0^{2 \pi} dq_\theta \, G_\lambda^{(1)}(q_r,q_\theta,J_\sigma).
\label{torusavg}
\ee
This zeroth order approximation describes the inspiralling motion of
the particle.  In Sec.\ \ref{sec:consdiss} below we show that only the
dissipative (ie half retarded minus half advanced) piece of the self
force contributes to the torus average (\ref{torusavg}).
Thus, the leading order motion depends only on the dissipative
self-force, as argued by Mino \cite{Mino2003}.
Our result extends slightly that of Mino, since he advocated using an
infinite time average on the right hand side of Eq.\ (\ref{minoeqn}),
instead of the phase space or torus average.
The two
averaging methods are equivalent for generic geodesics, but not for
geodesics for which the ratio of radial and azimuthal periods is a
rational number.
The time-average prescription is therefore correct for generic geodesics, while
the result (\ref{motionans}) is valid for all geodesics.
The effect of the nongeneric geodesics is discussed in detail in
Refs.\ \cite{FH08a,FH08b}.


Consider next the subleading, post-1-adiabatic corrections to the inspiral
given by the functions $\psi_\alpha^{(1)}$ and ${\cal J}_\lambda^{(1)}$.
These corrections
are important for assessing the
accuracy of the adiabatic approximation, and will be
needed for accurate data analysis of detected signals.
The differential equations determining $\psi_\alpha^{(1)}$ and ${\cal J}_\lambda^{(1)}$
are Eqs.\ (\ref{sec2:eq:Omega1ans}) and (\ref{sec2:eq:calJ1ans}) below.
These equations depend on (i) the oscillating (not averaged) piece
of the dissipative, first order self force; (ii) the conservative
piece of the first order self force, and (iii) the torus averaged,
dissipative piece of the second order self force.
Thus, all three of these quantities will be required to compute the
inspiral to subleading order, confirming arguments made in Refs.\
\cite{Burko:2002fd,Hughes:2005qb,scalar,tanaka}.  In particular,
knowledge of the
full first order
self force will not enable computation of more accurate inspirals until
the averaged, dissipative piece of the second order self force is
known.\footnote{This statement remains true when one takes into
  account resonances \cite{FH08b}.}

\subsection{Two timescale expansion of the Einstein equations and
  adiabatic waveforms}
\label{sec:intro:waves}

We now turn to a brief description of the two timescale expansion of
the Einstein equations; more details will be given in the forthcoming
paper \cite{FH08c}.  We focus attention on a region ${\cal R}$ of
spacetime defined by the conditions (i) The distance
from the particle is large compared to its mass $\mu$; (ii) The
distance $r$ from the large black hole is
small compared to the inspiral time, $r \ll M^2 / \mu$; and (iii)
The extent of the region in time covers the entire inspiral in the
relativistic regime.  In this domain we make an ansatz for the form of
the metric that is justified a posteriori order by order.

At distances $\sim \mu$ from the particle, one needs to use a
different type of analysis (eg black hole perturbation theory for a
small black hole), and to mesh that analysis with the solution in the
region ${\cal R}$ by matching in a domain of common validity.
This procedure is very well understood and is the standard method
for deriving the first order self force
\cite{Mino:1996nk,PhysRevLett.86.1931,Poisson:2003nc}.
It is valid for our metric ansatz (\ref{metricansatz})
below since that ansatz reduces,
locally in time at each instant, to standard black hole perturbation
theory.  Therefore we do not focus on this aspect of the problem here.
Similarly, at large distances, one needs to match the solution within
${\cal R}$ onto an outgoing wave solution in order to read off the
asymptotic waveforms.\footnote{This  matching is not necessary at the
leading, adiabatic order, for certain special choices of time
coordinate in the background spacetime, as argued in Ref.\
\cite{Minoscalar}.  It is needed to higher orders.}

Within the region ${\cal R}$, our ansatz for the form of the metric
in the non-resonant case is
\bea
g_{\alpha\beta}({\bar t}, {\bar x}^j;\varepsilon) &=& g_{\alpha\beta}^{(0)}({\bar
  x}^j)
+ \varepsilon
g_{\alpha\beta}^{(1)}(q_r,q_\theta,q_\phi,{\tilde t}, {\bar x}^j)
\nonumber \\
&&+ \varepsilon^2
g_{\alpha\beta}^{(1)}(q_r,q_\theta,q_\phi,{\tilde t}, {\bar x}^j) +
O(\varepsilon^3).\ \ \ \
\label{metricansatz}
\eea
Here $g_{\alpha\beta}^{(0)}$ is the background, Kerr metric.  The
coordinates $({\bar t}, {\bar x}^j)$ can be any set of coordinates in
Kerr, subject only to the restriction that $\partial / \partial {\bar
  t}$ is the timelike Killing vector.
On the right hand side, ${\tilde t}$ is the slow time variable ${\tilde
  t}= \varepsilon {\bar t}$, and the quantities $q_r$, $q_\theta$ and $q_\phi$
are the values of the orbit's angle variables at the intersection of
the inspiralling orbit with the hypersurface ${\bar t} = $ constant.
These are functions of ${\bar t}$ and of $\varepsilon$, and can be
obtained from the solutions (\ref{aux}) and (\ref{qalpha0}) of the inspiral
problem by eliminating the proper time $\tau$.  The result is of the
form
\be
q_i({\bar t}, \varepsilon) = \frac{1}{\varepsilon} f_i^{(0)}({\tilde
  t}) + f_i^{(1)}({\tilde t}) + O(\varepsilon),
\label{qiexpand}
\ee
for some functions $f_i^{(0)}$, $f_i^{(1)}$.
On the right hand side of Eq.\ (\ref{metricansatz}),
the $O(\varepsilon^3)$ refers to an asymptotic expansion
associated with the limit $\varepsilon \to 0$ at fixed $q_i$, ${\bar x}^k$ and
${\tilde t}$.  Finally the functions $g_{\alpha\beta}^{(1)}$ and
$g_{\alpha\beta}^{(2)}$ are assumed to be multiply periodic in $q_r$,
$q_\theta$ and $q_\phi$ with period $2\pi$ in each variable.

By inserting the ansatz (\ref{metricansatz}) into Einstein's
equations, one
obtains a set of equations that determines the free functions, order
by order.  At leading order we obtain an equation of the form
\be
{\cal D} g_{\alpha\beta}^{(0)} = 0,
\ee
where ${\cal D}$ is a linear differential operator on the six
dimensional manifold with coordinates $(q_r,q_\theta,q_\phi,{\bar
  x}^j)$.  In solving this equation, ${\tilde t}$ is treated as a
constant.  The solution that matches appropriately onto the worldline
source can be written as
\bea
g^{(1)}_{\alpha\beta} &=& \frac{ \partial
  g^{(0)}_{\alpha\beta}}{\partial M} \delta M({{\tilde t}})+
\frac{ \partial g^{(0)}_{\alpha\beta}}{\partial a} \delta a({{\tilde
    t}}) + \ldots \nonumber \\
&& + {\cal F}_{\alpha\beta}[{q_r},
{q_\theta},
{q_\phi},{\bar x}^j,E({{\tilde t}}),L_z({{\tilde t}}),
Q({{\tilde t}})]. \ \  \  \ \
\label{metricans}
\eea
Here the terms on the first line are the secular pieces of the
solution.  They arise since the variable ${\tilde t}$ is treated as a
constant, and so one can obtain a solution by taking the perturbation
to the metric generated by allowing the parameters of the black hole
(mass, spin, velocity, center of mass location) to vary as arbitrary
functions of ${\tilde t}$.  For example, the mass of the black hole
can be written as $M({\tilde t}) = M + \delta M({\tilde t})$, where $M
= M(0)$ is the initial mass.  The functions $\delta M({\tilde t})$,
$\delta a({\tilde t})$ etc. are freely specifiable at this order, and will
be determined at the next (post-1-adiabatic) order.

The second line of Eq.\ (\ref{metricans}) is the oscillatory piece of the
solution.  Here one obtains a solution by taking the function ${\cal F}_{\alpha\beta}$
to be the function
$$
{\cal F}_{\alpha\beta}({q_r}, {q_\theta},
{q_\phi},{\bar x}^j,E,L_z,Q)
$$
that one obtains from standard linear perturbation theory with a geodesic source.
This function is known analytically
in Boyer-Lindquist coordinates $(t,r,\theta,\phi)$
in terms of a mode
expansion.\footnote{In coordinates ${\bar t} = t-r$, $r$, $\theta$,
  $\phi$, the
  explicit form of the asymptotic solution can be obtained by taking
  Eq.\ (3.1) of Ref.\ \protect{\cite{Drasco:2006ws}}, eliminating the
  phases $\chi_{lmkn}$ using Eq.\ (8.29) of Ref.\
  \protect{\cite{scalar}}, and making the identifications
$q_r =  \Omega_r [ t - r - t_0 + {\hat t}_r(-\lambda_{r0}) - {\hat
  t}_\theta(-\lambda_{\theta0})] - \Upsilon_r \lambda_{r0}$,
$q_\theta =  \Omega_\theta [ t - r - t_0 + {\hat t}_r(-\lambda_{r0}) - {\hat
  t}_\theta(-\lambda_{\theta0})] - \Upsilon_\theta \lambda_{\theta0}$, and
$q_\phi =  \Omega_\phi [ t - r - t_0 + {\hat t}_r(-\lambda_{r0}) - {\hat
  t}_\theta(-\lambda_{\theta0})] + \phi_0 - {\hat
  \phi}_r(-\lambda_{r0}) + {\hat \phi}_\theta(-\lambda_{\theta0})$.}${}^{,}$
\footnote{The function ${\cal F}_{\alpha\beta}$ depends on $q_\phi$
  and $\phi$ only through the combination $q_\phi - \phi$.
This allows us to show that the two-timescale form (\ref{metricansatz}) of the
metric reduces to a standard Taylor series expansion, locally in time
near almost every value $\tt_0$ of $\tt$.  For equatorial orbits there is no
dependence on $q_\theta$, and the $\varepsilon$ dependence of the
metric has the standard form up to linear order, in coordinates
$(t',r',\theta',\phi')$ defined by
$t' = (\tt - \tt_0)/\varepsilon + [ f_r^{(0)}(\tt_0)/\varepsilon]
  / \omega_{r0}$, $\phi' = \phi + \omega_{\phi0} [
  f_r^{(0)}(\tt_0)/\varepsilon] / \omega_{r0} -
  [f^{(0)}_\phi(\tt_0)/\varepsilon]$, $r'=r$, $\theta'=\theta$,
where $\omega_{r0} = f^{(0)\prime}_r(\tt_0)$, $\omega_{\phi0} =
f^{(0)\prime}_\phi(\tt_0)$, and for any number $x$, $[x] \equiv x + 2 \pi
n$ where the integer $n$ is chosen so that $0 \le [x] < 2 \pi$.
A similar construction works for circular orbits for which there is no
dependence on $q_r$.  For generic orbits a slightly more involved construction works,
but only if $\omega_{r0}/\omega_{\phi0}$ is irrational \protect{\cite{FH08c}}, which occurs
for almost every value of $\tt_0$.}

The gauge freedom in this formalism consists of those one parameter
families of diffeomorphisms which preserve the form (\ref{metricansatz}) of the
metric ansatz.  To the leading order, these consist of
(i) gauge
transformations of the background coordinates that are independent of
$\varepsilon$, which preserve the timelike Killing vector, and (ii)
transformations of the form
\be
x^\alpha \to x^\alpha + \varepsilon
\xi^\alpha(q_r,q_\theta,q_\phi,{\tilde t}, x^j) + O(\varepsilon^2).
\ee
Note that this is {\it not} the standard gauge freedom of linear
perturbation theory, since $\xi^\alpha$ depends on 4 ``time
variables'' instead of one.
This modified gauge group allows the two timescale method to evade the
two problems discussed at the end of Sec.\ \ref{ssec:methods} above,
since the gradual evolution is described entirely by the $\tt$
dependence, and, at each fixed $\tt$, the leading order dependence on
the variables $q_r$, $q_\theta$, $q_\phi$, $r$, $\theta$ and $\phi$ is
the same as in standard perturbation theory with the same gauge
transformation properties.

\subsection{Organization of this Paper}

The organization of this paper is as follows. In Sec.\ \ref{sec:kerrapplication} we derive
the fundamental equations describing the inspiral of a point particle
into a Kerr black hole in terms of generalized action-angle variables.
In Sec.\ \ref{sec:generalsystem}
we define a class of general, weakly perturbed
dynamical systems of which the inspiral
motion in Kerr is a special case.
We then study the solutions of this class of systems using two-timescale expansions,
first for a single degree of freedom in Sec.\ \ref{sec:derivation_single}, and then for the general case
in Sec.\ \ref{sec2:manyvariables}.
Section \ref{sec:numerics} gives an example of a numerical integration of a system of this kind,
and Sec.\ \ref{sec:discussion} gives the final discussion and conclusions.

\subsection{Notation and Conventions}

Throughout this paper we use units with $G = c = 1$.  Lower case Roman
indices $a,b,c, \ldots $ denote abstract
indices in the sense of Wald \cite{Wald}.  We use these indices both
for tensors on spacetime and for tensors on the eight dimensional phase space.
Lower case Greek indices $\nu,\lambda, \sigma, \tau, \ldots$ from the
middle of the alphabet denote components of spacetime tensors on a
particular coordinate system; they thus transform under spacetime
coordinate transformations.  They run over $0,1,2,3$.
Lower case Greek indices $\alpha, \beta, \gamma \ldots$ from the
start of the alphabet label position or momentum coordinates on 8
dimensional phase space that are not associated with coordinates on
spacetime. They run over $0,1,2,3$ and do not transform
under spacetime coordinate transformations.
In Sec.\ \ref{sec2:manyvariables}, and just in that section,
indices $\alpha,\beta,\gamma,\delta,\varepsilon, \ldots$ from the
start of the Greek alphabet run over
$ 1 \ldots N$, and indices $\lambda,\mu,\nu,\rho,\sigma,\ldots$ from
the second half of the alphabet run over $1 \ldots M$.
Bold face quantities generally denote vectors, as in $\bfJ = (J_1,
\ldots, J_M)$, although in Sec.\ \ref{sec:kerrapplication} the bold
faced notation is used for differential forms.

\section{EXTREME-MASS RATIO INSPIRALS IN KERR
 FORMULATED USING ACTION-ANGLE VARIABLES}
\label{sec:kerrapplication}

In this section we derive the form of the fundamental equations
describing the inspiral of a point particle into a Kerr black hole,
using action-angle type variables.  Our final result is given in Eqs.\
(\ref{eq:eomsimplified}) below, and the properties of the solutions of
these equations are analyzed in detail in the remaining sections of
this paper.

The description of geodesic motion in Kerr in terms of action angle
variables was first given by Schmidt \cite{Schmidt}, and has been
reviewed by Glampedakis and Babak \cite{Glampedakis:2005cf}.
We follow closely Schmidt's treatment, except that we work in an eight
dimensional phase space instead of a six dimensional phase space, thus
treating the time and spatial variables on an equal footing.
We also clarify the extent to which the fundamental frequencies of
geodesic motion are uniquely determined and gauge invariant, as
claimed by Schmidt.

We start in subsection \ref{sec:actionanglereview} by reviewing the
geometric definition of action angle variables in Hamiltonian
mechanics, which is based on the Liouville-Arnold theorem \cite{Arnold}.
This definition does not apply to geodesic motion in Kerr, since the
level surfaces defined by the conserved quantities in the eight
dimensional phase space are non-compact.
In subsection \ref{sec:noncompact} we discuss how generalized action
angle variables can be defined for non-compact level surfaces, and in
subsection \ref{subsec:Kerrorbits} we apply this to give a
coordinate-independent construction of generalized action angle
variables for generic bound geodesics in Kerr.
Subsection \ref{sec:kerrexplicit} specializes to Boyer-Lindquist
coordinates on phase space, and describes explicitly, following Schmidt
\cite{Schmidt}, the explicit canonical transformation from those
coordinates to the generalized action angle variables.

We then turn to using these variables to describe a radiation-reaction
driven inspiral.  In subsection \ref{sec:eomderive} we derive the
equations of motion in terms of the generalized action angle
variables.  These equations define a flow on the eight dimensional
phase space, and do not explicitly exhibit the conservation of rest
mass.  In subsection \ref{sec:backreaction} we therefore switch to a
modified set of variables and equations in which the conservation of
rest mass is explicit.  We also augment the equations to describe the
backreaction of gravitational radiation passing through the horizon of
the black hole.

\subsection{Review of action-angle variables in geometric Hamiltonian mechanics}
\label{sec:actionanglereview}

We start by recalling the standard geometric framework for Hamiltonian
mechanics \cite{Arnold}.  A Hamiltonian system consists of a $2N$-dimensional
differentiable manifold ${\cal M}$ on which there is defined a smooth
function $H$ (the Hamiltonian), and a non-degenerate 2-form
$\Omega_{ab}$ which is closed, $\nabla_{[a} \Omega_{bc]} =0$.
Defining the tensor $\Omega^{ab}$ by $\Omega^{ab} \Omega_{bc} =
\delta^a_c$, the Hamiltonian vector field is defined as
\be
v^a = \Omega^{ab} \nabla_b H,
\ee
and the integral curves of this vector fields give the motion of the
system.  The two form $\Omega_{ab}$ is called the symplectic
structure.  Coordinates $(q_\alpha,p_\alpha)$ with $1 \le \alpha \le
N$ are called symplectic coordinates if the symplectic structure can
be written as
$
{\bf \Omega} = dp_\alpha \wedge dq_\alpha,
$
i.e. $\Omega_{ab} = 2 \nabla_{[a} p_\alpha \nabla_{b]} q_\alpha$.

We shall be interested in systems that possess $N-1$ first integrals of
motion which, together with the Hamiltonian $H$, form a complete set
of $N$ independent first integrals.
We denote these first integrals by $P_\alpha$, $1 \le \alpha
\le N$, where $P_1 = H$.  These quantities are functions on
${\cal M}$ for which the Poisson
brackets
\be
\{P_\alpha, H\} \equiv \Omega^{ab} (\nabla_a P_\alpha) (\nabla_b
H)
\label{eq:conserved0}
\ee
vanish for $1 \le \alpha \le N$.  If the first integrals satisfy
the stronger condition that all the Poisson brackets vanish,
\be
\{ P_\alpha, P_\beta\}=0
\label{eq:involution}
\ee
for $1 \le \alpha,\beta \le N$, then the first integrals are said to
be in involution.  If the 1-forms
$
\nabla_a P_\alpha
$
for $1 \le \alpha \le N$ are linearly independent, then the first
integrals are said to be independent.
A system is said to be {\it completely integrable} in some open region ${\cal U}$
in ${\cal M}$ if there exist $N$ first integrals which are
independent and in involution at every point of ${\cal U}$.

For completely integrable systems, the phase space ${\cal M}$ is
foliated by invariant level sets of the first integrals.
For a given set of values ${\bf p} = (p_1, \ldots, p_N)$, we define
the level set
\be
{\cal M}_{\bf p} = \left\{ x \in {\cal M} \right| \left. P_\alpha(x) =
p_\alpha, 1 \le \alpha \le N \right\},
\ee
which is an $N$-dimensional submanifold of ${\cal M}$.
The level sets are invariant under the Hamiltonian flow by Eq.\
(\ref{eq:conserved0}).
Also the pull back of the symplectic structure $\bfOmega$ to ${\cal
  M}_{\bf p}$ vanishes, since the vector fields ${\vec v}_\alpha$ defined by
\be
v^a_\alpha = \Omega^{ab} \nabla_b P_\alpha
\label{eq:vecdef}
\ee
for $1 \le \alpha \le N$
form a basis of the tangent space to ${\cal M}_{\bf p}$ at each point,
and satisfy $\Omega_{ab} v^a_{\alpha} v^b_\beta=0$ for $1 \le
\alpha,\beta \le N$ by Eq.\ (\ref{eq:involution}).

A classic theorem of mechanics, the Liouville-Arnold theorem
\cite{Arnold}, applies to systems which are completely integrable in
a neighborhood of some level set ${\cal M}_{\bf p}$ that is connected
and compact.  The theorem says that
\begin{itemize}
\item The level set ${\cal M}_{\bf p}$ is diffeomorphic to an $N$-torus
  $T^N$.  Moreover there is a neighborhood ${\cal V}$ of ${\cal
    M}_{\bf p}$ which is diffeomorphic to the product $T^N \times {\cal B}$
  where ${\cal B}$ is an open ball, such that the level sets are the
  $N$-tori.

\item There exist symplectic coordinates $(q_\alpha, J_\alpha)$ for $1 \le
  \alpha \le N$ (action-angle variables) on ${\cal V}$ for which the angle variables
  $q_\alpha$ are periodic, $$ q_\alpha + 2 \pi \equiv q_\alpha,$$ and
  for which the first integrals depend only on the action variables,
  $P_\alpha = P_\alpha(J_1, \ldots , J_N)$ for $1 \le \alpha \le N$.
\end{itemize}

An explicit and coordinate-invariant prescription for computing a set
of action variables $J_\alpha$ is as follows \cite{Arnold}.
A symplectic potential $\bfTheta$ is a
1-form which satisfies $d \bfTheta = {\bf \Omega}$.  Since the
2-form ${\bf \Omega}$ is closed, such 1-forms always exist locally.
For example, in any local symplectic coordinate system
$(q_\alpha,p_\alpha)$, the 1-form $\bfTheta = p_\alpha dq_\alpha$ is a
symplectic potential.
It follows from the hypotheses of the Liouville-Arnold theorem that
there exist symplectic potentials that are defined on a neighborhood
of ${\cal M}_{\bf p}$ \cite{ActionAngle}.
The first homotopy group $\Pi_1({\cal M}_{\bf p})$ is
defined to be the set of equivalence classes of loops on ${\cal
  M}_{\bf p}$, where two loops are equivalent if one can be
continuously deformed into the other.  Since ${\cal M}_{\bf p}$ is
diffeomorphic to the $N$-torus, this group
is isomorphic to $({\bf Z}^N,+)$, the group of $N$-tuples of integers
under addition.  Pick a set of generators $\gamma_1,
\ldots, \gamma_N$ of $\Pi_1({\cal M}_{\bf p})$, and for each loop
$\gamma_\alpha$ define
\be
J_\alpha = \frac{1}{2 \pi} \int_{\gamma_\alpha} \bfTheta.
\label{eq:actionvar}
\ee
This integral is independent of the choice of symplectic potential
$\bfTheta$.\footnote{The type of argument used
  in Ref.\ \protect{\cite{ActionAngle}} can be used to show that the
  pullback to ${\cal M}_{\rm p}$ of the difference between two
  symplectic potentials is exact since it is closed.}  It is also
independent of the
choice of loop $\gamma_\alpha$ in the equivalence class of the generator of
$\Pi_1({\cal M}_{\bf p})$, since if $\gamma_\alpha$ and
$\gamma_\alpha^\prime$ are two equivalent loops, we have
\be
\int_{\gamma_\alpha} \bfTheta - \int_{\gamma_\alpha^\prime} \bfTheta =
\int_{\partial {\cal R}} \bfTheta = \int_{\cal R} d \bfTheta = \int_
{\cal R} {\bf \Omega} =0.
\ee
Here ${\cal R}$ is a 2-dimensional surface in ${\cal M}_{\bf p}$ whose
boundary is $\gamma_\alpha - \gamma_\alpha^\prime$, we have used
Stokes theorem, and in the last equality we have used the fact that
the pull back of ${\bf \Omega}$ to the level set ${\cal M}_{\bf p}$
vanishes.

Action-angle variables for a given system are not unique
\cite{NonUnique}.  There is a
freedom to redefine the coordinates via
\be
q_\alpha \to A_{\alpha\beta} q_\beta, \ \ \ \ \ J_\alpha \to
B_{\alpha\beta} J_\beta,
\label{eq:redefine1}
\ee
where $A_{\alpha\beta}$ is a constant matrix of integers with
determinant $\pm 1$, and $A_{\alpha\beta} B_{\alpha\gamma} =
\delta_{\beta\gamma}$.  This is just the freedom present in choosing a
set of generators of the group $\Pi_1({\cal M}_{\bf p}) \sim ({\bf Z}^N,+)$.
Fixing this freedom requires the specification of some additional
information, such as a choice of coordinates on the torus;
once the coordinates $q_\alpha$ are chosen, one can take the loops $\gamma_\alpha$
to be the curves $q_\beta = $ constant for $\beta\ne\alpha$.
There is also a freedom to redefine the origin of the angle
variables separately on each torus:
\be
q_\alpha \to q_\alpha + \frac{ \partial
  Z(J_\beta) }{ \partial J_\alpha}, \ \ \ \ \ \ J_\alpha \to
J_\alpha.
\label{eq:redefine2}
\ee
Here $Z(J_\beta)$ can be an arbitrary function of the action
variables.

\subsection{Generalized action-angle variables for non-compact level sets}
\label{sec:noncompact}

One of the crucial assumptions in the Liouville-Arnold theorem is that
the level set ${\cal M}_{\bf p}$ is compact.  Unfortunately, this
assumption is not satisfied by the dynamical system of bound orbits
in Kerr which we discuss in Sec.\ \ref{subsec:Kerrorbits} below,
because we will work in the 8
dimensional phase space and the motion is not bounded in the time direction.
We shall therefore use instead a generalization of the Liouville-Arnold
theorem to non-compact level sets, due to
Fiorani, Giachetta and Sardanashvily \cite{ActionAngle}.

Consider a Hamiltonian system which is completely integrable in a
neighborhood ${\cal U}$ of a connected level set ${\cal M}_{\bf p}$,
for which the $N$ vector fields (\ref{eq:vecdef}) are complete on ${\cal U}$, and for which
the level sets ${\cal M}_{{\bf p}'}$ foliating ${\cal U}$ are all
diffeomorphic to one another.  For such systems Fiorani
et.\ al.\ \cite{ActionAngle} prove that

\begin{itemize}
\item There is an integer $k$ with $0 \le k \le N$ such that the
  level set ${\cal M}_{\bf p}$ is diffeomorphic to the product
$T^k \times {\bf R}^{N-k}$, where ${\bf R}$ is the set of real numbers.
Moreover there is a neighborhood ${\cal V}$ of ${\cal
    M}_{\bf p}$ which is diffeomorphic to the product $T^k \times {\bf
    R}^{N-k} \times {\cal B}$ where ${\cal B}$ is an open ball.

\item There exist symplectic coordinates $(q_\alpha, J_\alpha)$ for $1 \le
  \alpha \le N$ (generalized action-angle variables) on ${\cal V}$
  for which the first $k$ variables
  $q_\alpha$ are periodic, $$ q_\alpha + 2 \pi \equiv q_\alpha,\ \ \
  \ 1 \le \alpha \le k,$$ and
  for which the first integrals depend only on the action variables,
  $P_\alpha = P_\alpha(J_1, \ldots , J_N)$ for $1 \le \alpha \le N$.
\end{itemize}
\noindent
Thus, there are $k$ compact dimensions in the level sets, and $N-k$
non-compact dimensions.  In our application to Kerr below, the values
of these parameters will be $k=3$ and $N-k=1$.

The freedom in choosing generalized action-angle variables is larger
than the corresponding freedom for action-angle variables discussed above.
The first $k$ action variables can be computed in the same way as
before, via the integral (\ref{eq:actionvar}) evaluated on a set of
generators $\gamma_1, \ldots, \gamma_k$ of $\Pi_1({\cal M}_{\bf p})$,
which in this case is isomorphic to $({\bf Z}^k,+)$.
This prescription is unique up to a group of redefinitions of the form
[cf.\ Eq.\ (\ref{eq:redefine1})
above]
\be
q_\alpha \to \sum_{\beta=1}^k A_{\alpha\beta} q_\beta, \ \ \ \ \ J_\alpha \to
\sum_{\beta=1}^k B_{\alpha\beta} J_\beta,
\label{eq:redefine3}
\ee
for $1 \le \alpha \le k$, where the $k\times k$ matrix
$A_{\alpha\beta}$ is a constant matrix of integers with
determinant $\pm 1$, and $A_{\alpha\beta} B_{\alpha\gamma} =
\delta_{\beta\gamma}$.
There is additional freedom
present in the choice of the rest of the action variables $J_{k+1},
\ldots, J_N$.  As a consequence, the remaining freedom in choosing generalized
action-angle variables consists of the transformations
(\ref{eq:redefine2}) discussed earlier,
together with transformations of the form
\be
q_\alpha \to A_{\alpha\beta} q_\beta, \ \ \ \ \ J_\alpha \to
B_{\alpha\beta} J_\beta,
\label{eq:redefine4}
\ee
where $A_{\alpha\beta}$ and $B_{\alpha\beta}$ are constant real
$N\times N$ matrices with $A_{\alpha\beta} B_{\alpha\gamma} =
\delta_{\beta\gamma}$ such that $J_1, \ldots, J_k$ are preserved.

In generalized action-angle variables, the equations of motion take
the simple form
\be
{\dot q}_\alpha = \frac{\partial H(\bfJ) }{\partial J_\alpha}
\label{sec22:eq:q_unperturbedeom}
\ee
and
\be
{\dot J}_\alpha =- \frac{\partial H(\bfJ) }{ \partial q_\alpha} = 0.
\label{sec22:eq:J_unperturbedeom}
\ee
We define the quantities
\be
\Omega_\alpha(\bfJ) \equiv \frac{\partial H(\bfJ) }{ \partial J_\alpha},
\label{sec22:eq:omegadef0}
\ee
which are angular frequencies for $1 \le \alpha \le k$ but not for
$k+1 \le \alpha \le N$.
The solutions of the equations of motion are then
\bes
\bea
q_\alpha(t) &=& \Omega_\alpha(\bfJ_0) t + q_{\alpha0} \\
J_\alpha(t) &=& J_{\alpha0},
\eea
\ees
for some constants $\bfJ_0$ and $\bfq_0$.

\subsection{Application to bound geodesic motion in Kerr}
\label{subsec:Kerrorbits}

We now apply the general theory discussed above to give a
coordinate-invariant definition of action-angle variables
for a particle on a bound orbit in the Kerr spacetime.
We denote by $({\cal M}_{\rm K}, g_{ab}$) the Kerr spacetime, and we denote
by $\xi^a$ and $\eta^a$ the timelike and axial Killing vector fields.
The cotangent bundle over ${\cal M}_{\rm K}$ forms an 8-dimensional phase
space
$
{\cal M} = T^* {\cal M}_{\rm K}.
$
Given any coordinate system $x^\nu$ on the Kerr spacetime, we can
define a coordinate system $(x^\nu,p_\nu)$ on ${\cal M}$, such
that the point $(x^\nu,p_\nu)$ corresponds to the covector or one form
$p_\nu d x^\nu$ at $x^\nu$ in ${\cal
  M}_{\rm K}$.
The natural symplectic structure on ${\cal M}$
is then defined by demanding that all such coordinate systems
$(x^\nu,p_\nu)$ be symplectic \cite{Arnold}.
The Killing vector fields $\xi^a$ and $\eta^a$ on ${\cal M}_{\rm K}$
have natural extensions to vector fields on phase space which Lie
derive the symplectic structure.

Consider now a particle of mass $\mu$ on a bound geodesic orbit.
A Hamiltonian on ${\cal M}$ that generates geodesic motion is given by
\be
H(x^\nu,p_\nu)=\frac{1}{2}g^{\nu\sigma}(x^\rho)p_\nu p_\sigma;
\label{sec4:eq:hamiltonian}
\ee
this definition is independent of the choice of coordinate system
$x^\nu$.  If we interpret $p_\nu$ to be the 4-momentum of the
particle, then the conserved value of $H$ is $-\mu^2/2$, and the
evolution parameter is the affine parameter $\lambda = \tau / \mu$
where $\tau$ is proper time.

As is well known, geodesics on Kerr possess three first integrals,
the energy $E = - \xi^a p_a$, the z-component of angular
momentum $L_z = \eta^a p_a$, and Carter constant $Q = Q^{ab} p_a
p_b$ where $Q^{ab}$ is a Killing tensor \cite{Carter}.
Together with
the Hamiltonian we therefore have four first integrals:
\be
P_\alpha = (P_0,P_1,P_2,P_3) = (H,E,L_z,Q).
\label{eq:Palphadef}
\ee
An explicit computation of the 4-form $dH \wedge dE \wedge dL_z \wedge
dQ$ on ${\cal M}$ shows
that it is non vanishing for bound orbits except for the degenerate
cases of circular (i.e.\ constant Boyer-Lindquist radial coordinate)
and equatorial orbits.  Also the various Poisson brackets
$\{P_\alpha,P_\beta\}$ vanish: $\{E,H\}$ and $\{L_z,H\}$ vanish since $\xi^a$
and $\eta^a$ are Killing fields, $\{E,L_z\}$ vanishes since these
Killing fields commute, $\{Q,H\}$ vanishes since $Q^{ab}$ is a Killing
tensor, and finally $\{E,Q\}$ and $\{L_z,Q\}$ vanish since the
Killing tensor is invariant under the flows generated by $\xi^a$ and $\eta^a$.
Therefore for generic orbits the theorem due to Fiorani et.\ al.\ discussed in the last
subsection applies.\footnote{One can check that the two other
  assumptions in the theorem listed in the second paragraph of Sec.\
  \protect{\ref{sec:noncompact}} are satisfied.}  The relevant
parameter values are $k=3$ and $N=4$, since
the level sets ${\cal M}_{\bf p}$ are non-compact in
the time direction only.  Thus geodesic motion can be
parameterized in terms of generalized action-angle variables.

We next discuss how to resolve in this context the non-uniqueness in the
choice of generalized action angle
variables discussed in the last subsection.  Consider first the freedom
(\ref{eq:redefine3}) associated with the choice of generators of
$\Pi_1({\cal M}_{\bf p})$.  One of these generators can be chosen to
be an integral curve of the extension to ${\cal M}$ of the axial
Killing field $\eta^a$.
The other two can be
chosen as follows.  Let $\pi : {\cal M} \to {\cal M}_{\rm K}$ be the
natural projection from phase space ${\cal M}$ to spacetime ${\cal
  M}_{\rm K}$ that takes $(x^\nu,p_\nu)$ to $x^\nu$.  A loop
$(x^\nu(\lambda),p_\nu(\lambda) )$ in the level set ${\cal
  M}_{\rm p}$ then projects to the curve $x^\nu(\lambda)$ in
$\pi({\cal M}_{\rm p})$.  Requiring that this curve
intersect the boundary of $\pi({\cal
  M}_{\rm p})$ only twice determines the two other generators of
$\Pi_1({\cal M}_{\bf p})$.\footnote{This excludes, for example, loops
  which wind around twice in the $r$ direction and once in the $\theta
 $ direction.}  The resulting three generators coincide
with the generators obtained from the motions in the $r$, $\theta$ and
$\phi$ directions in Boyer-Lindquist coordinates \cite{Schmidt}.
We denote the resulting generalized action-angle variables by
$(q_t,q_r,q_\theta,q_\phi,J_t,J_r,J_\theta,J_\phi)$.

The remaining ambiguity (\ref{eq:redefine4}) is of the form
\be
J_i \to J_i,\ \ \ \ J_t \to \gamma J_t + v^i J_i,
\ee
where $i$ runs over $r$, $\theta$ and $\phi$ and the parameters
$\gamma$ and $v^i$ are arbitrary.  The corresponding transformation of
the frequencies (\ref{sec22:eq:omegadef0}) is
\be
\Omega_t \to \gamma^{-1} \Omega_t, \ \ \ \ \Omega_i \to \Omega_i -
\gamma^{-1} v^i \Omega_t.
\ee
A portion of this ambiguity (the portion given by $\gamma=1$, $v^r =
v^\theta =0$) is that associated with the choice of rotational frame,
$\phi \to \phi + \Omega t$ where $\Omega$ is an angular velocity.
It is not possible to eliminate this rotational-frame ambiguity using only
the spacetime geometry in a neighborhood of the orbit.
In this sense, the action angle variables are not
uniquely determined by local geometric information.
However, we can resolve the ambiguity using global geometric information, by
choosing
\be
J_t = \frac{1}{2\pi} \int_{\gamma_t} \bfTheta,
\label{eq:geometric_Jt_def}
\ee
where $\gamma_t$ is an integral curve of length $2 \pi$ of the extension to
${\cal M}$ of the timelike Killing field $\xi^a$.\footnote{The Killing
field $\xi^a$ encodes global geometric information since it is
defined to be timelike and of unit norm at spatial infinity.}
The definition (\ref{eq:geometric_Jt_def}) is independent of the
choice of such a curve $\gamma_t$ and of the choice of symplectic
potential $\bfTheta$.

To summarize, we have a given a coordinate-invariant definition of the
generalized action-angle variables
$(q_t,q_r,q_\theta,q_\phi,J_t,J_r,J_\theta,J_\phi)$ for generic
bound orbits in Kerr.  These variables are uniquely determined up to
relabeling and up to the residual ambiguity (\ref{eq:redefine2}).
A similar construction has been given by Schmidt \cite{Schmidt},
except that Schmidt first projects out the time direction of the level
sets, and then defines three action variables $(J_r,J_\theta,J_\phi)$
and three angle variables $(q_r,q_\theta,q_\phi)$.

\subsection{Explicit expressions in terms of Boyer-Lindquist coordinates}
\label{sec:kerrexplicit}

In Boyer-Lindquist coordinates $(t,r,\theta,\phi)$, the Kerr
metric is
\begin{eqnarray}
ds^2 &=&
- \left( 1-\frac{2Mr}{\Sigma} \right) ~dt^2
+ \frac{\Sigma}{\Delta}~dr^2
+ \Sigma~d\theta^2 \nonumber \\
&&+ \left( r^2+a^2 + \frac{2Ma^2r}{\Sigma}\sin^2\theta \right)\sin^2\theta~d\phi^2 \nonumber \\
&&- \frac{4Mar}{\Sigma}\sin^2\theta~dt~d\phi,
\label{sec4:eq:kerrmetric}
\end{eqnarray}
where
\begin{align}
& \Sigma = r^2 + a^2\cos^2\theta, & \Delta = r^2 - 2Mr + a^2, &
\end{align}
and $M$ and $a$ are the black hole
mass and spin parameters.
The timelike and axial Killing fields are ${\vec \xi} = \partial/\partial t$ and
${\vec \eta} = \partial / \partial \phi$, and so the energy and angular momentum are
\bes
\label{eq:capitalPs}
\be
E=-\vec \xi\cdot \vec p=-p_t
\label{EKilling}
\ee
and
\be
L_z =\vec \eta\cdot \vec p =p_{\phi}.
\label{LKilling}
\ee
The Carter constant is given by \cite{Carter}
\bea
Q&=&p_{\theta}^2+a^2 \cos^2\theta \left(\mu^2-p_t^2\right)+\cot^2
\theta p_\phi^2,
\eea
and the Hamiltonian (\ref{sec4:eq:hamiltonian}) is
\bea
H &=& \frac{\Delta}{2 \Sigma} p_r^2 + \frac{1}{2 \Sigma} p_\theta^2 +
\frac{(p_\phi + a \sin^2 \theta p_t)^2 }{ 2\Sigma \sin^2 \theta}
\nonumber \\
&&-
\frac{\left[ (r^2 + a^2) p_t + a p_\phi \right]^2}{2 \Sigma \Delta}.
\label{eq:HamiltonianKerr}
\eea
\ees

Following Schmidt \cite{Schmidt}, we can obtain an invertible
transformation from the Boyer-Lindquist phase space coordinates
$(x^\nu,p_\nu)$ to the generalized action angle variables
$(q_\alpha, J_\alpha)$ as follows.  Equations (\ref{eq:capitalPs}) can
be inverted to express the momenta $p_\nu$ in terms of $x^\nu$
and the four first integrals
\be
P_\alpha = (H,E,L_z,Q) = \left(- \frac{1}{2}\mu^2,E,L_z,Q\right)
\label{firstintegrals}
\ee
up to some signs
\cite{Carter}:
\be
p_t=-E,{\;}{\;}{\;}p_{\phi}=L_z,{\;}{\;}{\;}p_r=\pm\frac{\sqrt{V_r(r)}}{\Delta},{\;}{\;}{\;}p_{\theta}=\pm\sqrt{V_\theta(\theta)}.
\label{sec4:eq:momenta}
\ee
Here the potentials $V_r(r)$ and $V_\theta(\theta)$ are defined
by
\bes
\bea
V_r(r)&=&\left[(r^2+a^2)E-aL_z\right]^2 \nn \\
&&-\Delta\left[\mu^2r^2+(L_z-aE)^2+Q\right],\\
V_\theta(\theta) &=&
Q-\left[(\mu^2-E^2)a^2+\frac{L_z^2}{\sin^2\theta}\right]\cos^2\theta.\ \ \
\eea
\label{sec4:eq:potentials}\ees
Using these formulae together with the symplectic potential $\bfTheta
= p_\nu dx^\nu$ in the definitions (\ref{eq:actionvar}) and
(\ref{eq:geometric_Jt_def})
gives
\bes
\bea
J_r&=&\frac{1}{2\pi}\oint\frac{\sqrt{V_r}}{\Delta}dr\\
J_\theta&=&\frac{1}{2\pi}\oint\sqrt{V_\theta}d\theta\\
J_\phi&=&\frac{1}{2\pi}\oint p_{\phi}d\phi=L_z\\
J_t &=& \frac{1}{2\pi} \int_0^{2\pi} p_t dt = - E.
\label{eq:Jtdef}
\eea
\label{sec4:eq:BLactions}\ees
These expressions give the action variables as functions of the first
integrals, $J_\alpha = J_\alpha(P_\beta)$.  The theorem discussed in
Sec.\ \ref{sec:noncompact} above guarantees that these relations can be inverted to
give
\be
P_\alpha = P_\alpha(J_\beta).
\label{eq:JtoP}
\ee

Next, to obtain expressions for the corresponding generalized angle variables, we
use the canonical transformation from the symplectic coordinates
$(x^\nu,p_\nu)$ to $(q_\alpha,J_\alpha)$ associated with
a general solution of the Hamilton Jacobi
equation
\be
H\left[x^{\nu},\frac{\partial {\cal S}}{\partial
x^\nu}\right] + \frac{\partial {\cal S}}{\partial
\lambda} =0.
\label{sec4:eq:jacobi}
\ee
As shown by Carter \cite{Carter}, this equation is separable and
the general solution\footnote{As indicated by the $\pm$ signs in Eq.\
(\ref{sec4:eq:completeintegral}), there are actually four
different solutions, one on each of the four coordinate
patches on which $(x^\nu,P_\alpha)$ are good coordinates,
namely ${\rm sgn}(p_r) = \pm 1$, ${\rm sgn}(p_\theta) = \pm 1$.}
can be written in terms of the first
integrals $P_\alpha$
\be
{\cal S}(x^\nu,P_\alpha,\lambda)= - H \lambda + {\cal W}(x^\nu,P_\alpha)
\label{sec4:eq:completeintegral0}
\ee
where $H = - \mu^2/2$,
\be
{\cal W}(x^\nu,P_\alpha)= -Et+L_z\phi \pm {\cal W}_r(r) \pm {\cal W}_{\theta}(\theta),
\label{sec4:eq:completeintegral}
\ee
\be
{\cal W}_r(r) = \int^r dr \frac{ \sqrt{V_r}}{\Delta},
\ee
and
\be
{\cal W}_\theta(\theta) = \int^\theta d\theta  \sqrt{V_\theta}.
\ee
Using the relation (\ref{eq:JtoP}) the function ${\cal W}$ can be expressed in terms
of the Boyer-Lindquist coordinates $x^\nu$ and the action variables
$J_\alpha$:
\be
{\cal W} = {\cal W}(x^\nu,J_\alpha).
\ee
This is a type II generating function that generates the required
canonical transformation from $(x^\nu,p_\nu)$ to $(q_\alpha,J_\alpha)$:
\bes
\bea
\label{eq:ct1}
p_\nu &=& \frac{\partial {\cal W}}{\partial x^\nu}(x^\nu,J_\beta) \\
q_\alpha &=& \frac{\partial {\cal W}}{\partial J_\alpha}(x^\nu,J_\beta).
\label{eq:ct2}
\eea
\ees
Equation (\ref{eq:ct1})
is already satisfied by virtue of the
definition (\ref{sec4:eq:completeintegral}) of ${\cal W}$ together with Eqs.\
(\ref{sec4:eq:momenta}).  Equation (\ref{eq:ct2}) furnishes the
required formulae for the generalized angle variables
$q_\alpha$.\footnote{The freedom (\ref{eq:redefine2}) to redefine the
  origin of the angle
variables on each torus is just the freedom to add to ${\cal W}$ any
function of $P_\alpha$.  We choose to resolve this freedom by demanding
that $q_r=0$ at the minimum value of $r$, and $q_\theta=0$ at the
minimum value of $\theta$.}

Although it is possible in principle to express the first integrals
$P_\alpha$ in terms of the action variables $J_\alpha$ using Eqs.\
(\ref{sec4:eq:BLactions}), it is not possible to obtain explicit
analytic expressions for $P_\alpha(J_\beta)$.  However, as pointed out
by Schmidt \cite{Schmidt}, it is possible to obtain explicit
expressions for the partial derivatives $\partial P_\alpha / \partial
J_\beta$, and this is sufficient to compute the frequencies
$\Omega_\alpha$.  We review this in appendix \ref{appendix:frequencies}.

\subsection{Application to slow inspiral motion in Kerr}
\label{sec:eomderive}

The geodesic equations of motion in terms of the generalized action angle
variables $(q_\alpha,J_\alpha)$ are [cf.\ Eqs.\
(\ref{sec22:eq:q_unperturbedeom}) -- (\ref{sec22:eq:omegadef0}) above]
\bes
\bea
\label{eq:qdotnoac}
\frac{d q_\alpha}{d \lambda} &=& \Omega_\alpha(J_\beta), \\
\frac{d J_\alpha}{d \lambda} &=& 0,
\eea
\ees
for $0 \le \alpha \le 3$.
Here $\lambda = \tau/\mu$ where $\tau$ is proper time and $\mu$ is the mass of the particle.
In this section we derive the modifications to these equations
required to describe the radiation-reaction driven inspiral of a
particle in Kerr.
Our result is of the form
\bes
\bea
\label{qdotforced}
\frac{dq_\alpha}{d\lambda} &=&\Omega_\alpha(J_\beta) + \mu^2
f_\alpha(q_\beta,J_\beta), \\
\frac{d J_\alpha}{d\lambda} &=& \mu^2 F_\alpha( q_\beta,J_\beta).
\label{Jdotforced}
\eea
\label{eq:actionangleeqns}\ees
We will derive explicit expressions for the forcing terms $f_\alpha$
and $F_\alpha$ in these equations.

The equation of motion for a particle subject to a self-acceleration
$a^\nu$ is
\be
\frac{d^2x^\nu}{d\lambda^2}+\Gamma^\nu_{\sigma\rho}\frac{dx^\sigma}{d\lambda}
\frac{dx^\rho}{d\lambda}=\mu^2 a^{\nu}.
\label{aselfdefinition}
\ee
Rewriting this second order
equation as two first order equations
allows us to use the Jacobian of the coordinate transformation
$\{x^\nu,p_\nu\} \to \{q_\alpha, J_\alpha\}$ to relate the
forcing terms for the two sets of variables:
\bes
\bea
\label{geodesicI}
\frac{dx^{\nu}}{d\lambda}&=& g^{\nu\sigma} p_{\sigma},\\
\frac{dp_\nu}{d\lambda}&=&- \frac{1}{2}
g^{\sigma\rho}_{\ \ ,\nu} p_\sigma p_\rho
+\mu^2 a_\nu.
\eea
\label{geodesic}\ees

We start by deriving the equation of motion for the action variables
$J_\alpha$.  Taking a derivative with respect to $\lambda$ of the
relation $J_\alpha = J_\alpha(x^\nu,p_\nu)$
and using Eqs.\ (\ref{geodesic}) gives
\bea
\frac{dJ_\alpha}{d\lambda}&=&\frac{\partial
J_\alpha}{\partial x^\nu}p^\nu+
\frac{\partial J_\alpha}{\partial p_\nu}
\frac{dp_\nu}{d\lambda}\nonumber\\
&=&\left[\frac{\partial J_\alpha}{\partial x^\nu} g^{\nu\sigma} p_\sigma
- \frac{1}{2} \frac{\partial J_\alpha}{\partial p_\nu}
g^{\sigma\rho}_{\ \ ,\nu} p_\sigma p_\rho
\right] \nn \\
&& +\mu^2 \frac{\partial J_\alpha}{\partial p_\nu}a_\nu.
\label{Jdot}
\eea
The term in square brackets must vanish identically since $J_\alpha$
is conserved in the absence of any acceleration $a_\nu$.
Rewriting the second term using $J_\alpha = J_\alpha(P_\beta)$ and the
chain rule gives
an equation of motion of the form (\ref{Jdotforced}), where the forcing terms
$F_\alpha$ are
\be
F_\alpha
= \frac{\partial
J_\alpha}{\partial P_\beta} \left( \frac{\partial P_\beta}{\partial
p_\nu} \right)_x a_\nu.
\label{Jdot1}
\ee
Here the subscript $x$ on the round brackets means that the
derivative is to be taken holding $x^\nu$ fixed.
When the sum over $\beta$ is evaluated the contribution from $P_1 = H$
vanishes since $a_\nu p^\nu=0$, and we obtain using Eqs.\
(\ref{eq:Palphadef}) and (\ref{sec4:eq:BLactions})
\bes
\bea
F_t&=&a_t,\\
F_r&=&- \frac{\partial J_r}{\partial E}a_t+\frac{\partial
J_r}{\partial Q}a_Q+\frac{\partial
J_r}{\partial L_z}a_\phi,\\
F_\theta&=& - \frac{\partial
J_\theta}{\partial E}a_t+\frac{\partial J_\theta}{\partial
Q}a_Q+\frac{\partial
J_\theta}{\partial L_z}a_\phi,\\
F_\phi&=&a_\phi.
\eea
\ees
Here we have defined $a_Q = 2 Q^{\nu\sigma} p_\nu a_\sigma$ and the various
coefficients $\partial J_\alpha / \partial P_\beta$ are given
explicitly as functions of $P_\alpha$ in Appendix
\ref{appendix:frequencies}.

We use a similar procedure to obtain the equation of motion
(\ref{qdotforced}) for the
generalized angle variables $q_\alpha$.  Differentiating the relation
$q_\alpha=q_\alpha(x^\nu,p_\nu)$ with respect to $\lambda$ and
combining with the two first order equations of motion (\ref{geodesic}) gives
\bea
\frac{dq_\alpha}{d\lambda}
&=&\left[\frac{\partial q_\alpha}{\partial x^\nu} g^{\nu\sigma} p_\sigma
- \frac{1}{2} \frac{\partial q_\alpha}{\partial p_\nu}
g^{\sigma\rho}_{\ \ ,\nu} p_\sigma p_\rho
\right]
\nn \\ &&
+\mu^2 \frac{\partial q_\alpha}{\partial p_\nu}a_\nu.
\label{qdot}
\eea
By comparing with Eq.\ (\ref{eq:qdotnoac}) in the case of vanishing acceleration
we see that the term in square brackets is $\Omega_\alpha(J_\beta)$.
This gives an equation of motion of the form (\ref{qdotforced}), where
the where the forcing term $f_\alpha$ is
\be
f_\alpha = \left( \frac{\partial q_\alpha}{\partial p_\nu} \right)_x
a_\nu.
\label{eq:falpha0}
\ee
Using the expression (\ref{eq:ct2}) for the angle variable $q_\alpha$
together with $J_\alpha = J_\alpha(P_\beta)$ gives
\bea
\left( \frac{\partial q_\alpha}{\partial p_\nu} \right)_x  &=&
\left( \frac{ \partial P_\gamma} {\partial p_\nu} \right)_x
\left[
  \frac{\partial P_\beta}{\partial J_\alpha} \left( \frac{ \partial^2
      {\cal W}}{\partial P_\beta \partial P_\gamma} \right)_x
\right. \nn \\
&& \left. + \left( \frac{ \partial {\cal W}}{\partial P_\beta} \right)_x
\frac{\partial}{\partial P_\gamma} \left( \frac{\partial P_\beta}
  {\partial J_\alpha} \right)
\right].
\eea
This yields for the forcing term
\bea
f_\alpha &=&
a_\nu \left( \frac{ \partial P_\gamma} {\partial p_\nu} \right)_x
\frac{ \partial P_\delta}{\partial J_\alpha}
\left[   \left( \frac{ \partial^2
      {\cal W}}{\partial P_\delta \partial P_\gamma} \right)_x
\right. \nn \\
&& \left. - \left( \frac{ \partial {\cal W}}{\partial P_\beta} \right)_x
\frac{ \partial P_\beta}{\partial J_\varepsilon} \frac{ \partial^2
  J_\varepsilon}{ \partial P_\gamma \partial P_\delta}
\right].
\label{eq:Gdef}
\eea
In this expression the first two factors are the same as the factors
which appeared in the forcing term (\ref{Jdot1}) for the action
variables.  The quantities $\partial P_\delta / \partial J_\alpha$,
$\partial P_\beta /\partial J_\varepsilon$ and
$ \partial^2 J_\varepsilon / ( \partial P_\gamma \partial P_\delta)$
can be evaluated explicitly as functions of $P_\alpha$ using the
techniques discussed in Appendix \ref{appendix:frequencies}.
The remaining factors in Eq.\ (\ref{eq:Gdef}) can be evaluated by
differentiating the formula (\ref{sec4:eq:completeintegral})
for Hamilton's principal function ${\cal W}$ and using the formulae
(\ref{sec4:eq:potentials}) for the potentials $V_r$ and $V_\theta$.

\subsection{Rescaled variables and incorporation of backreaction on the black hole}
\label{sec:backreaction}

We now augment the action-angle equations of motion
(\ref{eq:actionangleeqns}) in order to describe the backreaction of
the gravitational radiation on the black hole.  We also modify the
equations to simplify and make explicit the dependence on the mass
$\mu$ of the particle.  The resulting modified equations of motion,
whose solutions we will analyze in the remainder of the paper, are
\bes
\bea
\frac{d q_\alpha }{d\tau} &=&
{\omega}_\alpha({\tilde P}_j,M_B)
+\varepsilon
g_\alpha^{(1)}(q_A,{\tilde P}_j,M_B) \nn \\ &&
+ \varepsilon^2 g_\alpha^{(2)}(q_A,{\tilde P}_j,M_B)
+ O(\varepsilon^3), \\
\frac{d {\tilde P}_i}{d\tau} &=&
\varepsilon G_i^{(1)}(q_A,{\tilde P}_j,M_A) + \varepsilon^2
G_i^{(2)}(q_A,{\tilde P}_j,M_B) \nn  \\
&&  + O(\varepsilon^3), \\
\frac{d M_A}{d\tau} &=&
\varepsilon^2 {\hat G}_A(q_A,{\tilde P}_j,M_B)   + O(\varepsilon^3).
\eea
\label{eq:eomsimplified}\ees
Here $\alpha$ runs over $0,1,2,3$, $i$, $j$ run over $1,2,3$, $A$, $B$ run over $1,2$,
$q_A = (q_r,q_\theta)$, $M_A = (M_1,M_2)$ and ${\tilde P}_i = ({\tilde P}_1, {\tilde P}_2, {\tilde P}_3)$.
Also all of the functions $\omega_\alpha$, $g^{(1)}_\alpha$, $g^{(2)}_\alpha$,
$G^{(1)}_i$, $G^{(2)}_i$ and ${\hat G}_A$ that appear on the right hand sides are smooth functions
of their arguments whose precise form will not be needed for this paper (and are currently unknown aside from $\omega_\alpha$).

Our final equations (\ref{eq:eomsimplified}) are similar in structure to the original
equations (\ref{eq:actionangleeqns}), but there are a number of
differences:
\begin{itemize}
\item We have switched the independent variable in the differential
  equations from affine parameter $\lambda$ to proper time $\tau = \mu
  \lambda$.

\item We have introduced the ratio
\be
\varepsilon = \frac{\mu}{M}
\ee
of the particle mass $\mu$ and black hole mass $M$, and have expanded
the forcing terms as a power series in $\varepsilon$.

\item The forcing terms
  $g_\alpha^{(1)}$, $g_\alpha^{(2)}$, $G_i^{(1)}$,
  $G_i^{(2)}$, and ${\hat G}_A$ depend only on the two angle variables $q_A \equiv
  (q_r,q_\theta)$, and are independent of $q_t$ and $q_\phi$.

\item Rather than evolving the action variables $J_\alpha$, we evolve
two different sets of variables, ${\tilde P}_i$ and $M_A$.  The first
of these sets consists of three of the first integrals of the motion, with
the dependence on the mass $\mu$ of the particle scaled out:
\be
{\tilde P}_i = ({\tilde P}_1,{\tilde P}_2, {\tilde P}_3) \equiv (E/\mu,
L_z/\mu, Q/\mu^2).
\ee
The second set consists of the mass and spin parameters of the black
hole, which gradually evolve due to absorption of
gravitational radiation by the black hole:
\be
M_A = (M_1,M_2) = (M,a).
\ee

\end{itemize}

We now turn to a derivation of the modified equations of motion
(\ref{eq:eomsimplified}).  The derivation consists of several steps.
First, since the mapping (\ref{sec4:eq:BLactions}) between the first
integrals $P_\alpha$ and the action variables $J_\alpha$ is a
bijection, we can use the $P_\alpha$ as dependent variables instead of
$J_\alpha$.\footnote{Note that since the variables $J_\alpha$ are adiabatic
invariants, so are the variables $P_\alpha$.}
Equation (\ref{qdotforced}) is unmodified except that the right
hand side is expressed as a function of $P_\alpha$ instead of
$J_\alpha$.  Equation (\ref{Jdotforced}) is replaced by
\bes
\bea
\frac{d P_\alpha}{d \lambda} &=& \mu^2 \left( \frac{\partial
  P_\alpha}{\partial p_\nu} \right)_x a_\nu.
\eea
\ees

Second, we switch to using modified versions ${\tilde P}_\alpha$ of the first
integrals $P_\alpha$ with the dependence on the mass $\mu$ scaled out.  These
rescaled first integrals are defined by
\bea
{\tilde P}_\alpha &=& ({\tilde H}, {\tilde E}, {\tilde L}_z, {\tilde Q}
) \nn \\
&\equiv& (H/\mu^2, E/\mu, L_z/\mu, Q/\mu^2).
\label{eq:Prescaled}
\eea
We also change the independent variable from affine parameter
$\lambda$ to proper time $\tau = \mu
\lambda$.  This gives
from Eqs.\ (\ref{eq:actionangleeqns}) and (\ref{eq:falpha0}) the system
of equations
\bes
\bea
\label{eq:qdot4}
\frac{d q_\alpha }{d\tau} &=& \frac{1}{\mu}
\Omega_\alpha(P_\beta)
+ \mu \left( \frac{ \partial q_\alpha}{\partial p_\nu} \right)_x a_\nu, \\
\frac{d {\tilde P}_\alpha}{d\tau} &=&
\mu^{1 - n_\alpha} \left( \frac{\partial P_\alpha}{\partial p_\nu} \right)_x a_\nu,
\label{eq:Pdot4}
\eea
\ees
where we have defined $n_\alpha = (2,1,1,2)$.

Third, we analyze the dependence on the mass $\mu$ of the right hand
sides of these equations.
Under the transformation $(x^\nu,p_\nu) \to (x^\nu,s p_\nu)$ for
$s>0$, we obtain the following transformation laws for the
first integrals (\ref{firstintegrals}),
the action variables (\ref{sec4:eq:BLactions}),
and Hamilton's principal function (\ref{sec4:eq:completeintegral}):
\bes
\bea
\label{eq:Pscaling}
P_\alpha &\to& s^{n_\alpha} P_\alpha \ \ \text{with} \ \ n_\alpha = (2,1,1,2),\\
J_\alpha &\to& s J_\alpha, \\
{\cal W} &\to& s {\cal W}.
\eea
\ees
From the definitions (\ref{sec22:eq:omegadef0}) and (\ref{eq:ct2}) of
the angular frequencies $\Omega_\alpha$ and the angle variables
$q_\alpha$ we also deduce
\bes
\bea
\label{Omegascaling}
\Omega_\alpha &\to& s \Omega_\alpha, \\
q_\alpha &\to& q_\alpha.
\label{eq:qscaling}
\eea
\ees
If we write the angular velocity $\Omega_\alpha$ as a function
$\omega_\alpha(P_\beta)$ of the first integrals $P_\beta$, then it follows
from the scalings (\ref{eq:Pscaling}) and (\ref{Omegascaling}) that
the first term on the right hand side of Eq.\ (\ref{eq:qdot4}) is
\be
\frac{\Omega_\alpha}{\mu} = \frac{ {
    \omega}_\alpha(P_\beta)}{\mu} = \frac{ {\omega}_\alpha( \mu^{n_\beta} {\tilde
  P}_\beta)}{\mu} = {\omega}_\alpha({\tilde P}_\beta).
\ee
This quantity is thus independent of $\mu$ at fixed ${\tilde
  P}_\beta$, as we would expect.

Similarly, if we write the angle variable $q_\alpha$ as a function
${\bar q}_\alpha(x^\nu,p_\nu)$ of $x^\nu$ and $p_\nu$, then the
scaling law (\ref{eq:qscaling}) implies that ${\bar q}_\alpha(x^\nu,
s p_\nu) = {\bar q}_\alpha(x^\nu,p_\nu)$, and it follows that
the coefficient of the 4-acceleration in Eq.\ (\ref{eq:qdot4}) is
\footnote{Note that
$
\mu \, {\partial }/{\partial p_\nu}
$
cannot be simplified to $ {\partial} / {\partial u_\nu}$ because we are
working in the eight dimensional phase space ${\cal M}$ where $\mu$ is
a coordinate and not a constant.}
\be
\mu \frac{ \partial {\bar q}_\alpha }{\partial p_\nu}(x^\sigma,p_\sigma)
= \mu \frac{ \partial {\bar q}_\alpha }{\partial p_\nu}(x^\sigma,\mu
u_\sigma) = \frac{ \partial {\bar q}_\alpha }{\partial p_\nu}(x^\sigma,
u_\sigma),
\label{eq:fcoeffdef}
\ee
where $u_\sigma$ is the 4-velocity.  This quantity is also independent of
$\mu$ at fixed ${\tilde P}_\beta$. We will denote this quantity by
$f_\alpha^\nu(q_\beta,{\tilde P}_\beta)$.  It can be obtained
explicitly by evaluating the coefficient of $a_\nu$ in Eq.\ (\ref{eq:Gdef})
at $P_\alpha = {\tilde P}_\alpha$, $p_\nu = u_\nu$.
A similar analysis shows that the driving term on the right hand side
of Eq.\ (\ref{eq:Pdot4}) can be written in the form
\be
F_\alpha^\nu(q_\beta,{\tilde P}_\beta) a_\nu \equiv (0, -a_t, a_\phi, 2
Q^{\nu\sigma} u_\nu a_\sigma).
\label{eq:Fcoeffdef}
\ee

The resulting rescaled equations of motion are
\bes
\bea
\label{eq:qdot5}
\frac{d q_\alpha }{d\tau} &=&
{\omega}_\alpha({\tilde P}_\beta)
+ f_\alpha^\nu(q_\beta,{\tilde P}_\beta) a_\nu, \\
\frac{d {\tilde P}_\alpha}{d\tau} &=&
F_\alpha^\nu(q_\beta,{\tilde P}_\beta) a_\nu.
\label{eq:Pdot5}
\eea
\label{eq:eoms5}\ees
Note that this formulation of the equations is completely independent
of the mass $\mu$ of the particle (except for the dependence on $\mu$
of the radiation reaction acceleration $a_\nu$ which we will discuss below).

Fourth, since $P_0 = H = - \mu^2/2$, the rescaled variable
is ${\tilde P}_0= -1/2$ from Eq.\ (\ref{eq:Prescaled}). Thus we can
drop the evolution equation for
${\tilde P}_0$, and retain only the equations for the remaining
rescaled first integrals
\be
{\tilde P}_i = ({\tilde P}_1,{\tilde P}_2,{\tilde P}_3) = ({\tilde
  E},{\tilde L}_z,{\tilde Q}).
\ee
We can also omit the dependence on ${\tilde P}_0$ in the right hand sides
of the evolution equations (\ref{eq:eoms5}), since ${\tilde P}_0$ is a
constant.  This yields
\bes
\bea
\label{eq:qdot6}
\frac{d q_\alpha }{d\tau} &=&
{\omega}_\alpha({\tilde P}_j)
+ f_\alpha^\nu(q_\beta,{\tilde P}_j) a_\nu, \\
\frac{d {\tilde P}_i}{d\tau} &=&
F_i^\nu(q_\beta,{\tilde P}_j) a_\nu.
\label{eq:Pdot6}
\eea
\label{eq:eom6}\ees

Fifth, the self-acceleration of the particle can be expanded in powers
of the
mass ratio $\varepsilon  = \mu/M$ as
\be
a_\nu = \varepsilon a_{\nu}^{(1)} + \varepsilon^2 a_\nu^{(2)} +
O(\varepsilon^3).
\label{selfacc}
\ee
Here $a_\nu^{(1)}$ is the leading order self-acceleration derived by
Mino, Sasaki and Tanaka \cite{Mino:1996nk} and by Quinn and Wald \cite{Quinn:1996am},
discussed in the introduction.  The subleading self-acceleration
$a_\nu^{(2)}$ has been computed in Refs.\
\cite{Rosenthal:2006nh,Rosenthal:2006iy,Chadthesis,Galley:2008ih,Chadtalk}.
The accelerations $a_{\nu}^{(1)}$ and $a_{\nu}^{(2)}$ are independent
of $\mu$ and thus depend only on $x^\nu$ and $u_\nu$, or, equivalently,
on $q_\alpha$ and ${\tilde P}_i$.
This yields the system of equations
\bes
\bea
\label{eq:qdot7}
\frac{d q_\alpha }{d\tau} &=&
{\omega}_\alpha({\tilde P}_j)
+\varepsilon
g_\alpha^{(1)}(q_\beta,{\tilde P}_j) + \varepsilon^2
g_\alpha^{(2)}(q_\beta,{\tilde P}_j) \nn \\
&& + O(\varepsilon^3), \\
\frac{d {\tilde P}_i}{d\tau} &=&
\varepsilon G_i^{(1)}(q_\beta,{\tilde P}_j) + \varepsilon^2
G_i^{(2)}(q_\beta,{\tilde P}_j) \nn \\
&&  + O(\varepsilon^3).
\label{eq:Pdot7}
\eea
\label{eq:eom7}\ees
Here the forcing terms are given by
\bes
\bea
g_\alpha^{(s)} &=& f^\nu_\alpha a^{(s)}_\nu, \\
G_i^{(s)} &=& F^\nu_i a^{(s)}_\nu,
\eea
\label{eq:forcingfnsformula}\ees
for $s=1,2$.

The formula (\ref{selfacc}) for the self-acceleration,
with the explicit formula for $a^{(1)}_\nu$ from Refs.\ \cite{Mino:1996nk,Quinn:1996am},
is valid when
one chooses the Lorentz gauge for the metric perturbation.
The form of Eq.\ (\ref{selfacc}) is also valid in a variety of other
gauges; see Ref.\ \cite{PhysRevD.64.124003} for a discussion of the gauge
transformation properties of the self force.
However, there exist gauge choices which are incompatible with
Eq.\ (\ref{selfacc}), which can be obtained by making
$\varepsilon$-dependent gauge transformations.  We shall restrict
attention to classes of gauges which
are consistent with our ansatz (\ref{metricansatz}) for the metric, as
discussed in Sec.\ \ref{sec:intro:waves} above.
This class of gauges has the properties that
(i) the
deviation of the metric from Kerr is $\alt \varepsilon$ over the
entire inspiral, and (ii)
the expansion (\ref{selfacc}) of the self-acceleration is valid.
These restrictions exclude, for example, the gauge
choice which makes $a^{(1)}_\nu \equiv 0$, since in that gauge the
particle does not inspiral, and the metric perturbation must therefore
become of order unity over an inspiral time.
We note that alternative classes of gauges have been suggested and
explored by Mino \cite{2005PThPh.113..733M,2006PThPh.115...43M,2005CQGra..22S.375M,2005CQGra..22S.717M}.

Sixth, from the
formula
(\ref{eq:ct2})
for the generalized angle variables $q_\alpha$
together with Eqs.\
(\ref{sec4:eq:completeintegral}) and (\ref{eq:Jtdef}) it follows that $q_t$
can be written as
\be
q_t = t + f_t(r,\theta,P_\alpha)
\ee
for some function $f_t$.  All of the other angle and action variables
are independent of $t$.  Therefore the vector field $\partial/\partial
t$ on phase space is just $\partial / \partial q_t$; the symmetry $t
\to t + \Delta t$ with $x^i$, $p_\mu$ fixed is the same as the
symmetry $q_t \to q_t + \Delta
t$ with $q_r,q_\theta,q_\phi$ and $J_\alpha$ fixed.  Since the
self-acceleration as
well as the background geodesic motion respect this symmetry, all of
the terms on the right hand side of Eqs.\ (\ref{eq:eom7}) must be independent
of $q_t$.  A similar argument shows that they are independent of $q_\phi$.
This gives
\bes
\bea
\label{eq:qdot8}
\frac{d q_\alpha }{d\tau} &=&
{\omega}_\alpha({\tilde P}_j)
+\varepsilon
g_\alpha^{(1)}(q_A,{\tilde P}_j) + \varepsilon^2
g_\alpha^{(2)}(q_A,{\tilde P}_j) \nn \\
&& + O(\varepsilon^3), \\
\frac{d {\tilde P}_i}{d\tau} &=&
\varepsilon G_i^{(1)}(q_A,{\tilde P}_j) + \varepsilon^2
G_i^{(2)}(q_A,{\tilde P}_j) \nn \\
&&  + O(\varepsilon^3),
\label{eq:Pdot8}
\eea
\label{eq:eom8}\ees
where $q_A \equiv (q_r,q_\theta)$.

Seventh, consider the evolution of the black hole background.
So far in our analysis we have assumed that the particle moves in a
fixed Kerr background, and is subject to a self-force $a_\nu =
\varepsilon a_\nu^{(1)} + \varepsilon^2 a_\nu^{(2)} +
O(\varepsilon^3)$.  In reality, the center of mass, 4-momentum and
spin angular momentum of the black hole will gradually evolve due to the
gravitational radiation passing through the event horizon.
The total change in the mass $M$ of the black hole over the inspiral
timescale $\sim M/\varepsilon$ is $\sim M \varepsilon$.  It follows
that the timescale for the black hole mass to change by a factor of
order unity is $\sim M /\varepsilon^2$.  The same timescale governs the evolution of the other
black hole parameters.

This effect of evolution of the black hole background will alter the
inspiral at the first subleading order (post-1-adiabatic order) in our
two-timescale expansion.  
A complete calculation of the inspiral to
this order requires solving simultaneously for the motion of the
particle and the gradual evolution of the background.
We introduce the extra variables
\be
M_A = (M_1,M_2) = (M,a),
\label{eq:newvarlist}
\ee
the mass and spin parameters of the black hole.
We modify the equations of motion (\ref{eq:eom8}) by
showing explicitly the dependence of the
frequencies $\omega_\alpha$ and the forcing
functions $g_\alpha^{(n)}$ and $G_i^{(n)}$ on these parameters (the
dependence has up to now been implicit).
We also add to the system of equations the following evolution
equations for the black hole parameters:
\be
\frac {d M_A}{d \tau} = \varepsilon^2 {\hat G}_A(q_B,{\tilde P}_j,M_B)
+ O(\varepsilon^3),
\label{eq:MAeqn}
\ee
where $A = 1,2$.  Here ${\hat G}_A$ are some functions describing the fluxes of energy and angular momentum down the horizon, whose explicit
form will not be important for our analyses.  They can in principle be computed
using, for example, the techniques developed in Ref.\
\cite{Poisson1}.\footnote{These techniques naturally furnish the
  derivatives of $M_A$ with respect to Boyer Lindquist time $t$, not
  proper time $\tau$ as in Eq.\ (\protect{\ref{eq:MAeqn}}).
However this difference is unimportant; one can easily convert from
one variable to the other by multiplying the functions ${\hat G}_A$ by
the standard expression for $dt/d\tau$ \cite{MTW},
$$
\frac{dt}{d\tau} = \frac{{\tilde E}}{\Sigma} \left(
  \frac{\varpi^4}{\Delta} - a^2 \sin^2 \theta \right) + \frac{a
  {\tilde L}_z}{\Sigma} \left( 1 - \frac{\varpi^2}{\Delta} \right),
$$
where $\varpi = \sqrt{r^2 + a^2}$.
This expression can be written in terms of
of $q_A,{\tilde P}_i$ and $M_A$, and is valid for accelerated
motion as well as geodesic motion
by Eqs.\ (\ref{sec4:eq:momenta}) and (\ref{geodesicI}).}  The reason
for the prefactor of $\varepsilon^2$ is that
the evolution timescale for the black hole parameters is $\sim
M/\varepsilon^2$, as discussed above.  The functions ${\hat G}_A$ are
independent of $q_t$ and $q_\phi$ for the reason discussed near Eq.\
(\ref{eq:eom8}): the fluxes through the horizon respect the symmetries of
the background spacetime.
Finally, we have omitted
in the set of new variables (\ref{eq:newvarlist})
the orientation of the total angular momentum, the location of
the center of mass, and the total linear momentum of the system, since these parameters
are not coupled to the inspiral motion at the leading order.  However, it would be possible
to enlarge the set of variables $M_A$ to include these parameters
without modifying in any way the
analyses in the rest of this paper.

These modifications result in the final system of equations
(\ref{eq:eomsimplified}).

Finally we note that an additional effect arises due to the fact that the
action-angle variables we use are defined, at each instant, to be the
action-angle variables associated with the black hole background at
that time.  In other words the coordinate transformation on phase
space from $(x^\nu,p_\nu) \to (q_\alpha,J_\alpha)$ acquires an
additional dependence on time.  Therefore the Jacobian of this
transformation, which was used in deriving the evolution equations
(\ref{eq:actionangleeqns}), has an extra term.
However, the corresponding correction to the evolution equations can
be absorbed into a redefinition of the forcing term $g^{(2)}_\alpha$.

\subsection{Conservative and dissipative pieces of the forcing terms}
\label{sec:consdiss}

In this subsection we define a splitting of the forcing terms
$g_\alpha$ and $G_i$
in the equations of motion (\ref{eq:eomsimplified})
into conservative and dissipative pieces, and
review some properties of this decomposition derived by Mino
\cite{Mino2003}.

We start by defining some notation.
Suppose that we have a particle at a point ${\cal P}$
with four velocity $u^\mu$, and that we are given a
linearized metric perturbation
$h_{\mu\nu}$ which is a solution (not necessarily the retarded solution) of the
linearized Einstein equation equation for which the source is
a delta function on the geodesic determined by ${\cal P}$ and
$u^\mu$.  The self-acceleration of the particle is then some
functional of ${\cal P}$, $u^\mu$, $h_{\mu\nu}$ and of the spacetime
metric $g_{\mu\nu}$, which we write as
\be
a^\mu\left[ {\cal P}, u^\mu, g_{\mu\nu}, h_{\mu\nu} \right].
\ee
Note that this functional does not depend on a choice of time
orientation for the manifold, and also it is invariant under $u^\mu
\to - u^\mu$.
The retarded self-acceleration is defined as
\be
a^\mu_{\rm ret}\left[ {\cal P}, u^\mu, g_{\mu\nu} \right]
= a^\mu\left[ {\cal P}, u^\mu, g_{\mu\nu},
  h^{\rm ret}_{\mu\nu} \right],
\label{eq:selfret}
\ee
where $h^{\rm ret}_{\mu\nu}$ is the retarded solution to the
linearized Einstein equation obtained using the time orientation that is
determined by demanding that $u^\mu$ be future directed.
This is the physical self-acceleration which is denoted by $a^\mu$
throughout the rest of this paper.
Similarly, the advanced self-acceleration is
\be
a^\mu_{\rm adv}\left[ {\cal P}, u^\mu, g_{\mu\nu} \right]
= a^\mu\left[ {\cal P}, u^\mu, g_{\mu\nu},
  h^{\rm adv}_{\mu\nu} \right],
\label{eq:selfadv}
\ee
where $h^{\rm adv}_{\mu\nu}$ is the advanced solution.
It follows from these definitions that
\be
a^\mu_{\rm ret}\left[ {\cal P}, -u^\mu, g_{\mu\nu} \right]
= a^\mu_{\rm adv}\left[ {\cal P}, u^\mu, g_{\mu\nu} \right].
\label{eq:flip}
\ee

We define the conservative and dissipative self-accelerations to be
\be
a^\mu_{\rm cons} = \frac{1}{2} \left( a^\mu_{\rm ret} + a^\mu_{\rm
  adv} \right),
\label{eq:aconsdef}
\ee
and
\be
a^\mu_{\rm diss} = \frac{1}{2} \left( a^\mu_{\rm ret} - a^\mu_{\rm
  adv} \right).
\label{eq:adissdef}
\ee
The physical self-acceleration can then be decomposed as
\be
a^\mu = a^\mu_{\rm ret} = a^\mu_{\rm cons} + a^\mu_{\rm diss}.
\ee
A similar decomposition applies to the forcing functions
(\ref{eq:forcingfnsformula}):
\bes
\bea
g_\alpha^{(s)} &=& g^{(s)}_{\alpha\,{\rm cons}} +  g^{(s)}_{\alpha\,{\rm diss}}, \\
G_i^{(s)} &=& G_{i\,{\rm cons}}^{(s)} + G_{i\,{\rm diss}}^{(s)},
\eea
\label{eq:forcingfnsdecomposition}\ees
for $s=1,2$.

Next, we note that if $\psi$ is any diffeomorphism from the spacetime to itself, then
the self acceleration satisfies the covariance relation
\be
a^\nu_{\rm ret}[\psi({\cal P}), \psi^* u^\nu, \psi^* g_{\mu\nu}
 ] = \psi^* a^\nu_{\rm ret}[{\cal P}, u^\nu, g_{\mu\nu}].
\label{eq:covariance}
\ee
Taking the point ${\cal P}$ to be $(t_0,r_0,\theta_0,\phi_0)$ in
Boyer-Lindquist coordinates, and choosing $\psi$ to be $t \to 2 t_0 -
t$, $\phi \to 2 \phi_0 - \phi$, then $\psi$ is an isometry, $\psi^*
g_{\mu\nu} = g_{\mu\nu}$.
It follows that
\be
a^\nu_{\rm ret}(-u_t,u_r,u_\theta,-u_\phi) = -\epsilon_\nu a^\nu_{\rm
  ret}(u_t,u_r,u_\theta,u_\phi),
\ee
where
\be
\epsilon_\nu = (1,-1,-1,1)
\label{epsilondef}
\ee
and there is no summation over
$\nu$ on the right hand side.  Combining this with the identity
(\ref{eq:flip}) gives
\be
a^\nu_{\rm adv}(u_t,u_r,u_\theta,u_\phi) = -\epsilon_\nu a^\nu_{\rm
  ret}(u_t,-u_r,-u_\theta,u_\phi).
\label{aadvans}
\ee

Now, under the transformation $p_r \to - p_r$, $p_\theta \to - p_\theta$ with
other quantities fixed, the action variables and the quantities $P_\alpha$ are invariant, the
angle variables $q_r$ and $q_\theta$ transform as $q_r \to 2 \pi - q_r$, $q_\theta \to 2
\pi - q_\theta$, while $q_t-t$ and $q_\phi - \phi$ flip sign.
This can be seen from the definitions (\ref{sec4:eq:completeintegral}) and (\ref{eq:ct2}).
Explicitly we have
\bes
\bea
{\bar q}_t(x^\gamma, \epsilon_\delta p_\delta) -t &=&
-  [{\bar q}_t(x^\gamma,p_\delta) -t],\\
{\bar q}_\phi(x^\gamma, \epsilon_\delta p_\delta) -\phi &=&
-  [{\bar q}_\phi(x^\gamma,p_\delta) -\phi],\\
{\bar q}_A(x^\gamma, \epsilon_\delta p_\delta) &=&
2 \pi  - {\bar
  q}_A(x^\gamma,p_\delta),\\
P_i(x^\gamma, \epsilon_\delta p_\delta) &=&
P_i(x^\gamma, p_\delta),
\eea
\ees
where we use the values (\ref{epsilondef}) of
$\epsilon_\alpha$,
the functions ${\bar q}_\alpha$
are defined before Eq.\ (\ref{eq:fcoeffdef}), and $q_A = (q_r,q_\theta)$.
If we now differentiate with respect to $p_\alpha$ holding $x^\alpha$ fixed and use the
definitions (\ref{eq:fcoeffdef}), (\ref{eq:Pdot4}) and (\ref{eq:Pdot5}) of the
functions $f^\nu_\alpha$ and $F^\nu_i$ we obtain
\bes
\bea
f^\nu_\alpha(x^\beta, \epsilon_\gamma u_\gamma) &=&
- \epsilon_\nu f^\nu_\alpha( x^\beta, u_\gamma),\\
F^\nu_i(x^\beta, \epsilon_\gamma u_\gamma) &=&
\epsilon_\nu  F^\nu_i( x^\beta, u_\gamma).
\eea
\label{tf}\ees

We now compute the conservative and dissipative pieces of the forcing
functions $g^{(1)}_\alpha$ and $G^{(1)}_i$, using the definitions
(\ref{eq:forcingfnsformula}) and (\ref{eq:forcingfnsdecomposition}).
Using the results (\ref{aadvans}) and (\ref{tf})
we obtain
\bea
g^{(1)}_{\alpha\,{\rm adv}}(u_\gamma) &=& f^\nu_\alpha(u_\gamma) \,
a^{(1)}_{\nu\,{\rm adv}}(u_\gamma) \nonumber \\
&=& \left[ - \epsilon_\nu f^\nu_\alpha(\epsilon_\gamma
  u_\gamma) \right] \,
\left[ - \epsilon_\nu a^{(1)}_{\nu\,{\rm ret}}(\epsilon_\gamma
  u_\gamma) \right] \ \ \ \ \nonumber \\
&=&  g^{(1)}_{\alpha\,{\rm ret}}(\epsilon_\gamma u_\gamma).
\eea
A similar computation gives
\bea
G^{(1)}_{i\,{\rm adv}}(u_\gamma) &=&
- G^{(1)}_{i\,{\rm ret}}(\epsilon_\gamma u_\gamma),
\eea
and using that the mapping $x^\nu \to x^\nu$, $u_\mu \to \epsilon_\mu
u_\mu$ corresponds to ${\tilde P}_j \to {\tilde P}_j$, $q_r \to 2 \pi
- q_r$, $q_\theta \to 2 \pi - q_\theta$
finally yields the identities
\begin{widetext}
\bes
\label{eq:ident450a}
\bea
\label{eq:ident45a}
g^{(1)}_{\alpha\,{\rm cons}}(q_A,{\tilde P}_j) &=&
\left[ g^{(1)}_{\alpha}(q_r,q_\theta,{\tilde P}_j) +
g^{(1)}_{\alpha}(2\pi-q_r,2\pi-q_\theta,{\tilde P}_j) \right]/2,\\
g^{(1)}_{\alpha\,{\rm diss}}(q_A,{\tilde P}_j) &=&
\left[ g^{(1)}_{\alpha}(q_r,q_\theta,{\tilde P}_j)
  - g^{(1)}_{\alpha}(2\pi-q_r,2 \pi-q_\theta,{\tilde P}_j) \right]/2,
\eea
\ees
and
\bes
\label{eq:ident450}
\bea
\label{eq:ident45}
G^{(1)}_{i\,{\rm cons}}(q_A,{\tilde P}_j) &=&
\left[ G^{(1)}_{i}(q_r,q_\theta,{\tilde P}_j) -
  G^{(1)}_{i}(2\pi-q_r,2\pi-q_\theta,{\tilde P}_j) \right]/2,\\
G^{(1)}_{i\,{\rm diss}}(q_A,{\tilde P}_j) &=&
\left[ G^{(1)}_{i}(q_r,q_\theta,{\tilde P}_j) +
  G^{(1)}_{i}(2\pi-q_r,2\pi-q_\theta,{\tilde P}_j) \right]/2.
\eea
\ees
\end{widetext}
Here we have used the fact that the forcing functions are independent
of $q_t$ and $q_\phi$, as discussed in the last subsection.
Similar equations apply with $g^{(1)}_\alpha$ and $G^{(1)}_i$ replaced
by the higher order forcing terms $g^{(s)}_\alpha$ and $G^{(s)}_i$, $s \ge 2$.

It follows from the identity (\ref{eq:ident45}) that, for the
action-variable forcing functions
$G^{(1)}_i$, the average over the 2-torus parameterized by $q_r$ and
$q_\theta$ of the conservative piece vanishes.
For generic orbits (for which $\omega_r$ and $\omega_\theta$ are incommensurate), the torus-average is equivalent to a time average, and so it follows that the time average vanishes, a result first
derived by Mino \cite{Mino2003}.
Similarly from Eqs.\ (\ref{eq:ident450a}) it follows that the torus-average of the
dissipative pieces of $g^{(1)}_\alpha$ vanish.

\section{A GENERAL WEAKLY PERTURBED DYNAMICAL SYSTEM}
\label{sec:generalsystem}

In the remainder of this paper we will study in detail the behavior
of a one-parameter family of dynamical systems parameterized by
a dimensionless parameter $\varepsilon$.  We shall be interested in
the limiting behavior of the systems as $\varepsilon \to 0$.
The system contains $N+M$ dynamical variables
\bes
\bea
\bfq(t) &=& \big(q_1(t), q_2(t), \ldots, q_N(t) \big), \\
\bfJ(t) &=& \big(J_1(t), J_2(t), \ldots, J_M(t) \big),
\eea
\ees
and is defined by the equations
\bes
\label{sec2:eq:eom1}
\bea
\label{sec2:eq:eom1a}
\frac{d q_\alpha}{dt} &=& \omega_\alpha(\bfJ,\tt) + \varepsilon
g_\alpha(\bfq,\bfJ,{\tilde t},\varepsilon), \ \ \ 1 \le \alpha \le N,
\ \ \ \\
\frac{d J_\lambda}{dt} &=& \varepsilon
G_\lambda(\bfq,\bfJ,\tt,\varepsilon), \ \ \ 1 \le \lambda \le M.
\eea
\ees
Here the variable
${\tilde t}$ is the ``slow time'' variable defined by
\be
\tt = \varepsilon t.
\label{sec2:eq:slowtimedef}
\ee
We assume that the functions $g_\alpha$ and $G_\lambda$ can be expanded as
\begin{eqnarray}
g_\alpha(\bfq,\bfJ,\tt,\varepsilon) &=& \sum_{s=1}^\infty
g_\alpha^{(s)}(\bfq,\bfJ,\tt) \varepsilon^{s-1} \nonumber \\
&=& g^{(1)}_\alpha(\bfq,\bfJ,\tt)
+ g^{(2)}_\alpha(\bfq,\bfJ,\tt) \varepsilon + O(\varepsilon^2), \nonumber \\
\label{sec2:eq:Gexpand}
\end{eqnarray}
and
\begin{eqnarray}
G_\lambda(\bfq,\bfJ,\tt,\varepsilon) &=& \sum_{s=1}^\infty
G^{(s)}_\lambda(\bfq,\bfJ,\tt) \varepsilon^{s-1} \nonumber \\
&=& G^{(1)}_\lambda(\bfq,\bfJ,\tt)
+ G^{(2)}_\lambda(\bfq,\bfJ,\tt) \varepsilon +
O(\varepsilon^2). \nonumber \\
\label{sec2:eq:gexpand}
\end{eqnarray}
These series are assumed to be asymptotic
series in $\varepsilon$ as $\varepsilon \to 0$ that are uniform in
${\tilde t}$.\footnote{In other words, there exists ${\tilde T}>0$
  such that for
  every $\bfq$, $\bfJ$, every integer $N$, and every  $\delta
  >0$ there exists $\epsilon_1= \epsilon_1(\bfq,\bfJ,N,\delta)$ such
  that
$$
\left| g_\alpha(\bfq,\bfJ,\tt,\varepsilon)  - \sum_{s=1}^N
g_\alpha^{(s)}(\bfq,\bfJ,\tt) \varepsilon^{s-1}\right| < \delta \varepsilon^{N-1}
$$
for all $\tt$ with $0 < \tt < {\tilde T}$ and for all $\varepsilon$
with $0 < \varepsilon < \epsilon_1$.}
We assume that the functions $\omega_\alpha$, $g_\alpha^{(s)}$ and
$G_\lambda^{(s)}$ are smooth functions of their arguments, and that the
frequencies $\omega_\alpha$ are nowhere vanishing.
Finally the functions $g_\alpha$ and $G_\lambda$ are assumed to be
periodic in each variable $q_\alpha$ with period $2 \pi$:
\bes
\bea
g_\alpha(\bfq+2\pi\bfk,\bfJ,\tt) &=& g_\alpha(\bfq,\bfJ,\tt), \ \ \ 1 \le \alpha \le N,\ \ \ \ \\
G_\lambda(\bfq+2\pi\bfk,\bfJ,\tt) &=& G_\lambda(\bfq,\bfJ,\tt), \ \ \ 1 \le \lambda \le M,\ \ \ \
\eea
\label{eq:periodic0}\ees
where $\bfk = (k_1, \ldots, k_N)$ is an arbitrary $N$-tuple of integers.

The equations (\ref{eq:eomsimplified}) derived in the previous section
describing the
inspiral of a point particle into a Kerr black hole are a special case
of the dynamical system (\ref{sec2:eq:eom1}).
This can be seen using the identifications $t = \tau$,
${\bf q} = (q_t,q_r,q_\theta,q_\phi)$, ${\bf J} =
({\tilde P}_2,{\tilde P}_3,{\tilde P}_4,M_1,M_2)$, $G_\lambda^{(1)} = (G_2^{(1)},
G_3^{(1)},G_4^{(1)},0,0)$ and $G_\lambda^{(2)} = (G_2^{(2)},
G_3^{(2)},G_4^{(2)},{\hat G}_1,{\hat G}_2)$.
The forcing functions
$g_\alpha^{(s)}$ and $G_\lambda^{(s)}$ are periodic
functions of $q_\alpha$ since they depend only on the variables
$q_A = (q_r,q_\theta)$ which are angle variables; they do not depend
on the variable $q_t$ which is not an angle variable.
Note that the system (\ref{sec2:eq:eom1})
allows the forcing functions $g_\alpha^{(s)}$, $G_\lambda^{(s)}$ and
frequencies $\omega_\alpha$ to depend in an arbitrary way on the slow
time ${\tilde t}$, whereas no such dependence is seen in the Kerr
inspiral system (\ref{eq:eomsimplified}).
The system studied here is thus slightly more general than is required
for our specific application.  We include the dependence on ${\tilde
  t}$ for greater generality and because it does not require any
additional complexity in the analysis.

Another special case of the system (\ref{sec2:eq:eom1}) is when $N=M$ and when
there exists a function $H(\bfJ,\tt)$ such that
\be
\omega_\alpha(\bfJ,\tt) = \frac{\partial H(\bfJ,\tt)}{\partial
  J_\alpha}
\ee
for $1 \le \alpha \le N$.
In this case the system (\ref{sec2:eq:eom1}) represents a Hamiltonian
system with slowly varying Hamiltonian $H(\bfJ,\tt)$, with action angle
variables $(q_\alpha,J_\alpha)$, and subject to
arbitrary weak perturbing forces that vary slowly with time.
The perturbed system is not necessarily Hamiltonian.

Because of the periodicity conditions (\ref{eq:periodic0}), we can
without loss of
generality interpret the variables $q_\alpha$ to be coordinates on the
$N$-torus $T^N$, and take the equations (\ref{sec2:eq:eom1}) to be
defined on the product of this N-torus with an open set.
This interpretation will useful below.

In the next several sections we will study in detail the behavior of
solutions of the system (\ref{sec2:eq:eom1}) in the limit $\varepsilon
\to 0$ using a two timescale expansion.  We follow closely the
exposition in the book by Kevorkian and Cole \cite{Kevorkian}, except
that we generalize their analysis and also correct some errors (see
Appendix \ref{app:Kevorkian}).
For clarity we treat first, in Sec.\ \ref{sec:derivation_single}, the
simple case of a single degree of freedom, $N=M=1$.
Section \ref{sec2:manyvariables} treats the case of general $N$ and
$M$, but with the restriction that the forcing functions $g_\alpha$
and $G_\lambda$ contain no resonant pieces (this is defined in Sec.\
\ref{sec:noresonanceassumption}).  The general case with resonances is
treated in the forthcoming papers \cite{FH08a,FH08b}.
Finally in Sec. \ref{sec:numerics}
we present a numerical integration of a particular example
of a dynamical system, in order to illustrate and validate the general theory of
Secs.\ \ref{sec:derivation_single} and \ref{sec2:manyvariables}.

\section{SYSTEMS WITH A SINGLE DEGREE OF FREEDOM}
\label{sec:derivation_single}

\subsection{Overview}

For systems with a single degree of freedom the general equations of
motion (\ref{sec2:eq:eom1}) discussed in Sec.\ \ref{sec:generalsystem}
reduce to
\bes
\label{eq:eom1}
\bea
{\dot q}(t) &=& \omega(J,\tt) + \varepsilon g(q,J,{\tilde
t},\varepsilon), \\
{\dot J}(t) &=& \varepsilon G(q,J,\tt,\varepsilon),
\eea
\label{eq:eom11}\ees
for some functions $G$ and $g$, where $\tt = \varepsilon t$ is the
slow time variable.
The asymptotic expansions (\ref{sec2:eq:Gexpand}) and (\ref{sec2:eq:gexpand})
of the forcing functions reduce to
\begin{eqnarray}
g(q,J,\tt,\varepsilon) &=& \sum_{s=1}^\infty
g^{(s)}(q,J,\tt) \varepsilon^{s-1} \nonumber \\
&=& g^{(1)}(q,J,\tt)
+ g^{(2)}(q,J,\tt) \varepsilon + O(\varepsilon^2), \nonumber \\
\label{eq:Gexpand}
\end{eqnarray}
and
\begin{eqnarray}
G(q,J,\tt,\varepsilon) &=& \sum_{s=1}^\infty
G^{(s)}(q,J,\tt) \varepsilon^{s-1} \nonumber \\
&=& G^{(1)}(q,J,\tt)
+ G^{(2)}(q,J,\tt) \varepsilon + O(\varepsilon^2). \nonumber \\
\label{eq:gexpand}
\end{eqnarray}
Also the periodicity conditions (\ref{eq:periodic0}) reduce to
\bes
\bea
g(q+2\pi,J,\tt) &=& g(q,J,\tt), \\
G(q+2\pi,J,\tt) &=& G(q,J,\tt).
\eea
\label{eq:periodic1}\ees

In this section we apply two-timescale expansions to study classes of
solutions of Eqs.\ (\ref{eq:eom11}) in the limit $\varepsilon \to 0$.
We start in Sec.\ \ref{sec:fourier} by defining our conventions and
notations for Fourier decompositions of the perturbing forces.  The
heart of the method is the ansatz we make for the form of the
solutions, which is given in Sec.\ \ref{sec:solnansatz}.
Sec.\ \ref{sec:results} summarizes the results we obtain at each order in the
expansion, and the derivations are given in Sec.\ \ref{sec:derivation}.
Although the results of this section are not directly applicable to
the Kerr inspiral problem, the analysis of this section gives an
introduction to the method of analysis, and is considerably simpler
than the multivariable case treated in Sec.\ \ref{sec2:manyvariables} below.

\subsection{Fourier expansions of the perturbing forces}
\label{sec:fourier}

The periodicity conditions (\ref{eq:periodic1}) apply at each order in the expansion in powers
of $\varepsilon$:
\bes
\bea
g^{(s)}(q+2\pi,J,\tt) &=& g^{(s)}(q,J,\tt), \\
G^{(s)}(q+2\pi,J,\tt) &=& G^{(s)}(q,J,\tt).
\eea
\ees
It follows that these functions can be expanded as Fourier series:
\bes
\bea
g^{(s)}(q,J,\tt) &=& \sum_{k=-\infty}^\infty g^{(s)}_k(J,{\tilde
t}) e^{i k q}, \\
G^{(s)}(q,J,\tt) &=& \sum_{k=-\infty}^\infty G^{(s)}_k(J,{\tilde
t}) e^{i k q},
\eea
\label{eq:ftdefs0}\ees
where
\bes
\label{eq:Gkdef}
\bea
g^{(s)}_k(J,\tt) &=& \frac{1 }{ 2 \pi} \int_0^{2 \pi} dq \ e^{-i k
q} \, g^{(s)}(q,J,\tt), \\
G^{(s)}_k(J,\tt) &=& \frac{1 }{ 2 \pi} \int_0^{2 \pi} dq \ e^{-i k
q} \, G^{(s)}(q,J,\tt).
\eea
\ees
For any periodic function $f=f(q)$, we introduce the notation
\be
\langle f \rangle = \frac{1}{2 \pi} \int_0^{2 \pi} f(q) dq
\ee
for the average part of $f$, and
\be
{\hat f}(q) = f(q) - \langle f \rangle
\label{eq:hatfdef}
\ee
for the remaining part of $f$.
It follows from these definitions that
\be
\langle g^{(s)}(q,J,\tt) \rangle = g^{(s)}_0(J,\tt),\ \
\
\langle G^{(s)}(q,J,\tt) \rangle = G^{(s)}_0(J,\tt),
\ee
and that
\bes
\bea
{\hat g}^{(s)}(q,J,\tt) &=&  \sum_{k\ne 0}
g^{(s)}_k(J,\tt) e^{i k q}, \\
{\hat G}^{(s)}(q,J,\tt) &=& \sum_{k\ne0}
G^{(s)}_k(J,\tt) e^{i k q}.
\label{eq:hatgformula}
\eea
\ees
We also have the identities
\bes
\label{eq:ident1}
\bea
\langle f_{,q} \rangle &=& \langle {\hat f} \rangle =0 \\
\langle f g \rangle &=& \langle {\hat f} {\hat g} \rangle +
\langle f \rangle \langle g \rangle
\eea
\ees
for any periodic functions $f(q)$, $g(q)$.

For any periodic function $f$, we also define a particular
anti-derivative ${\cal I}{\hat f}$ of ${\hat f}$ by
\be
({\cal I}{\hat f})(q) \equiv  \sum_{k\ne0} \frac{f_k}{i
k} e^{i k q},
\label{eq:calIdef}
\ee
where $f_k = \int dq e^{-i k q} f(q)/(2\pi)$ are the Fourier
coefficients of $f$.  This operator satisfies the identities
\bes
\label{eq:ident2}
\bea
({\cal I} {\hat f})_{,q} &=& {\hat f}, \\
\langle ( {\cal I} {\hat f}) {\hat g} \rangle &=& - \langle {\hat f}
({\cal I} {\hat g}) \rangle, \\
\langle {\hat f} ({\cal I} {\hat f}) \rangle &=&0.
\eea
\ees

\subsection{Two timescale ansatz for the solution}
\label{sec:solnansatz}

We now discuss the {\it ansatz} we use for the form of the solutions
of the equations of motion.  This ansatz will be justified a
posteriori order by order in $\varepsilon$.  The method used here is
sometimes called the ``method of strained coordinates'' \cite{Kevorkian}.

We assume that $q$ and $J$ have asymptotic expansions in $\varepsilon$
as functions of two different variables, the slow time parameter
$\tt = \varepsilon t$, and a phase variable $\Psi$ (also called
a ``fast-time parameter''), the dependence on which is periodic with
period $2 \pi$.   Thus we assume
\bes
\bea
\label{eq:ansatz1a}
q(t,\varepsilon) &=& \sum_{s=0}^\infty \varepsilon^s
q^{(s)}(\Psi,\tt)  \nonumber \\
&=& q^{(0)}(\Psi,\tt) + \varepsilon q^{(1)}(\Psi,\tt)
+ O(\varepsilon^2), \\
J(t,\varepsilon) &=& \sum_{s=0}^\infty \varepsilon^s
J^{(s)}(\Psi,\tt)  \nonumber \\
&=& J^{(0)}(\Psi,\tt) + \varepsilon J^{(1)}(\Psi,\tt)
+ O(\varepsilon^2).
\label{eq:ansatz1}
\eea
\ees
These asymptotic expansions are assumed to be uniform in $\tt$.
The expansion coefficients $J^{(s)}$ are each periodic
in the phase variable $\Psi$ with period $2 \pi$:
\be
J^{(s)}(\Psi + 2 \pi,\tt) = J^{(s)}(\Psi,\tt).
\ee
The phase variable $\Psi$ is chosen so that angle variable $q$
increases by $2\pi$ when $\Psi$ increases by $2\pi$; this
implies that the expansion coefficients $q^{(s)}$ satisfy
\bes
\bea
\label{eq:q0periodic}
q^{(0)}(\Psi + 2 \pi,\tt) &=& q^{(0)}(\Psi,\tt) + 2 \pi,\\
q^{(s)}(\Psi + 2 \pi,\tt) &=& q^{(s)}(\Psi,\tt), \ \ \ s \ge 1.
\eea
\ees

The angular velocity $\Omega = d \Psi/dt$ associated with the phase
$\Psi$ is assumed to depend only on the slow time variable ${\tilde
t}$ (so it can vary slowly with time), and on $\varepsilon$.  We
assume that it has an asymptotic expansion in $\varepsilon$ as
$\varepsilon \to 0$ which is uniform in $\tt$:
\begin{eqnarray}
\frac{d \Psi }{d t} &=& \Omega(\tt,\varepsilon) = \sum_{s=0}^\infty \varepsilon^s \Omega^{(s)}(\tt)
\\
&=& \Omega^{(0)}(\tt) + \varepsilon \Omega^{(1)}(\tt) +
O(\varepsilon^2).
\label{eq:Omegaexpand}
\end{eqnarray}
Equation (\ref{eq:Omegaexpand}) serves to define the phase variable
$\Psi$ in terms the angular velocity variables $\Omega^{(s)}(\tt)$, $s
=0,1,2 \ldots$, up to constants of integration.  One constant of
integration arises at each order in $\varepsilon$.
Without loss of generality we choose these
constants of integration so that
\be
q^{(s)}(0,\tt) =0
\label{eq:ansatz5}
\ee
for all $s$, $\tt$.  Note that this does not restrict the final solutions
$q(t,\varepsilon)$ and $J(t,\varepsilon)$, as we show explicitly
below, because there are additional constants of integration that
arise when solving for the functions $q^{(s)}(\Psi,\tt)$ and
$J^{(s)}(\Psi,\tt)$.

Roughly speaking, the meaning of these assumptions is the following.
The solution of the equations of motion consists of a mapping from
$(t,\varepsilon)$ to $(q,J)$.  That mapping contains dynamics on two
different timescales, the dynamical timescale $\sim 1$ and the slow
timescale $\sim 1/\varepsilon$.  The mapping can be uniquely
written the composition of two mappings
\be
(t,\varepsilon) \ \ \ \to \ \ \ (\Psi,\tt,\varepsilon) \ \ \ \to \ \ \
(q,J),
\ee
such that the first mapping contains all the fast dynamics, and is
characterized by the slowly evolving frequency
$\Omega(\tt,\varepsilon)$, and the second mapping contains dynamics
only on the slow timescale.

\subsection{Results of the two-timescale analysis}
\label{sec:results}

By substituting the ansatz (\ref{eq:ansatz1}) -- (\ref{eq:ansatz5})
into the equations of motion (\ref{eq:eom1}) we find that all of the
assumptions made in the ansatz can be satisfied, and that all of the
expansion coefficients are uniquely determined, order by order in
$\varepsilon$.  This derivation is given in Sec.\ \ref{sec:derivation}
below.  Here we list the results obtained for the various expansion
coefficients up to the leading and sub-leading orders.

\subsubsection{Terminology for various orders of the approximation}

We can combine the definitions just summarized to obtain an explicit
expansion for the quantity of most interest, the angle variable $q$ as a
function of time.
From the periodicity condition (\ref{eq:q0periodic}) it follows that the function
$q^{(0)}(\Psi,\tt)$ can be written as $\Psi + {\bar
  q}^{(0)}(\Psi,\tt)$ where ${\bar q}^{(0)}$ is a periodic function of
$\Psi$.  [We shall see that ${\bar q}^{(0)}$ in fact vanishes,
cf.\ Eq.\ (\ref{eq:q0ans}) below.]
From the definitions (\ref{sec2:eq:slowtimedef})
and (\ref{eq:Omegaexpand}), we can write the
phase variable $\Psi$ as
\be
\Psi = \frac{1}{\varepsilon} \psi^{(0)}(\tt) + \psi^{(1)}(\tt) +
\varepsilon \psi^{(2)}(\tt) + O(\varepsilon^2),
\ee
where the functions $\psi^{(s)}(\tt)$ are defined by
\be
\psi^{(s)}(\tt) = \int^\tt d \tt^\prime \Omega^{(s)}(\tt^\prime).
\label{psisdef}
\ee
Inserting this into the expansion (\ref{eq:ansatz1a}) of $q$ and using the
above expression for $q^{(0)}$ gives
\begin{eqnarray}
q(t,\varepsilon) &=& \frac{1}{\varepsilon} \psi^{(0)}(\tt) + \left[
  \psi^{(1)}(\tt) + {\bar q}^{(0)}(\Psi,\tt) \right] \nonumber \\
&& +
\varepsilon \left[ \psi^{(2)}(\tt) + q^{(1)}(\Psi,\tt) \right] +
O(\varepsilon^2).
\label{eq:qfinalans}
\end{eqnarray}
We will call the leading order, $O(1/\varepsilon)$ term in Eq.\
(\ref{eq:qfinalans}) the {\it adiabatic approximation}, the
sub-leading $O(1)$ term the {\it post-1-adiabatic} term, the next
$O(\varepsilon)$ term the {\it post-2-adiabatic} term, etc.
This choice of terminology is motivated by the terminology used in
post-Newtonian theory.

It is important to note that the expansion in powers of $\varepsilon$ in
Eq.\ (\ref{eq:qfinalans}) is {\it not} a straightforward power series
expansion at fixed $\tt$,
since the variable $\Psi$ depends on $\varepsilon$.
[The precise definition of the expansion of the solution which we are using is given
by Eqs.\ (\ref{eq:ansatz1a}) -- (\ref{eq:ansatz5}).]  Nevertheless,
the expansion (\ref{eq:qfinalans}) as written correctly captures the
$\varepsilon$ dependence of the secular
pieces of the solution, since the functions ${\bar q}^{(0)}$ and
$q^{(1)}$ are periodic functions of $\Psi$ and so have no secular pieces.

\subsubsection{Adiabatic Order}

First, the zeroth order action variable is given by
\be
J^{(0)}(\Psi,\tt) = {\cal J}^{(0)}(\tt),
\label{eq:J0ans}
\ee
where ${\cal J}^{(0)}$ satisfies the differential equation
\be
\frac{ d {\cal J}^{(0)}(\tt)} {d \tt} = G^{(1)}_0[ {\cal
J}^{(0)}(\tt), \tt ].
\label{eq:calJ0ans}
\ee
Here the right hand side denotes the average over $q$ of the
forcing term $G^{(1)}[q,{\cal J}^{(0)}(\tt),\tt]$, cf.\
Eqs.\ (\ref{eq:ftdefs0}) above.
The zeroth order angle variable is given by
\be
q^{(0)}(\Psi,\tt) = \Psi,
\label{eq:q0ans}
\ee
and the angular velocity $\Omega$ that defines the phase variable $\Psi$
is given to zeroth order by
\be
\Omega^{(0)}(\tt) = \omega[ {\cal J}^{(0)}(\tt), {\tilde
t}].
\label{eq:Omega0ans}
\ee
Note that this approximation is equivalent to the following simple
prescription: (i) Truncate the equations of motion (\ref{eq:eom1}) to
the leading order in $\varepsilon$:
\bes
\label{eq:eom1aa}
\bea
{\dot q}(t) &=& \omega(J,\tt) + \varepsilon g^{(1)}(q,J,{\tilde
t}), \\
{\dot J}(t) &=& \varepsilon G^{(1)}(q,J,\tt);
\eea
\ees
(ii) Omit the driving term $g^{(1)}$ in the equation for the angle
variable; and (iii) Replace the driving term $G^{(1)}$ in the equation for the action
variable with its average over $q$.

\subsubsection{Post-1-adiabatic Order}

Next, the first order action variable is given by
\be
J^{(1)}(\Psi,\tt) = \frac{ {\cal I}{\hat G}^{(1)}[\Psi,{\cal
J}^{(0)}(\tt),\tt] }{\Omega^{(0)}(\tt) } + {\cal J}^{(1)}(\tt),
\label{eq:J1ans}
\ee
where the symbol ${\cal I}$ on the right hand side denotes the integration
operator (\ref{eq:calIdef}) with respect to $\Psi$.
In Eq.\ (\ref{eq:J1ans}) the quantity ${\cal J}^{(1)}(\tt)$ satisfies
the differential equation
\bea
&&\frac{ d {\cal J}^{(1)}(\tt)}{d \tt} - \frac{ \partial G^{(1)}_0}
{\partial J}[{\cal J}^{(0)}(\tt),\tt] {\cal J}^{(1)}(\tt) \nn \\
&=&\frac{ \langle \frac{ \partial {\hat G}^{(1)}}{\partial J} {\cal I}
{\hat G}^{(1)} \rangle}{\Omega^{(0)}(\tt)}
- \frac{ \langle {\hat G}^{(1)} {\hat g}^{(1)} \rangle}{\Omega^{(0)}(\tt)}
 +  G^{(2)}_0.
\label{eq:calJ1ans}
\eea
Here it is understood that the quantities on the right hand side are
evaluated at $q = q^{(0)} = \Psi$ and $J = {\cal J}^{(0)}(\tt)$.  The sub-leading correction to
the phase variable $\Psi$ is given by
\be
\Omega^{(1)}(\tt) = \frac{\partial \omega}{\partial J}[ {\cal
J}^{(0)}(\tt),\tt] {\cal J}^{(1)}(\tt) + g^{(1)}_0[ {\cal
J}^{(0)}(\tt),\tt].
\label{eq:Omega1ans}
\ee

Finally, the sub-leading term in the angle variable is
\be
q^{(1)}(\Psi,\tt) = {\hat q}^{(1)}(\Psi,\tt) + {\cal Q}^{(1)}(\tt),
\label{eq:q1ans}
\ee
where
\bea
\label{eq:hatq1ans}
{\hat q}^{(1)}(\Psi,\tt) &=& \frac{1}{\Omega^{(0)}(\tt)^2 }
\frac{\partial \omega}{\partial J}[{\cal J}^{(0)}(\tt),\tt] \,
{\cal I}^2{\hat G}^{(1)}[\Psi,{\cal J}^{(0)}(\tt),\tt] \nn \\
&& + \frac{1}{\Omega^{(0)}(\tt) }
{\cal I}{\hat g}^{(1)}[\Psi,{\cal J}^{(0)}(\tt),\tt]
\eea
and
\be
{\cal Q}^{(1)}(\tt) = - {\hat q}^{(1)}(0,\tt).
\label{eq:calQ1ans}
\ee

\subsubsection{Discussion}
\label{resultsdiscussion}

One of the key results of the general analysis of this section
is the identification of which pieces of the external forces are
required to compute the adiabatic and post-1-adiabatic solutions.
From Eqs.\ (\ref{eq:calJ0ans}),
(\ref{eq:Omega0ans}) and (\ref{eq:qfinalans}),
the adiabatic solution depends only on the
averaged piece
$G^{(1)}_0(J,\tt) = \langle G^{(1)}(q,J,\tt) \rangle$ of the leading
order external force $G^{(1)}$.
This quantity is purely dissipative, as can be seen in the Kerr
inspiral context from Eqs.\ (\ref{eq:ident450}) and
(\ref{eq:ident450a}).  More generally,
if the perturbing forces $g$ and $G$ arise from
  a perturbation $\varepsilon \Delta H = \sum_s \varepsilon^s \Delta
  H^{(s)}$ to the Hamiltonian, then the forcing function $G^{(s)}$ is
$$
G^{(s)}(q,J,\tt) = - \frac{ \partial \Delta H^{(s)}(q,J,\tt)}{\partial
  q},
$$
and it follows that the average over $q$ of $G^{(s)}$ vanishes.

At the next order,
the post-1-adiabatic term $\psi^{(1)}(\tt)$ depends on the averaged piece
$G^{(2)}_0(J,\tt) = \langle G^{(2)}(q,J,\tt) \rangle$ of the sub-leading
force $G^{(2)}$, again purely dissipative, as well as the remaining
conservative and dissipative pieces of the leading order forces
$G^{(1)}(q,J,\tt)$ and $g^{(1)}(q,J,\tt)$.  This can be seen from
Eqs.\ (\ref{eq:calJ1ans}) and (\ref{eq:Omega1ans}).
These results have been previously discussed briefly in the EMRI
context in Refs.\ \cite{Hughes:2005qb,scalar}.
For circular, equatorial orbits,
the fact that there is a
post-1-adiabatic order contribution from the second order self-force
was first argued by Burko \cite{Burko:2002fd}.

\subsubsection{Initial conditions and the generality of our ansatz}
\label{sec:initialconditions}

We will show in the next subsection that our ansatz
(\ref{eq:ansatz1a}) -- (\ref{eq:ansatz5})
is compatible with the
one parameter family of differential equations (\ref{eq:eom1}).
However, it does not necessarily follow that our ansatz is compatible
with the most general one parameter family $[q(t,\varepsilon),
J(t,\varepsilon)]$ of solutions, because of the possibility of
choosing arbitrary, $\varepsilon$-dependent initial conditions
$q(0,\varepsilon)$ and $J(0,\varepsilon)$ at the
initial time $t =0$.\footnote{More generally we could consider
  specifying initial conditions at some time $t=t_0$.  In that case we
would modify the definition of the rescaled time coordinate to
  $\tt = \varepsilon( t - t_0)$.}  In general, the $\varepsilon$
dependence of the solutions arises from both the $\varepsilon$
dependence of the initial conditions
and the $\varepsilon$ dependence of the differential equations.  It is
possible to choose initial conditions which are incompatible with our
ansatz.

To see this explicitly, we evaluate the expansions
(\ref{eq:qfinalans}) and (\ref{eq:J1ans}) at $t = \tt = 0$.  This gives
\bes
\label{initialc}
\bea
q(0,\varepsilon) &=& \varepsilon^{-1} \psi^{(0)}(0) + \psi^{(1)}(0) +
O(\varepsilon), \\
J(0,\varepsilon) &=& {\cal J}^{(0)}(0) + \varepsilon {\cal J}^{(1)}(0)
\nn \\
&&+ \varepsilon
\frac{ {\cal I}{\hat G}^{(1)}[\varepsilon^{-1} \psi^{(0)}(0) + \psi^{(1)}(0),
{\cal J}^{(0)}(0),0] }{\omega[{\cal J}^{(0)},0] } \nn \\
&&+ O(\varepsilon^2).
\eea
\ees
Recalling that parameters $\psi^{(0)}(0)$, $\psi^{(1)}(0)$, ${\cal
  J}^{(0)}(0)$ and ${\cal J}^{(1)}(0)$ are assumed to be
independent of $\varepsilon$, we see that the conditions
(\ref{initialc}) strongly constrain the allowed $\varepsilon$
dependence of the initial conditions.
We note, however, that the choice of constant ($\varepsilon$
independent) initial conditions
\be
q(0,\varepsilon) = q_0, \ \ \ \ \ J(0,\varepsilon) = J_0
\ee
can be accommodated, which is
sufficient for most applications of the formalism.
To achieve this one chooses
\be
\psi^{(0)}(0) = 0, \ \ \ \psi^{(1)}(0) = q_0, \ \ \ {\cal J}^{(0)}(0) = J_0,
\ee
and
\be
{\cal J}^{(1)}(0) = -\frac{ {\cal I}{\hat G}^{(1)}[q_0,J_0,0]
}{\omega[J_0,0] }.
\ee

\subsection{Derivation}
\label{sec:derivation}

In this subsection we give the derivation of the results
(\ref{eq:J0ans}) -- (\ref{eq:calQ1ans}) summarized above.
At each order $s$ we introduce the notation ${\cal J}^{(s)}(\tt)$ for
the average part of $J^{(s)}(\Psi,\tt)$:
\be
{\cal J}^{(s)}(\tt) \equiv \langle J^{(s)}(\Psi,\tt)  \rangle
= \frac{1}{2\pi} \int_0^{2 \pi} J^{(s)}(\Psi,\tt) d\Psi.
\ee
We denote by ${\hat J}^{(s)}$ the remaining part of $J^{(s)}$, as in
Eq.\ (\ref{eq:hatfdef}).  This gives the decomposition
\be
J^{(s)}(\Psi,\tt) = {\cal J}^{(s)}(\tt) + {\hat J}^{(s)}(\Psi,\tt)
\label{eq:Jdecomposition}
\ee
for all $s \ge 0$.  Similarly for the angle variable we have the
decomposition
\be
q^{(s)}(\Psi,\tt) = {\cal Q}^{(s)}(\tt) + {\hat q}^{(s)}(\Psi,\tt)
\label{eq:qdecomposition}
\ee
for all $s \ge 1$.  [We do not use this notation for the $s=0$ case
for the angle variable, since $q^{(0)}$ is not a periodic function of
$\Psi$, by Eq.\ (\ref{eq:q0periodic})].

Using the expansions (\ref{eq:ansatz1a}) and (\ref{eq:ansatz1})
of $q$ and $J$ together with the expansion (\ref{eq:Omegaexpand}) of
$d\Psi/dt$, we obtain
\bea
\frac{dq}{dt} &=& \Omega^{(0)} q^{(0)}_{,\Psi} + \varepsilon \left[
\Omega^{(1)} q^{(0)}_{,\Psi} + \Omega^{(0)} q^{(1)}_{,\Psi} +
q^{(0)}_{,\tt} \right] \nn \\
&& + \varepsilon^2 \left[ \Omega^{(2)} q^{(0)}_{,\Psi} + \Omega^{(0)}
q^{(2)}_{,\Psi} + \Omega^{(1)} q^{(1)}_{,\Psi} + q^{(1)}_{,\tt}
\right] \nn \\
&& + O(\varepsilon^3).
\eea
Here we use commas to denote partial derivatives.
We now insert this expansion together with a similar expansion for
$dJ/dt$ into the equations of motion (\ref{eq:eom1}) and use the
expansions (\ref{eq:Gexpand}) and (\ref{eq:gexpand}) of the external
forces $g$ and $G$.   Equating coefficients\footnote{As is well known,
this procedure is valid for asymptotic series as well as normal power
series.} of powers of
$\varepsilon$ then gives at zeroth order
\bes
\label{eq:1var0order}
\bea
\Omega^{(0)} q^{(0)}_{,\Psi} &=& \omega, \\
\Omega^{(0)} J^{(0)}_{,\Psi} &=& 0,
\eea
\ees
at first order
\bes
\label{eq:1var1order}
\bea
\label{eq:1var1orderA}
\Omega^{(0)} q^{(1)}_{,\Psi} - \omega_{,J} J^{(1)} &=&
- \Omega^{(1)} q^{(0)}_{,\Psi} - q^{(0)}_{,\tt} + g^{(1)},\ \ \ \ \\
\Omega^{(1)} J^{(0)}_{,\Psi} + \Omega^{(0)} J^{(1)}_{,\Psi} &=& -
J^{(0)}_{,\tt} + G^{(1)},
\label{eq:1var1orderB}
\eea
\ees
and at second order
\bes
\label{eq:1var2order}
\bea
\Omega^{(0)} q^{(2)}_{,\Psi} - \omega_{,J} J^{(2)} &=&
\frac{1}{2} \omega_{,JJ} (J^{(1)})^2 + g^{(1)}_{,q} q^{(1)} +
g^{(1)}_{,J} J^{(1)} \nn \\
&& + g^{(2)} - \Omega^{(2)} q^{(0)}_{,\Psi}
- \Omega^{(1)} q^{(1)}_{,\Psi} \nn \\
&& - q^{(1)}_{,\tt}, \\
\Omega^{(2)} J^{(0)}_{,\Psi} + \Omega^{(0)} J^{(2)}_{,\Psi} &=&
G^{(1)}_{,q} q^{(1)} + G^{(1)}_{,J} J^{(1)} - \Omega^{(1)}
J^{(1)}_{,\Psi}
\nn \\
&& - J^{(1)}_{,\tt} + G^{(2)}.
\label{eq:1var2orderB}
\eea
\ees
Here it is understood that all functions of $q$ and $J$ are evaluated
at $q^{(0)}$ and $J^{(0)}$.

\subsubsection{Zeroth order analysis}

The zeroth order equations (\ref{eq:1var0order}) can be written more
explicitly as
\bes
\bea
\Omega^{(0)}(\tt) q^{(0)}_{,\Psi}(\Psi,\tt) &=&
\omega[J^{(0)}(\Psi,\tt),\tt], \\
\Omega^{(0)}(\tt) J^{(0)}_{,\Psi}(\Psi,\tt) &=& 0.
\eea
\ees
The second of these equations implies that $J^{(0)}$ is independent of $\Psi$,
so we obtain $J^{(0)}(\Psi,\tt) = {\cal J}^{(0)}(\tt)$.  The first
equation then implies that $q^{(0)}_{,\Psi}$ is independent of $\Psi$,
and integrating with respect to $\Psi$ gives
\be
q^{(0)}(\Psi,\tt) = \frac{ \omega[ {\cal
J}^{(0)}(\tt),\tt]}{\Omega^{(0)}(\tt) } \Psi + {\cal Q}^{(0)}(\tt),
\label{eq:v1}
\ee
where ${\cal Q}^{(0)}$ is some function of $\tt$.
The periodicity condition (\ref{eq:q0periodic}) now implies that the
coefficient of $\Psi$ in Eq.\ (\ref{eq:v1}) must be unity, which gives the
formula (\ref{eq:Omega0ans}) for the angular velocity $\Omega^{(0)}(\tt)$.
Finally, the assumption (\ref{eq:ansatz5}) forces ${\cal
Q}^{(0)}(\tt)$ to vanish, and we recover the formula (\ref{eq:q0ans}) for
$q^{(0)}(\Psi,\tt)$.

\subsubsection{First order analysis}

The first order equation (\ref{eq:1var1orderB}) can be written more
explicitly as
\bea
\label{eq:1var1orderBexplicit}
\Omega^{(0)}(\tt) J^{(1)}_{,\Psi}(\Psi,\tt) &=&
-{\cal J}^{(0)}_{,\tt}(\tt) \nn \\
&&+ G^{(1)}[\Psi,{\cal J}^{(0)}(\tt),\tt],
\eea
where we have simplified using the zeroth order solutions
(\ref{eq:J0ans}) and (\ref{eq:q0ans}). We now take the average with
respect to $\Psi$ of this equation.  The left hand side vanishes since
it is a total derivative, and we obtain using the definition
(\ref{eq:Gkdef}) the differential equation (\ref{eq:calJ0ans}) for
${\cal J}^{(0)}(\tt)$.
Next, we subtract from Eq.\ (\ref{eq:1var1orderBexplicit}) its
averaged part, and use the decomposition (\ref{eq:Jdecomposition}) of
$J^{(1)}$.  This gives
\bea
\label{eq:1var1orderBexplicit1}
\Omega^{(0)}(\tt) {\hat J}^{(1)}_{,\Psi}(\Psi,\tt) = {\hat G}^{(1)}[\Psi,{\cal J}^{(0)}(\tt),\tt].
\eea
We solve this equation using the Fourier decomposition
(\ref{eq:hatgformula}) of ${\hat G}^{(1)}$ to obtain
\be
{\hat J}^{(1)}(\Psi,\tt) = \sum_{k\ne 0}
\frac{ G^{(1)}_k[{\cal J}^{(0)}(\tt),\tt] e^{i k \Psi}} {i k \Omega^{(0)}(\tt)}.
\ee
This yields the first term in the result (\ref{eq:J1ans}) for
$J^{(1)}$ when we use the notation (\ref{eq:calIdef}).

Next, we simplify the first order equation (\ref{eq:1var1orderA})
using the zeroth order solutions (\ref{eq:J0ans}) and
(\ref{eq:q0ans}), to obtain
\bea
\label{eq:1var1orderAexplicit}
&&\Omega^{(0)}(\tt) q^{(1)}_{,\Psi}(\Psi,\tt) - \omega_{,J}[{\cal
J}^{(0)}(\tt),\tt] J^{(1)}[\Psi,\tt] \nn \\
&& \ \ = - \Omega^{(1)}(\tt)  + g^{(1)}[\Psi,{\cal J}^{(0)}(\tt),\tt].
\eea
Averaging with respect to $\Psi$ and using the decompositions
(\ref{eq:Jdecomposition}) and (\ref{eq:qdecomposition}) of
$J^{(1)}$ and $q^{(1)}$ now gives the formula (\ref{eq:Omega1ans}) for
$\Omega^{(1)}(\tt)$.  Note however that the function ${\cal
J}^{(1)}(\tt)$ in that formula has not yet been determined; it will be
necessary to go to one higher order to compute this function.

Finally, we subtract from Eq.\ (\ref{eq:1var1orderAexplicit}) its average
over $\Psi$ using the decompositions (\ref{eq:Jdecomposition}) and
(\ref{eq:qdecomposition}) and then integrate with respect to $\Psi$
using the notation (\ref{eq:calIdef}).  This gives
\bea
\label{eq:hatq1ansA}
{\hat q}^{(1)}(\Psi,\tt) &=& \frac{1}{\Omega^{(0)}(\tt) } \bigg\{
\omega_{,J}[{\cal J}^{(0)}(\tt),\tt] \,
{\cal I}{\hat J}^{(1)}[\Psi,\tt] \nn \\
&& + {\cal I}{\hat g}^{(1)}[\Psi,{\cal J}^{(0)}(\tt),\tt] \bigg\}.
\eea
Using the result for ${\hat J}^{(1)}$ given by the first term in Eq.\
(\ref{eq:J1ans}) now yields
the formula (\ref{eq:hatq1ans}) for ${\hat q}^{(1)}(\Psi,\tt)$, and
the result (\ref{eq:q1ans}) for $q^{(1)}$ then follows from the decomposition
(\ref{eq:qdecomposition}) together with the initial condition
(\ref{eq:ansatz5}).

\subsubsection{Second order analysis}

We simplify the second order equation (\ref{eq:1var2orderB})
using the zeroth order solutions (\ref{eq:J0ans})
and (\ref{eq:q0ans}), average
over $\Psi$, and simplify using the
decompositions (\ref{eq:Jdecomposition}) and
(\ref{eq:qdecomposition}) and the identities (\ref{eq:ident1}).
The result is
\bea
 {\cal J}_{,\tt}^{(1)}(\tt) &=&   G^{(1)}_{0,J}
[{\cal J}^{(0)}(\tt),\tt] {\cal J}^{(1)}(\tt)
+ G^{(2)}_0[ {\cal J}^{(0)}(\tt), \tt ]
\nn \\
&& + \left\langle {\hat q}^{(1)}(\Psi,\tt) \, {\hat G}^{(1)}_{,q}[ \Psi, {\cal
J}^{(0)}(\tt),\tt] \right\rangle \nn \\
&& + \left\langle {\hat J}^{(1)}(\Psi,\tt) \, {\hat G}^{(1)}_{,J}[ \Psi, {\cal
J}^{(0)}(\tt),\tt] \right\rangle.
\label{eq:calJ1ansA}
\eea
Using the expressions (\ref{eq:hatq1ans}) and (\ref{eq:J1ans}) for ${\hat
q}^{(1)}$ and ${\hat J}^{(1)}$ and simplifying using the identities
(\ref{eq:ident2}) now gives the differential equation (\ref{eq:calJ1ans}) for
${\cal J}^{(1)}$.

\subsubsection{Extension to arbitrary order}
\label{subsec:induction}

In this subsection we prove by induction that solutions are uniquely
determined at each order in $\varepsilon$.  Our inductive hypothesis
is that, given the equations up to order $s$, we can compute all of
the expansion coefficients $q^{(u)}(\Psi,\tt)$, $J^{(u)}(\Psi,\tt)$
and $\Omega^{(u)}(\tt)$ for $0 \le u \le s$, except for the averaged
piece ${\cal J}^{(s)}(\tt)$ of $J^{(s)}(\Psi,\tt)$, and except for
$\Omega^{(s)}(\tt)$, which is assumed to be determined by
${\cal J}^{(s)}(\tt)$.  From the preceding subsections this
hypothesis is true for $s=0$ and for $s=1$.  We shall assume it is true
at order $s-1$ and prove it is true at order $s$.

The equations of motion at order $s$, when simplified using the zeroth
zeroth order solutions (\ref{eq:J0ans}) and (\ref{eq:q0ans}), can be
written as
\bes
\label{eq:1varsorder}
\bea
\label{eq:1varsorderA}
\Omega^{(0)} q^{(s)}_{,\Psi} + \Omega^{(s)} - \omega_{,J} J^{(s)} &=&
\omega_{,JJ} J^{(1)} J^{(s-1)} + g^{(1)}_{,q} q^{(s-1)}
\nn \\ &&
+g^{(1)}_{,J} J^{(s-1)}
- \Omega^{(1)} q^{(s-1)}_{,\Psi} \nn \\
&& - \Omega^{(s-1)} q^{(1)}_{,\Psi}
- q^{(s-1)}_{,\tt}, \nn \\
&& + {\cal S}  \\
\Omega^{(0)} J^{(s)}_{,\Psi} &=&
G^{(1)}_{,q} q^{(s-1)} + G^{(1)}_{,J} J^{(s-1)} \nn \\
&&
 - \Omega^{(s-1)} J^{(1)}_{,\Psi} - \Omega^{(1)} J^{(s-1)}_{,\Psi}
\nn \\
&& - J^{(s-1)}_{,\tt} + {\cal T}.
\label{eq:1varsorderB}
\eea
\ees
Here ${\cal S} = {\cal S}(\Psi,\tt)$ and ${\cal T} = {\cal
T}(\Psi,\tt)$ are expressions involving the forces $G^{(u)}$ and
$g^{(u)}$ for $0 \le u \le s$ evaluated at $q = q^{(0)} = \Psi$ and
$J = J^{(0)} = {\cal J}^{(0)}$, and involving the coefficients
$q^{(u)}$, $J^{(u)}$ and $\Omega^{(u)}$ for $0 \le u \le s-2$ which by
the inductive hypothesis are known.  Therefore we can treat ${\cal S}$
and ${\cal T}$ as known functions.

Averaging Eq.\ (\ref{eq:1varsorderB}) over $\Psi$ yields the differential
equation
\bea
{\cal J}^{(s-1)}_{,\tt} - G^{(1)}_{0,J} {\cal J}^{(s-1)} &=& \langle
{\cal T} \rangle
+ \langle {\hat G}^{(1)}_{,q} {\hat q}^{(s-1)} \rangle
 \nn \\
&&
+ \langle {\hat G}^{(1)}_{,J} {\hat J}^{(s-1)} \rangle.
\eea
By the inductive hypothesis all the terms on the right hand side are
known, so we can solve this differential equation to determine ${\cal
J}^{(s-1)}$.

Next, averaging Eq.\ (\ref{eq:1varsorderA}) yields
\bea
\Omega^{(s)} - \omega_{,J} {\cal J}^{(s)} &=& \omega_{,JJ} \langle
{\hat J}^{(1)} {\hat J}^{(s-1)} \rangle
+ \omega_{,JJ} {\cal J}^{(1)} {\cal J}^{(s-1)}
\nn \\
&& + \langle {\hat g}^{(1)}_{,q} {\hat q}^{(s-1)} \rangle +
\langle {\hat g}^{(1)}_{,J} {\hat J}^{(s-1)} \rangle \nn \\
&& + g^{(1)}_{0,J} {\cal J}^{(s-1)} - {\cal Q}^{(s-1)}_{,\tt} +
\langle {\cal S} \rangle.
\eea
Since ${\cal J}^{(s-1)}$ has already been determined, the right hand
side of this equation is known and therefore the equation can be used
to solve for $\Omega^{(s)}$ once ${\cal J}^{(s)}$ is specified, in
accord with the inductive hypothesis.
Next, Eq.\ (\ref{eq:1varsorderB}) with the average part subtracted
can be used to solve for ${\hat J}^{(s)}$, and once ${\hat J}^{(s)}$
is known Eq.\ (\ref{eq:1varsorderA}) with the average part subtracted
can be used to solve for ${\hat q}^{(s)}$.  Finally, the averaged
piece ${\cal Q}^{(s)}(\tt)$ of $q^{(s)}(\Psi,\tt)$ can be computed
from ${\hat q}^{(s)}$ using the initial condition (\ref{eq:ansatz5}) and the
decomposition (\ref{eq:qdecomposition}). Thus the inductive hypothesis
is true at order $s$ if it is true at order $s-1$.

\section{SYSTEMS WITH SEVERAL DEGREES OF FREEDOM SUBJECT TO
  NON-RESONANT FORCING}
\label{sec2:manyvariables}

\subsection{Overview}

In this section we generalize the analysis of the preceding section to the general
system of equations (\ref{sec2:eq:eom1})
with several degrees of freedom.
For convenience we reproduce those equations here:
\bes
\label{sec5:eq:eom1}
\bea
\label{sec5:eq:eom1a}
\frac{d q_\alpha}{dt} &=& \omega_\alpha(\bfJ,\tt) + \varepsilon
g_\alpha^{(1)}(\bfq,\bfJ,{\tilde t})
+ \varepsilon^2 g_\alpha^{(2)}(\bfq,\bfJ,{\tilde t}) \nonumber \\
&& + O(\varepsilon^3) , \ \ \ 1 \le \alpha \le N,
\ \ \ \\
\frac{d J_\lambda}{dt} &=& \varepsilon
G^{(1)}_\lambda(\bfq,\bfJ,\tt)
+ \varepsilon^2 G^{(2)}_\lambda(\bfq,\bfJ,\tt) \nonumber \\
&& + O(\varepsilon^3), \ \ \ 1 \le \lambda \le M.
\eea
\ees
For the remainder of this paper, unless otherwise specified,
indices $\alpha,\beta,\gamma,\delta,\varepsilon, \ldots$ from the
start of the Greek alphabet will run over
$ 1 \ldots N$, and indices $\lambda,\mu,\nu,\rho,\sigma,\ldots$ from
the second half of the alphabet
will run over $1 \ldots M$.

The generalization from one to several variables is straightforward
except for the treatment of resonances \cite{Kevorkian}.
The key new feature in the $N$ variable case is that the asymptotic expansions now
have additional terms proportional to $\sqrt{\varepsilon}$,
$\varepsilon^{3/2}, \ldots$ as well as the integer powers of
$\varepsilon$.  The coefficients of these half-integer powers of
$\varepsilon$ obey source-free differential equations, except at
resonances.  Therefore, in the absence of resonances, all of these
coefficients can be set to zero without loss of generality.
In this paper we develop the general theory with both types of terms
present, but we specialize to the case where no resonances occur.
Subsequent papers \cite{FH08a,FH08b} will extend the treatment to include
resonances, and derive
the form of the
source terms for the half-integer power coefficients.

We start in Sec.\ \ref{sec2:fourier} by defining our conventions and
notations for Fourier decompositions of the perturbing forces.
In Sec.\ \ref{sec:noresonanceassumption} we discuss the assumptions we make that prevent the
occurrence of resonances in the solutions.
The heart of the method is the ansatz we make for the form of the
solutions, which is given in Sec.\ \ref{sec2:solnansatz}.
Section \ref{sec2:results} summarizes the results we obtain at each order in the
expansion, and the derivations are given in Sec.\ \ref{sec2:sec:derivation}.
The implications of the results are discussed in detail in Sec.\
\ref{sec:discussion} below.

\subsection{Fourier expansions of perturbing forces}
\label{sec2:fourier}

The periodicity condition (\ref{eq:periodic0})
applies at each order in the expansion in powers
of $\varepsilon$, so we obtain
\bes
\bea
g^{(s)}_\alpha(\bfq+2\pi\bfk,\bfJ,\tt) &=& g^{(s)}_\alpha(\bfq,\bfJ,\tt), \\
G^{(s)}_\lambda(\bfq+2\pi\bfk,\bfJ,\tt) &=& G^{(s)}_\lambda(\bfq,\bfJ,\tt),
\eea
\label{eq:periodic33}\ees
for $s \ge 1$, $1 \le \alpha \le N$, and $1 \le \lambda \le M$.
Here $\bfk = (k_1, \ldots, k_N)$ can be an arbitrary $N$-tuple of integers.
It follows from Eqs.\ (\ref{eq:periodic33}) that these functions can
be expanded as multiple Fourier series:
\bes
\bea
g^{(s)}_\alpha(\bfq,\bfJ,\tt) &=& \sum_{\bfk}
g^{(s)}_{\alpha\,\bfk}(\bfJ,{\tilde t}) e^{i \bfk \cdot \bfq}, \\
G^{(s)}_\lambda(\bfq,\bfJ,\tt) &=& \sum_{\bfk}
G^{(s)}_{\lambda\,\bfk}(\bfJ,{\tilde t}) e^{i \bfk \cdot \bfq},
\eea
\ees
where
\bes
\label{sec2:eq:Gkdef}
\bea
g^{(s)}_{\alpha\,\bfk}(\bfJ,\tt) &=& \frac{1 }{ (2 \pi)^N} \int d^Nq \ e^{-i \bfk
\cdot \bfq} \, g^{(s)}_\alpha(\bfq,\bfJ,\tt), \ \ \ \ \ \\
G^{(s)}_{\lambda\,\bfk}(\bfJ,\tt) &=& \frac{1 }{ (2 \pi)^N} \int d^Nq \ e^{-i \bfk
\cdot \bfq} \, G^{(s)}_\lambda(\bfq,\bfJ,\tt).\ \ \ \ \
\eea
\label{sec2:ftdefs}\ees
Here we adopt the usual notations
\be
\sum_\bfk \equiv \sum_{k_1=-\infty}^\infty \ldots
\sum_{k_N=-\infty}^\infty,
\ee
\be
\int d^N q \equiv \int_0^{2\pi} dq_1 \ldots \int_0^{2\pi} dq_N.
\ee
and
\be
\bfk \cdot \bfq \equiv \sum_{\alpha=1}^N k_\alpha q_\alpha.
\ee

For any multiply periodic function $f=f(\bfq)$, we introduce the notation
\be
\langle f \rangle = \frac{1 }{ (2 \pi)^N} \int d^Nq f(\bfq)
\label{sec2:averagingdef}
\ee
for the average part of $f$, and
\be
{\hat f}(\bfq) = f(\bfq) - \langle f \rangle
\label{sec2:eq:hatfdef}
\ee
for the remaining part of $f$.
It follows from these definitions that
\be
\langle g^{(s)}_\alpha(\bfq,\bfJ,\tt) \rangle = g^{(s)}_{\alpha\,\bfzero}(\bfJ,\tt),\ \
\
\langle G^{(s)}_\lambda(\bfq,\bfJ,\tt) \rangle = G^{(s)}_{\lambda\,\bfzero}(\bfJ,\tt),
\ee
and that
\bes
\bea
{\hat g}^{(s)}_\alpha(\bfq,\bfJ,\tt) &=&  \sum_{\bfk\ne \bfzero}
g^{(s)}_{\alpha\,\bfk}(\bfJ,\tt) e^{i \bfk \cdot \bfq}, \\
{\hat G}^{(s)}_\lambda(\bfq,\bfJ,\tt) &=& \sum_{\bfk \ne \bfzero}
G^{(s)}_{\lambda\,\bfk}(\bfJ,\tt) e^{i \bfk \cdot \bfq}.
\label{sec2:eq:hatgformula}
\eea
\ees
We also have the identities
\bes
\label{sec2:eq:ident1}
\bea
\left\langle \frac{\partial f}{\partial q_\alpha} \right \rangle &=& \langle {\hat
f} \rangle =0 \\
\langle f g \rangle &=& \langle {\hat f} {\hat g} \rangle +
\langle f \rangle \langle g \rangle
\eea
\ees
for any multiply periodic functions $f(\bfq)$, $g(\bfq)$.

For any multiply periodic function $f$ and for any vector ${\bf v} =
(v_1, \ldots, v_N)$,  we also define the quantity
${\cal I}_{\bf v}{\hat f}$ by
\be
({\cal I}_{\bf v}{\hat f})(\bfq) \equiv  \sum_{\bfk \ne \bfzero}
\frac{f_\bfk}{i
\bfk \cdot {\bf v}} e^{i \bfk \cdot \bfq},
\label{sec2:eq:calIdef}
\ee
where $f_\bfk = \int d^Nq e^{-i \bfk \cdot \bfq} f(\bfq)/(2\pi)^N$ are
the Fourier coefficients of $f$.  The operator ${\cal I}_\bfv$
satisfies the identities
\bes
\label{sec2:eq:ident2}
\bea
{\cal I}_\bfv (\bfv \cdot \bfnabla {\hat f}) &=& {\hat f}, \\
\langle ( {\cal I}_\bfv {\hat f}) {\hat g} \rangle &=& - \langle {\hat f}
({\cal I}_\bfv {\hat g}) \rangle, \\
\langle {\hat f} ({\cal I}_\bfv {\hat f}) \rangle &=&0.
\eea
\ees

\subsection{The no-resonance assumption}
\label{sec:noresonanceassumption}

For each set of action variables ${\bf J}$ and for each time ${\tilde
  t}$, we will say that an N-tuple of integers $\bfk \ne 0$
is a {\it resonant N-tuple} if
\be
\bfk \cdot \bfomega(\bfJ,\tt)=0.
\label{resonantNdef}
\ee
where $\bfomega = (\omega_1, \ldots, \omega_N)$ are the frequencies
that appear on the right hand side of the equation of motion
(\ref{sec2:eq:eom1a}).
This condition governs the occurrence of
resonances in our perturbation expansion, as is well known in context
of perturbations of multiply periodic Hamiltonian systems \cite{Arnold}.
We will assume that for a given $\bfk$, the set of values of $\tt$
at which the quantity
\be
\sigma_\bfk(\tt) \equiv \bfk \cdot \bfomega[\bfcalJ^{(0)}(\tt), \tt]
\ee
vanishes (i.e.\ the resonant values) consists of isolated points.
Here $\bfcalJ^{(0)}(\tt)$ is the leading order solution for $\bfJ$ given by
Eq.\ (\ref{sec2:eq:calJ0ans}) below.
This assumption excludes persistent resonances that last for a finite interval in $\tt$.
Generically we expect this to be true because of the time dependence
of $\bfcalJ^{(0)}(\tt)$.

Our no-resonance assumption is essentially that the
Fourier components of the forcing terms vanish for resonant N-tuples.
More precisely, for each fixed $\bfk$ and for each time $\tt_{\rm r}$
for which $\sigma_\bfk(\tt_{\rm r}) =0$,
we assume that
\bes
\label{eq:noresonance}
\bea
g^{(s)}_{\alpha\,\bfk}\left[\bfcalJ^{(0)}(\tt),\tt\right] &=& 0, \\
G^{(s)}_{\lambda\,\bfk}\left[\bfcalJ^{(0)}(\tt),\tt\right] &=& 0,
\eea
\ees
for $s \ge 1$ and for all $\tt$ in a neighborhood of $\tt_{\rm r}$.
Our no-resonance assumption will
be relaxed in the forthcoming papers \cite{FH08a,FH08b}.

In our application to inspirals in Kerr black holes, the no-resonance
condition will be automatically satisfied for two classes of orbits:
circular and equatorial orbits.
This is because for these orbits there is either no radial motion, or
no motion in $\theta$, and so the two-dimensional torus
$(q_r,q_\theta)$ reduces to a one-dimensional circle.  The resonance
condition $k_r \omega_r + k_\theta \omega_\theta=0$ reduces to $k_r
\omega_r=0$ for equatorial orbits, or
$k_\theta \omega_\theta =0$ for circular orbits, and these conditions
can never be satisfied since the fundamental frequencies $\omega_r$
and $\omega_\theta$ are positive.


\subsection{Two timescale ansatz for the solution}
\label{sec2:solnansatz}

We now discuss the two-timescale {\it ansatz} we use for the form of the solutions
of the equations of motion.  This ansatz will be justified a
posteriori order by order in $\sqrt{\varepsilon}$.
Our ansatz essentially consists of the assumption that
the mapping from $(t,\varepsilon)$ to
$(\bfq,\bfJ)$ can be written as an asymptotic expansion in
$\sqrt{\varepsilon}$, each term of which
is the composition of two maps, the first
from $(t,\varepsilon)$ to an abstract $N$-torus with coordinates $\bfPsi
= (\Psi_1, \ldots, \Psi_N)$, and the second from $(\bfPsi,\tt,\varepsilon)$ to
$(\bfq,\bfJ)$.  Here $\tt = \varepsilon t$ is the slow time parameter.
All the fast timescale dynamics is encapsulated in the first mapping.
More precisely, we assume
\bes
\label{sec2:eq:ansatzg}
\bea
\label{sec2:eq:ansatz1a}
q_\alpha(t,\varepsilon) &=& \sum_{n=0}^\infty \varepsilon^{n/2}
q^{(n/2)}_\alpha(\bfPsi,\tt)  \nonumber \\
&=& q^{(0)}_\alpha(\bfPsi,\tt)
+ \sqrt{\varepsilon} q^{(1/2)}_\alpha(\bfPsi,\tt)
+ \varepsilon q^{(1)}_\alpha(\bfPsi,\tt) \nonumber \\
&&+ \varepsilon^{3/2} q^{(3/2)}_\alpha(\bfPsi,\tt)
+ O(\varepsilon^{2}), \\
J_\lambda(t,\varepsilon) &=& \sum_{n=0}^\infty \varepsilon^{n/2}
J^{(n/2)}_\lambda(\bfPsi,\tt)  \nonumber \\
&=& J^{(0)}_\lambda(\bfPsi,\tt) + \sqrt{\varepsilon} J^{(1/2)}_\lambda(\bfPsi,\tt)
+ \varepsilon J^{(1)}_\lambda(\bfPsi,\tt) \nonumber \\
&& + \varepsilon^{3/2} J^{(3/2)}_\lambda(\bfPsi,\tt) +
O(\varepsilon^2).
\label{sec2:eq:ansatz1}
\eea
\ees
These asymptotic expansions are assumed to be uniform in $\tt$.
The expansion coefficients $J^{(s)}_\lambda$, where $s = 0,1/2,1,\ldots$, are multiply periodic
in the phase variables $\Psi_\alpha$ with period $2 \pi$ in each variable:
\be
J^{(s)}_\lambda(\bfPsi + 2 \pi \bfk,\tt) = J^{(s)}_\lambda(\bfPsi,\tt).
\label{sec2:Jsperiodic}
\ee
Here $\bfk = (k_1, \ldots, k_N)$ is an arbitrary $N$-tuple of integers.
The mapping of the abstract $N$-torus with coordinates $\bfPsi$ into the
torus in phase space parameterized by $\bfq$ is assumed to have a
trivial wrapping, so that the angle variable $q_\alpha$
increases by $2\pi$ when $\Psi_\alpha$ increases by $2\pi$; this
implies that the expansion coefficients $q^{(s)}$ satisfy
\bes
\bea
\label{sec2:eq:q0periodic}
q^{(0)}_\alpha(\bfPsi + 2 \pi \bfk,\tt) &=& q^{(0)}_\alpha(\bfPsi,\tt)
+ 2 \pi k_\alpha,\\
q^{(s)}_\alpha(\bfPsi + 2 \pi \bfk,\tt) &=&
q^{(s)}_\alpha(\bfPsi,\tt), \ \ \ s \ge 1/2,
\eea
\ees
for arbitrary $\bfk$.  The variables $\Psi_1, \ldots, \Psi_N$ are
sometimes called ``fast-time parameters''.

The angular velocity
\be
\Omega_\alpha = d \Psi_\alpha/dt
\label{sec2:omegadef}
\ee
associated
with the phase $\Psi_\alpha$ is assumed to depend only on the slow
time variable ${\tilde t}$ (so it can vary slowly with time), and on
$\varepsilon$.  We assume that it has an asymptotic expansion in
$\sqrt{\varepsilon}$ as $\varepsilon \to 0$ which is uniform in $\tt$:
\begin{eqnarray}
\Omega_\alpha(\tt,\varepsilon) &=& \sum_{n=0}^\infty \varepsilon^{n/2}
\Omega^{(n/2)}_\alpha(\tt)\\
&=& \Omega^{(0)}_\alpha(\tt) + \sqrt{\varepsilon} \Omega^{(1/2)}_\alpha(\tt)
+ \varepsilon \Omega^{(1)}_\alpha(\tt) \nn \\
&& + \varepsilon^{3/2} \Omega^{(3/2)}_\alpha(\tt)
+ O(\varepsilon^2).
\label{sec2:eq:Omegaexpand}
\end{eqnarray}
Equations (\ref{sec2:omegadef}) and (\ref{sec2:eq:Omegaexpand}) serve
to define the phase variable
$\Psi_\alpha$ in terms the angular velocity variables
$\Omega^{(s)}_\alpha(\tt)$, $s = 0,1/2,1 \ldots$, up to constants of
integration.  One constant of
integration arises at each order in $\sqrt{\varepsilon}$, for each $\alpha$.
Without loss of generality we choose these
constants of integration so that
\be
q^{(s)}_\alpha(\bfzero,\tt) =0
\label{sec2:eq:ansatz5}
\ee
for all $\alpha$, $s$ and $\tt$.  Note that this does not restrict the
final solutions $q_\alpha(t,\varepsilon)$ and
$J_\lambda(t,\varepsilon)$, as we show explicitly below, because there
are additional constants of integration that arise when solving for
the functions $q^{(s)}_\alpha(\bfPsi,\tt)$ and $J^{(s)}_\lambda(\bfPsi,\tt)$.

\subsection{Results of the two-timescale analysis}
\label{sec2:results}

By substituting the ansatz (\ref{sec2:eq:ansatz1}) -- (\ref{sec2:eq:ansatz5})
into the equations of motion (\ref{sec2:eq:eom1}) we find that all of the
assumptions made in the ansatz can be satisfied, and that all of the
expansion coefficients are uniquely determined, order by order in
$\sqrt{\varepsilon}$.  This derivation is given in Sec.\ \ref{sec2:sec:derivation}
below.  Here we list the results obtained for the various expansion
coefficients up to the first three orders.

\subsubsection{Terminology for various orders of the approximation}
\label{sec:p1a}

We can combine the definitions just summarized to obtain an explicit
expansion for the quantity of most interest, the angle variables $q_\alpha$ as a
function of time.
From the periodicity condition (\ref{eq:q0periodic}) it follows that the function
$q^{(0)}_\alpha(\bfPsi,\tt)$ can be written as $\Psi_\alpha + {\bar
  q}_\alpha^{(0)}(\bfPsi,\tt)$ where ${\bar q}_\alpha^{(0)}$ is a
multiply periodic function of
$\bfPsi$.
From the definitions (\ref{sec2:eq:slowtimedef})
and (\ref{sec2:eq:Omegaexpand}), we can write the
phase variables $\Psi_\alpha$ as
\bea
\Psi_\alpha &=& \frac{1}{\varepsilon} \psi_\alpha^{(0)}(\tt)
+ \frac{1}{\sqrt{\varepsilon}} \psi_\alpha^{(1/2)}(\tt)
+ \psi^{(1)}_\alpha(\tt) +
\sqrt{\varepsilon} \psi_\alpha^{(3/2)}(\tt) \nn \\
&&
+ \varepsilon \psi_\alpha^{(2)}(\tt) + O(\varepsilon^{3/2}),
\eea
where the functions $\psi_\alpha^{(s)}(\tt)$ are defined by
\be
\psi_\alpha^{(s)}(\tt) = \int^\tt d \tt^\prime \Omega_\alpha^{(s)}(\tt^\prime).
\label{jks}
\ee
Inserting this into the expansion (\ref{sec2:eq:ansatz1a}) of
$q_\alpha$
gives
\bea
q_\alpha(t,\varepsilon) &=& \frac{1}{\varepsilon} \psi_\alpha^{(0)}(\tt)
+ \frac{1}{\sqrt{\varepsilon}} \psi_\alpha^{(1/2)}(\tt)
\nonumber \\ &&
+ \left[ \psi_\alpha^{(1)}(\tt) + {\bar q}_\alpha^{(0)}(\bfPsi,\tt)
\right]
\nonumber \\ &&
+ \sqrt{\varepsilon} \left[ \psi_\alpha^{(3/2)}(\tt) +
  q_\alpha^{(1/2)}(\bfPsi,\tt) \right]
\nn \\
&&
+ \varepsilon \left[ \psi_\alpha^{(2)}(\tt) + q_\alpha^{(1)}(\bfPsi,\tt) \right] +
O(\varepsilon^{3/2}). \ \ \ \
\label{sec2:eq:qfinalans}
\eea
We will call the leading order, $O(1/\varepsilon)$ term in Eq.\
(\ref{sec2:eq:qfinalans}) the {\it adiabatic approximation}, the
sub-leading $O(1/\sqrt{\varepsilon})$ term the {\it post-1/2-adiabatic} term, the next
$O(1)$ term the {\it post-1-adiabatic} term, etc.
Below we will see that the functions
${\bar q}_\alpha^{(0)}$ and $q^{(1/2)}_\alpha$
in fact vanish identically,
and so the oscillatory, $\bfPsi$-dependent terms in the expansion (\ref{sec2:eq:qfinalans})
arise only at post-2-adiabatic and higher orders.

As before we note that the expansion in powers of $\varepsilon$ in
Eq.\ (\ref{sec2:eq:qfinalans}) is {\it not} a straightforward power
series expansion at fixed $\tt$,
since the variable $\Psi$ depends on $\varepsilon$.
[The precise definition of the expansion of the solution which we are using is given
by Eqs.\ (\ref{sec2:eq:ansatz1a}) -- (\ref{sec2:eq:ansatz5}).]  Nevertheless,
the expansion (\ref{sec2:eq:qfinalans}) as written correctly captures the
$\varepsilon$ dependence of the secular
pieces of the solution, since the functions ${\bar q}^{(0)}$,
$q^{(1/2)}_\alpha$ and $q^{(1)}_\alpha$ are multiply periodic
functions of $\Psi$ and so have no secular pieces.

\subsubsection{Adiabatic Order}

The zeroth order action variables are given by
\be
J^{(0)}_\lambda(\bfPsi,\tt) = {\cal J}^{(0)}_\lambda(\tt),
\label{sec2:eq:J0ans}
\ee
where $\bfcalJ^{(0)}(\tt) = \big({\cal J}^{(0)}_1(\tt), \ldots,
{\cal J}^{(0)}_M(\tt) \big)$
satisfies the set of coupled ordinary differential equations
\be
\frac{ d {\cal J}^{(0)}_\lambda(\tt)} {d \tt} = G^{(1)}_{\lambda\,\bfzero}[
{\bfcalJ}^{(0)}(\tt), \tt ].
\label{sec2:eq:calJ0ans}
\ee
Here the right hand side denotes the average over ${\bf q}$ of the
forcing term $G^{(1)}_\lambda[\bfq,\bfcalJ^{(0)}(\tt),\tt]$, cf.\
Eqs.\ (\ref{sec2:ftdefs}) above.
The zeroth order angle variables are given by
\be
q^{(0)}_\alpha(\bfPsi,\tt) = \Psi_\alpha,
\label{sec2:eq:q0ans}
\ee
and the angular velocity $\Omega_\alpha$ that defines the phase
variable $\Psi_\alpha$ is given to zeroth order by
\be
\Omega^{(0)}_\alpha(\tt) = \omega_\alpha[ {\bfcalJ}^{(0)}(\tt), \tt].
\label{sec2:eq:Omega0ans}
\ee
Note that this approximation is equivalent to the following simple
prescription: (i) Truncate the equations of motion (\ref{sec5:eq:eom1}) to
the $O(\varepsilon)$;
(ii) Omit the driving terms $g^{(1)}_\alpha$ in the equations for the angle
variables; and (iii) Replace the driving terms $G^{(1)}_\lambda$ in the equations for the action
variables with their averages over ${\bf q}$.

\subsubsection{Post-1/2-adiabatic order}

Next, the $O(\sqrt{\varepsilon})$ action variables are given by
\be
J^{(1/2)}_\lambda(\bfPsi,\tt) = {\cal J}^{(1/2)}_\lambda(\tt),
\label{sec2:eq:J0.5ans}
\ee
where $\bfcalJ^{(1/2)}(\tt) = \big({\cal J}^{(1/2)}_1(\tt), \ldots,
{\cal J}^{(1/2)}_M(\tt) \big)$
satisfies the set of coupled, source-free ordinary differential equations
\be
\frac{ d {\cal J}^{(1/2)}_\lambda(\tt)} {d \tt} - \frac{\partial G^{(1)}_{\lambda\,\bfzero}}{\partial J_\mu}[{\bfcalJ}^{(0)}(\tt), \tt ] {\cal J}^{(1/2)}_\mu(\tt)=0.
\label{sec2:eq:calJ0.5ans}
\ee
Equation (\ref{sec2:eq:calJ0.5ans})
will acquire a source term in Ref.\ \cite{FH08b} where
we include the effects of resonances.
The $O(\sqrt{\varepsilon})$ angle variables are given by
\be
q^{(1/2)}_\alpha(\bfPsi,\tt) = 0,
\label{sec2:eq:q0.5ans}
\ee
and the angular velocity $\Omega_\alpha$ that defines the phase
variable $\Psi_\alpha$ is given to $O(\sqrt{\varepsilon})$ by
\be
\Omega^{(1/2)}_\alpha(\tt) = \frac{ \partial \omega_\alpha}{\partial
  J_\lambda}[ {\bfcalJ}^{(0)}(\tt), \tt] {\cal J}^{(1/2)}_\lambda(\tt).
\label{sec2:eq:Omega0.5ans}
\ee
Note that Eqs.\ (\ref{sec2:eq:calJ0.5ans}) and (\ref{sec2:eq:Omega0.5ans}) can be obtained simply by
linearizing Eqs.\ (\ref{sec2:eq:calJ0ans}) and
(\ref{sec2:eq:Omega0ans}) about the zeroth order solution.
That is, $\bfcalJ^{(0)} + \sqrt{\varepsilon} \bfcalJ^{(1/2)}$ and
$\bfOmega^{(0)} + \sqrt{\varepsilon} \bfOmega^{(1/2)}$ satisfy the
zeroth order equations (\ref{sec2:eq:calJ0ans}) and
(\ref{sec2:eq:Omega0ans}) to $O(\sqrt{\varepsilon})$.
This means that setting $\bfcalJ^{(1/2)}$ and $\bfOmega^{(1/2)}$ to
zero does not cause any loss of generality in the solutions
(under the no-resonance assumption of this paper), as long as we
allow initial conditions to have sufficiently general dependence on
$\varepsilon$.

\subsubsection{Post-1-adiabatic order}

The first order action variable is given by
\be
J^{(1)}_\lambda(\bfPsi,\tt) = {\cal I}_{\bfOmega^{(0)}(\tt)}{\hat
  G}^{(1)}_\lambda[\bfPsi,\bfcalJ^{(0)}(\tt),\tt]
+ {\cal J}^{(1)}_\lambda(\tt),
\label{sec2:eq:J1ans}
\ee
where the symbol ${\cal I}$ on the right hand side denotes the integration
operator (\ref{sec2:eq:calIdef}) with respect to $\bfPsi$,
${\hat G}^{(1)}_\lambda$ is the non-constant piece of $G^{(1)}_\lambda$
as defined in Eq.\ (\ref{sec2:eq:hatfdef}), and $\bfOmega^{(0)}$ is
given by Eq.\ (\ref{sec2:eq:Omega0ans}).
In Eq.\ (\ref{sec2:eq:J1ans}) the quantity $\bfcalJ^{(1)}(\tt)$ satisfies
the differential equation
\bea
&&\frac{ d {\cal J}^{(1)}_\lambda(\tt)}{d \tt} - \frac{ \partial G^{(1)}_{\lambda\,\bfzero}}
{\partial J_\mu}[\bfcalJ^{(0)}(\tt),\tt] {\cal J}^{(1)}_\mu(\tt) \nn \\
&=& G^{(2)}_{\lambda\,\bfzero}
+ \frac{1}{2} \frac{ \partial^2 G^{(1)}_{\lambda\,\bfzero}}
{\partial J_\mu \partial J_\sigma} {\cal J}^{(1/2)}_\mu {\cal J}^{(1/2)}_\sigma
\nn \\ &&
+
 \left<
\frac{\partial {\hat G}^{(1)}_\lambda } {\partial J_\mu} \,
{\cal I}_{\bfOmega^{(0)}} {\hat G}^{(1)}_\mu
\right>
+ \left<
\frac{\partial {\hat G}^{(1)}_\lambda } {\partial q_\alpha} \,
{\cal I}_{\bfOmega^{(0)}} {\hat g}^{(1)}_\alpha
\right>
\nn \\
&& +
\frac{ \partial \omega_\alpha} {\partial J_\mu}
\left< \frac{\partial {\hat G}^{(1)}_\lambda}{\partial q_\alpha}
{\cal I}_{\bfOmega^{(0)}} {\cal I}_{\bfOmega^{(0)}}
{\hat G}^{(1)}_\mu \right>.
\label{sec2:eq:calJ1ans}
\eea
Here it is understood that the quantities on the right hand side are
evaluated at $\bfJ = \bfcalJ^{(0)}(\tt)$ and $\bfq = \bfq^{(0)} = \bfPsi$.  The last three terms on the right hand side of Eq.\
(\ref{sec2:eq:calJ1ans}) can be written more explicitly
using the definition (\ref{sec2:eq:calIdef}) of ${\cal I}$
and the definition (\ref{sec2:averagingdef}) of the averaging $\langle
\ldots \rangle$ as
\bea
&& \sum_{\bfk \ne \bfzero} \frac{1}{\bfOmega^{(0)} \cdot \bfk}
\left\{ i k_\alpha \frac{ \partial \omega_\alpha}{\partial J_\mu}
\frac{ G^{(1)\,*}_{\lambda\,\bfk} \, G^{(1)}_{\mu\,\bfk} }
{\bfOmega^{(0)} \cdot \bfk} - k_\alpha G^{(1)\,*}_{\lambda\,\bfk}
g^{(1)}_{\alpha\,\bfk} \right. \nn \\
&& \left.  - i G^{(1)}_{\mu\,\bfk} \, \frac{ \partial
  G^{(1)\,*}_{\lambda\,\bfk}} {\partial J_\mu} \right\}.
\eea

The $O(\varepsilon)$ correction to
the angular velocity $\Omega_\alpha$ is given by
\bea
\Omega^{(1)}_\alpha(\tt) &=&
g^{(1)}_{\alpha\,\bfzero}[ {\bfcalJ}^{(0)}(\tt),\tt]
+ \frac{\partial \omega_\alpha}{\partial
J_\lambda}[ {\bfcalJ}^{(0)}(\tt),\tt] {\cal J}^{(1)}_\lambda(\tt)
\nn \\ &&
+ \frac{1}{2} \frac{\partial^2 \omega_\alpha}{\partial
J_\lambda \partial J_\mu}[ {\bfcalJ}^{(0)}(\tt),\tt] {\cal
J}^{(1/2)}_\lambda(\tt) {\cal J}^{(1/2)}_\mu(\tt). \nn \\
&&
\label{sec2:eq:Omega1ans}
\eea

Finally, the sub-leading term in the angle variable is
\be
q^{(1)}_\alpha(\bfPsi,\tt) = {\hat q}^{(1)}_\alpha(\bfPsi,\tt) + {\cal
  Q}^{(1)}_\alpha(\tt),
\label{sec2:eq:q1ans}
\ee
where
\bea
\label{sec2:eq:hatq1ans}
{\hat q}^{(1)}_\alpha(\bfPsi,\tt) &=&
\frac{\partial \omega_\alpha}{\partial J_\lambda}
[\bfcalJ^{(0)}(\tt),\tt]
\nn \\ && \times
{\cal I}_{\bfOmega^{(0)}(\tt)}
{\cal I}_{\bfOmega^{(0)}(\tt)}
{\hat G}^{(1)}_\lambda[\bfPsi,\bfcalJ^{(0)}(\tt),\tt] \nn \\
&& +
{\cal I}_{\bfOmega^{(0)}(\tt)} {\hat
  g}^{(1)}_\alpha[\bfPsi,\bfcalJ^{(0)}(\tt),\tt]
\eea
and
\be
{\cal Q}^{(1)}_\alpha(\tt) = - {\hat q}^{(1)}_\alpha(\bfzero,\tt).
\label{sec2:eq:calQ1ans}
\ee

\subsubsection{Discussion}
\label{sec:discuss}

One of the key results of the general analysis of this section
is the identification of which pieces of the external forces are
required to compute the adiabatic, post-1/2-adiabatic and post-1-adiabatic solutions.
From Eqs.\ (\ref{sec2:eq:calJ0ans}),
(\ref{sec2:eq:Omega0ans}) and (\ref{sec2:eq:qfinalans}),
the adiabatic solution depends only on the
averaged piece $G^{(1)}_{\lambda\,\bfzero}(\bfJ,\tt) = \langle
G^{(1)}_\lambda(\bfq,\bfJ,\tt) \rangle$ of the leading
order external force $G^{(1)}_\lambda$.
Only the dissipative piece of the force $G^{(1)}_\lambda$ normally
contributes to this average.
For our application to inspirals in Kerr, this follows from the identity
(\ref{eq:ident45}) which shows that the average of the conservative
piece of $G^{(1)}_\lambda$ vanishes.  For a Hamiltonian system with
$N=M$, if the perturbing forces $g_\alpha$ and $G_\beta$ arise from
a perturbation $\varepsilon \Delta H = \sum_s \varepsilon^s \Delta
H^{(s)}$ to the Hamiltonian, then the forcing function $G^{(s)}_\beta$ is
$$
G_\beta^{(s)}(\bfq,\bfJ,\tt) = - \frac{ \partial \Delta H^{(s)}(\bfq,\bfJ,\tt)}{\partial
  q_\beta},
$$
and it follows that the average over $\bfq$ of $G^{(s)}_\beta$ vanishes.

At the next, post-1/2-adiabatic order, it follows from Eqs.\ (\ref{sec2:eq:calJ0.5ans})
and (\ref{sec2:eq:Omega0.5ans}) that the term $\psi^{(1/2)}_\alpha(\tt)$ depends again
only on the averaged, dissipative piece $G^{(1)}_{\lambda\,\bfzero}$  of the leading order force.
However, we shall see in the forthcoming paper \cite{FH08b} that when the effects of
resonances are included, additional dependencies on the remaining
(non-averaged) pieces of the first order self forces will arise.

At the next, post-1-adiabatic order, the term $\psi^{(1)}_\alpha(\tt)$
in Eq.\ (\ref{sec2:eq:qfinalans}) depends on the averaged piece
$G^{(2)}_{\lambda\,\bfzero}(\bfJ,\tt) = \langle
G^{(2)}_\lambda(\bfq,\bfJ,\tt) \rangle$ of the sub-leading
force $G^{(2)}_\lambda$, again normally purely dissipative, as well as the remaining
conservative and dissipative pieces of the leading order forces
$G^{(1)}_\lambda(\bfq,\bfJ,\tt)$ and $g^{(1)}_\alpha(\bfq,\bfJ,\tt)$.
This can be seen from
Eqs.\ (\ref{sec2:eq:calJ1ans}) and (\ref{sec2:eq:Omega1ans}).
These results have been previously discussed briefly in the EMRI
context in Refs.\ \cite{Hughes:2005qb,scalar}.
For circular, equatorial orbits,
the fact that there is a
post-1-adiabatic order contribution from the second order self-force
was first argued by Burko \cite{Burko:2002fd}.

Finally, we consider the choice of initial conditions for the
approximate differential equations we have derived.  The discussion
and conclusions here parallel those in the single variable
case, given in Sec.\ \ref{sec:initialconditions} above, and the
results are summarized in Sec.\ \ref{sec:radiative} below

\subsection{Derivation}
\label{sec2:sec:derivation}

We will denote by ${\cal R}(\tt)$ the set of resonant N-tuples $\bfk$
at time $\tt$, and
by ${\cal R}^{\rm c}(\tt)$ the remaining non-resonant nonzero N-tuples.  The
set of all N-tuples can therefore be written as the disjoint union
\be
{\bf Z}^N = \{ {\bf 0} \} \, {\dot \cup} \, {\cal R}(\tt) \, {\dot \cup} \,
{\cal R}^{\rm c}(\tt).
\ee

At each order $s$ we introduce the notation ${\cal J}^{(s)}_\lambda(\tt)$ for
the average part of $J^{(s)}_\lambda(\bfPsi,\tt)$:
\bea
{\cal J}^{(s)}_\lambda(\tt) &\equiv& \langle J^{(s)}_\lambda(\bfPsi,\tt) \rangle
 \\ &=&
\frac{1}{(2\pi)^N} \int_0^{2 \pi} d\Psi_1 \ldots
\int_0^{2\pi} d\Psi_N  J^{(s)}_\lambda(\bfPsi,\tt). \nn
\eea
We denote by ${\hat J}^{(s)}_\beta$ the remaining part of
$J^{(s)}_\beta$, as in Eq.\ (\ref{sec2:eq:hatfdef}).  This gives the
decomposition
\be
J^{(s)}_\lambda(\bfPsi,\tt) = {\cal J}^{(s)}_\lambda(\tt) + {\hat
J}^{(s)}_\lambda(\bfPsi,\tt)
\label{sec2:eq:Jdecomposition}
\ee
for all $s \ge 0$.
Similarly for the angle variable we have the
decomposition
\be
q^{(s)}_\alpha(\bfPsi,\tt) = {\cal Q}^{(s)}_\alpha(\tt) + {\hat
q}^{(s)}_\alpha(\bfPsi,\tt)
\label{sec2:eq:qdecomposition}
\ee
for all $s \ge 1/2$.
For the case $s=0$ we use the fact that $q^{(0)}_\alpha(\bfPsi,\tt) -
\Psi_\alpha$ is a multiply periodic function of $\bfPsi$, by Eq.\
(\ref{sec2:eq:q0periodic}), to obtain the decomposition
\be
q^{(0)}_\alpha(\bfPsi,\tt) = \Psi_\alpha + {\cal Q}^{(0)}_\alpha(\tt)
+ {\hat q}^{(0)}_\alpha(\bfPsi,\tt),
\label{sec2:eq:qdecomposition0}
\ee
where ${\hat q}^{(0)}_\alpha(\bfPsi,\tt)$ is multiply periodic in
$\bfPsi$ with zero average.

Using the expansions (\ref{sec2:eq:ansatz1a}) and (\ref{sec2:eq:ansatz1})
of $q_\alpha$ and $J_\beta$ together with the expansion
(\ref{sec2:eq:Omegaexpand}) of $d\Psi_\alpha/dt$, we obtain
\bea
\frac{dq_\alpha}{dt} &=& \Omega^{(0)}_\beta
q^{(0)}_{\alpha,\Psi_\beta} + \sqrt{\varepsilon} \left[
\Omega^{(1/2)}_\beta q^{(0)}_{\alpha,\Psi_\beta} + \Omega^{(0)}_\beta
q^{(1/2)}_{\alpha,\Psi_\beta} \right] \nn \\
&&+ \varepsilon \left[
\Omega^{(1)}_\beta q^{(0)}_{\alpha,\Psi_\beta} +
\Omega^{(1/2)}_\beta q^{(1/2)}_{\alpha,\Psi_\beta}
+ \Omega^{(0)}_\beta q^{(1)}_{\alpha,\Psi_\beta} + q^{(0)}_{\alpha,\tt} \right] \nn \\
&&+ \varepsilon^{3/2} \left[
\Omega^{(3/2)}_\beta q^{(0)}_{\alpha,\Psi_\beta} +
\Omega^{(1)}_\beta q^{(1/2)}_{\alpha,\Psi_\beta} +
\Omega^{(1/2)}_\beta q^{(1)}_{\alpha,\Psi_\beta} \right. \nn \\
&& \left.
+ \Omega^{(0)}_\beta q^{(3/2)}_{\alpha,\Psi_\beta} +
q^{(1/2)}_{\alpha,\tt} \right]
+ \varepsilon^{2} \left[
\Omega^{(2)}_\beta q^{(0)}_{\alpha,\Psi_\beta}
\right. \nn \\ && \left.
+ \Omega^{(3/2)}_\beta q^{(1/2)}_{\alpha,\Psi_\beta}
+ \Omega^{(1)}_\beta q^{(1)}_{\alpha,\Psi_\beta} +
\Omega^{(1/2)}_\beta q^{(3/2)}_{\alpha,\Psi_\beta}
\right. \nn \\ && \left.
+ \Omega^{(0)}_\beta q^{(2)}_{\alpha,\Psi_\beta} +
q^{(1)}_{\alpha,\tt} \right]
 + O(\varepsilon^{5/2}).
\eea
We now insert this expansion together with a similar expansion for
$dJ_\lambda/dt$ into the equations of motion (\ref{sec2:eq:eom1}) and use the
expansions (\ref{sec2:eq:Gexpand}) and (\ref{sec2:eq:gexpand}) of the external
forces $g_\alpha$ and $G_\lambda$.   Equating coefficients of powers\footnote{This is justified since both sides are asymptotic expansions in powers of $\sqrt{\varepsilon}$ at fixed $\bfPsi$, $\tt$.} of
$\sqrt{\varepsilon}$ then gives at zeroth order
\bes
\label{sec2:eq:1var0order}
\bea
\Omega^{(0)}_\beta q^{(0)}_{\alpha,\Psi_\beta} &=& \omega_\alpha, \\
\Omega^{(0)}_\beta J^{(0)}_{\lambda,\Psi_\beta} &=& 0,
\eea
\ees
at order $O(\sqrt{\varepsilon})$
\bes
\label{sec2:eq:1var0.5order}
\bea
\label{sec2:eq:1var0.5orderA}
\Omega^{(0)}_\beta q^{(1/2)}_{\alpha,\Psi_\beta} &=&
- \Omega^{(1/2)}_\beta q^{(0)}_{\alpha,\Psi_\beta}
+\omega_{\alpha,J_\lambda} J^{(1/2)}_\lambda
, \ \ \ \ \ \\
\Omega^{(0)}_\beta J^{(1/2)}_{\lambda,\Psi_\beta} &=& - \Omega^{(1/2)}_\beta J^{(0)}_{\lambda,\Psi_\beta} ,
\label{sec2:eq:1var0.5orderB}
\eea
\ees
at order $O(\varepsilon)$
\bes
\label{sec2:eq:1var1order}
\bea
\label{sec2:eq:1var1orderA}
\Omega^{(0)}_\beta q^{(1)}_{\alpha,\Psi_\beta} &=&
- \Omega^{(1/2)}_\beta q^{(1/2)}_{\alpha,\Psi_\beta}
- \Omega^{(1)}_\beta q^{(0)}_{\alpha,\Psi_\beta}
- q^{(0)}_{\alpha,\tt} + g^{(1)}_\alpha
\nn \\ &&
+ \omega_{\alpha,J_\lambda} J^{(1)}_\lambda
+ \frac{1}{2} \omega_{\alpha,J_\lambda J_\mu} J^{(1/2)}_\lambda J^{(1/2)}_\mu
,\ \ \ \ \ \ \ \ \\
\Omega^{(0)}_\beta
J^{(1)}_{\lambda,\Psi_\beta} &=&
-\Omega^{(1/2)}_\beta J^{(1/2)}_{\lambda,\Psi_\beta}
-\Omega^{(1)}_\beta J^{(0)}_{\lambda,\Psi_\beta}
- J^{(0)}_{\lambda,\tt} \nn \\ &&
+ G^{(1)}_\lambda,
\label{sec2:eq:1var1orderB}
\eea
\ees
at order $O(\varepsilon^{3/2})$
\bes
\label{sec2:eq:1var1.5order}
\bea
\label{sec2:eq:1var1.5orderA}
\Omega^{(0)}_\beta q^{(3/2)}_{\alpha,\Psi_\beta} &=&
- \Omega^{(1/2)}_\beta q^{(1)}_{\alpha,\Psi_\beta}
- \Omega^{(1)}_\beta q^{(1/2)}_{\alpha,\Psi_\beta}
- \Omega^{(3/2)}_\beta q^{(0)}_{\alpha,\Psi_\beta}
\nn \\ &&
- q^{(1/2)}_{\alpha,\tt} + g^{(1)}_{\alpha,q_\beta} q^{(1/2)}_\beta
+ g^{(1)}_{\alpha,J_\lambda} J^{(1/2)}_\lambda
\nn \\ &&
+ \omega_{\alpha,J_\lambda} J^{(3/2)}_\lambda
+  \omega_{\alpha,J_\lambda J_\mu} J^{(1/2)}_\lambda J^{(1)}_\mu
\nn \\ &&
+  \frac{1}{6} \omega_{\alpha,J_\lambda J_\mu J_\sigma } J^{(1/2)}_\lambda J^{(1/2)}_\mu J^{(1/2)}_\sigma, \\
\Omega^{(0)}_\beta
J^{(3/2)}_{\lambda,\Psi_\beta} &=&
-\Omega^{(1/2)}_\beta J^{(1)}_{\lambda,\Psi_\beta}
-\Omega^{(1)}_\beta J^{(1/2)}_{\lambda,\Psi_\beta}
-\Omega^{(3/2)}_\beta J^{(0)}_{\lambda,\Psi_\beta}
 \nn \\ &&
- J^{(1/2)}_{\lambda,\tt}
+ G^{(1)}_{\lambda,q_\beta} q^{(1/2)}_\beta
+ G^{(1)}_{\lambda,J_\mu} J^{(1/2)}_\mu,
\nn \\ &&
\label{sec2:eq:1var1.5orderB}
\eea
\ees
and at order $O(\varepsilon^2)$
\bes
\label{sec2:eq:1var2order}
\bea
\label{sec2:eq:1var2orderA}
\Omega^{(0)}_\beta q^{(2)}_{\alpha,\Psi_\beta} &=&
- \Omega^{(1/2)}_\beta q^{(3/2)}_{\alpha,\Psi_\beta}
- \Omega^{(1)}_\beta q^{(1)}_{\alpha,\Psi_\beta}
- \Omega^{(3/2)}_\beta q^{(1/2)}_{\alpha,\Psi_\beta}
\nn \\ &&
- \Omega^{(2)}_\beta q^{(0)}_{\alpha,\Psi_\beta}
- q^{(1)}_{\alpha,\tt}
+ g^{(2)}_\alpha
+ g^{(1)}_{\alpha,q_\beta} q^{(1)}_\beta
\nn \\ &&
+ g^{(1)}_{\alpha,J_\lambda} J^{(1)}_\lambda
+ \frac{1}{2} g^{(1)}_{\alpha,q_\beta q_\gamma} q^{(1/2)}_\beta q^{(1/2)}_\gamma
\nn \\ &&
+ \frac{1}{2} g^{(1)}_{\alpha,J_\lambda J_\mu} J^{(1/2)}_\lambda J^{(1/2)}_\mu
+  g^{(1)}_{\alpha,q_\beta J_\lambda} q^{(1/2)}_\beta J^{(1/2)}_\lambda
\nn \\ &&
+ \omega_{\alpha,J_\lambda} J^{(2)}_\lambda
+  \frac{1}{2} \omega_{\alpha,J_\lambda J_\mu J_\sigma } J^{(1)}_\lambda J^{(1/2)}_\mu J^{(1/2)}_\sigma
\nn \\ &&
+  \frac{1}{2} \omega_{\alpha,J_\lambda J_\mu} J^{(1)}_\lambda
J^{(1)}_\mu
+  \omega_{\alpha,J_\lambda J_\mu} J^{(1/2)}_\lambda J^{(3/2)}_\mu
\nn \\ &&
+  \frac{1}{24} \omega_{\alpha,J_\lambda J_\mu J_\sigma J_\tau} J^{(1/2)}_\lambda J^{(1/2)}_\mu J^{(1/2)}_\sigma J^{(1/2)}_\tau, \  \ \ \ \ \  \nn \\
&& \\
\Omega^{(0)}_\beta
J^{(2)}_{\lambda,\Psi_\beta} &=&
-\Omega^{(1/2)}_\beta J^{(3/2)}_{\lambda,\Psi_\beta}
-\Omega^{(1)}_\beta J^{(1)}_{\lambda,\Psi_\beta}
-\Omega^{(3/2)}_\beta J^{(1/2)}_{\lambda,\Psi_\beta}
 \nn \\ &&
-\Omega^{(2)}_\beta J^{(0)}_{\lambda,\Psi_\beta}
- J^{(1)}_{\lambda,\tt}
+ G^{(2)}_\lambda
+ G^{(1)}_{\lambda,q_\beta} q^{(1)}_\beta
\nn \\ &&
+ G^{(1)}_{\lambda,J_\mu} J^{(1)}_\mu
+ \frac{1}{2} G^{(1)}_{\lambda,q_\beta q_\gamma} q^{(1/2)}_\beta q^{(1/2)}_\gamma
\nn \\ &&
+ \frac{1}{2} G^{(1)}_{\lambda,J_\mu J_\sigma} J^{(1/2)}_\mu J^{(1/2)}_\sigma
+  G^{(1)}_{\lambda,q_\beta J_\mu} q^{(1/2)}_\beta J^{(1/2)}_\mu.
\nn \\ &&
\label{sec2:eq:1var2orderB}
\eea
\ees
Here it is understood that all functions of $\bfq$ and $\bfJ$ are evaluated
at $\bfq^{(0)}$ and $\bfJ^{(0)}$.

\subsubsection{Zeroth order analysis}
\label{sec:zeroorderanalysis}

The zeroth order equations (\ref{sec2:eq:1var0order}) can be written more
explicitly as
\bes
\bea
\label{sec2:first}
\Omega^{(0)}_\beta(\tt) q^{(0)}_{\alpha,\Psi_\beta}(\bfPsi,\tt) &=&
\omega_\alpha[\bfJ^{(0)}(\bfPsi,\tt),\tt], \\
\Omega^{(0)}_\beta(\tt) J^{(0)}_{\lambda,\Psi_\beta}(\bfPsi,\tt) &=& 0.
\label{sec2:second}
\eea
\ees
Since $\bfJ^{(0)}$ is a multiply periodic function of $\bfPsi$ by
Eq.\ (\ref{sec2:Jsperiodic}), we can rewrite Eq.\ (\ref{sec2:second})
in terms of the Fourier components $J^{(0)}_{\lambda\,\bfk}(\tt)$ of
$J^{(0)}_\lambda$ as
\be
\sum_\bfk \, \left[i \bfOmega^{(0)}(\tt) \cdot \bfk \right]
J^{(0)}_{\lambda\,\bfk}(\tt) \, e^{i \bfk \cdot \bfPsi} =0.
\ee
For non-resonant N-tuples $\bfk$ we have
\be
\bfOmega^{(0)}(\tt) \cdot \bfk \ne 0
\label{eq:noresonanceassumption}
\ee
by Eqs.\ (\ref{resonantNdef}) and (\ref{sec2:eq:Omega0ans})
unless $\bfk = \bfzero$.  This implies that $J^{(0)}_{\lambda\,\bfk}(\tt)
=0$ for all nonzero non-resonant $\bfk$.

It follows that, for a given $\bfk$, $J^{(0)}_{\lambda\,\bfk}(\tt)$ must
vanish except at those values of $\tt$ at which $\bfk$ is resonant.
Since we assume that $J^{(0)}_{\lambda\,\bfk}(\tt)$ is a continuous function of
$\tt$, and since the set of resonant values of $\tt$ for a given
$\bfk$ consists of isolated points
(cf.\ Sec.\ \ref{sec:noresonanceassumption} above), it follows that
$J^{(0)}_{\lambda\,\bfk}(\tt)$ vanishes for all nonzero $\bfk$.
The formula (\ref{sec2:eq:J0ans}) now follows from the decomposition
(\ref{sec2:eq:Jdecomposition}).

Next, substituting the formula (\ref{sec2:eq:J0ans}) for
$\bfJ^{(0)}$ and the decomposition (\ref{sec2:eq:qdecomposition0}) of
$\bfq^{(0)}$ into Eq.\ (\ref{sec2:first}) gives
\bea
\Omega^{(0)}_\alpha(\tt) &+&
\sum_\bfk \, \left[i \bfOmega^{(0)}(\tt) \cdot \bfk \right]
{\hat q}^{(0)}_{\alpha\,\bfk}(\tt) \, e^{i \bfk \cdot \bfPsi} \nn \\
&=& \omega_\alpha[\bfcalJ^{(0)}(\tt),\tt],
\eea
where ${\hat q}^{(0)}_{\alpha\,\bfk}(\tt)$ are the Fourier components
of ${\hat q}^{(0)}_\alpha(\bfPsi,\tt)$.
The $\bfk=0$ Fourier component of this equation gives the formula
(\ref{sec2:eq:Omega0ans}) for the zeroth order angular velocity $\bfOmega^{(0)}$.
The $\bfk \ne 0$ Fourier components imply, using
an argument similar to that just given
for Eq.\ (\ref{sec2:second}),
that ${\hat q}^{(0)}_{\alpha\,\bfk}(\tt) =0$ for all nonzero $\bfk$.
The decomposition
(\ref{sec2:eq:qdecomposition0}) then gives
\be
q^{(0)}_\alpha(\bfPsi,\tt) = \Psi_\alpha + {\cal Q}^{(0)}_\alpha(\tt).
\label{sec2:eq:v1}
\ee
Finally, the assumption (\ref{sec2:eq:ansatz5}) forces ${\cal
Q}^{(0)}_\alpha(\tt)$ to vanish, and we recover the formula
(\ref{sec2:eq:q0ans}) for $q^{(0)}_\alpha(\bfPsi,\tt)$.

\subsubsection{Order $O(\sqrt{\varepsilon})$ analysis}

The $O(\sqrt{\varepsilon})$ equation (\ref{sec2:eq:1var0.5orderB}) can be written more
explicitly as
\be
\label{sec2:eq:1var0.5orderBexplicit}
\Omega^{(0)}_\beta(\tt) J^{(1/2)}_{\lambda,\Psi_\beta}(\bfPsi,\tt) =0,
\ee
where we have simplified using the zeroth order solution
(\ref{sec2:eq:J0ans}).
An argument similar to that given in Sec.\ \ref{sec:zeroorderanalysis} now forces the
$\bfPsi$ dependent piece of $\bfJ^{(1/2)}$ to vanish, and so we obtain
the formula (\ref{sec2:eq:J0.5ans}).

Next, we simplify the order $O(\sqrt{\varepsilon})$ equation (\ref{sec2:eq:1var0.5orderA})
using the solutions (\ref{sec2:eq:J0ans}), (\ref{sec2:eq:q0ans}) and (\ref{sec2:eq:J0.5ans}) to obtain
\bea
\label{sec2:eq:1var0.5orderAexplicit}
\Omega^{(0)}_\beta(\tt) q^{(1/2)}_{\alpha,\Psi_\beta}(\bfPsi,\tt) &=&
\omega_{\alpha,J_\lambda}[\bfcalJ^{(0)}(\tt),\tt] {\cal J}^{(1/2)}_\lambda(\tt) \nn \\
&& \ \  - \Omega^{(1/2)}_\alpha(\tt).
\eea
After averaging with respect to $\bfPsi$, the term on the left hand
side vanishes since it is a total derivative, and we obtain
the formula (\ref{sec2:eq:Omega0.5ans}) for
$\Omega^{(1/2)}(\tt)$.  Note however that the function
$\bfcalJ^{(1/2)}(\tt)$ in that formula has not yet been determined; it
will be necessary to go to two higher orders in $\sqrt{\varepsilon}$ to compute this function.

Next, we subtract from Eq.\ (\ref{sec2:eq:1var0.5orderAexplicit}) its
averaged part and use the decomposition (\ref{sec2:eq:qdecomposition}) of $q_\alpha^{(1/2)}$
to obtain
\be
\label{sec2:eq:1var0.5orderAexplicita}
\Omega^{(0)}_\beta(\tt) {\hat q}^{(1/2)}_{\alpha,\Psi_\beta}(\bfPsi,\tt) =0.
\ee
An argument similar to that given in Sec.\ \ref{sec:zeroorderanalysis}
now shows that ${\hat {\bf q}}^{(1/2)}=0$, and the result (\ref{sec2:eq:q0.5ans})
then follows from the decomposition (\ref{sec2:eq:qdecomposition})
together with the initial condition condition (\ref{sec2:eq:ansatz5}).

\subsubsection{Order $O(\varepsilon)$ analysis}

The first order equation (\ref{sec2:eq:1var1orderB}) can be written more
explicitly as
\bea
\label{sec2:eq:1var1orderBexplicit}
\Omega^{(0)}_\beta(\tt) J^{(1)}_{\lambda,\Psi_\beta}(\bfPsi,\tt) &=&
-{\cal J}^{(0)}_{\lambda,\tt}(\tt) \nn \\
&&+ G^{(1)}_\lambda[\bfPsi,\bfcalJ^{(0)}(\tt),\tt],
\eea
where we have simplified using the zeroth order solutions
(\ref{sec2:eq:J0ans}) and (\ref{sec2:eq:q0ans})
and the $O(\sqrt{\varepsilon})$ solution (\ref{sec2:eq:J0.5ans}).
We now take the average with
respect to $\bfPsi$ of this equation.  The left hand side vanishes since
it is a derivative, and we obtain using the definition
(\ref{sec2:eq:Gkdef}) the differential equation (\ref{sec2:eq:calJ0ans}) for
$\bfcalJ^{(0)}(\tt)$.
Next, we subtract from Eq.\ (\ref{sec2:eq:1var1orderBexplicit}) its
averaged part, and use the decomposition (\ref{sec2:eq:Jdecomposition}) of
$\bfJ^{(1)}$.  This gives
\bea
\label{sec2:eq:1var1orderBexplicit1}
\Omega^{(0)}_\beta(\tt) {\hat J}^{(1)}_{\lambda,\Psi_\beta}(\bfPsi,\tt)
= {\hat G}^{(1)}_\lambda[\bfPsi,\bfcalJ^{(0)}(\tt),\tt].
\eea
We solve this equation using the Fourier decomposition
(\ref{sec2:eq:hatgformula}) of ${\hat G}^{(1)}_\lambda$
to obtain
\bea
{\hat J}^{(1)}_\lambda(\bfPsi,\tt) &=& \sum_{\bfk \in {\cal R}^{\rm c}(\tt)}
\frac{ G^{(1)}_{\lambda\,\bfk}[\bfcalJ^{(0)}(\tt),\tt] }
{i \bfk \cdot \bfOmega^{(0)}(\tt)} e^{i \bfk \cdot \bfPsi} \nn \\
&&+ \sum_{\bfk \in {\cal R}(\tt)}
J^{(1)}_{\lambda\bfk}(\tt) e^{i \bfk \cdot \bfPsi}.
\label{eq:J1ans0}
\eea
Here the first term is a sum over non-resonant N-tuples, and the
second term is a sum over resonant N-tuples, for which the
coefficients are unconstrained by Eq.\ (\ref{sec2:eq:1var1orderBexplicit1}).
However for each fixed $\bfk$, the values of $\tt$ that correspond to
resonances are isolated, and furthermore by the
the no-resonance assumption (\ref{eq:noresonanceassumption})
we have $G^{(1)}_{\beta\,\bfk}[\bfcalJ^{(0)}(\tt),\tt] =0$ in the
vicinity of those values of $\tt$.  Therefore using the assumed continuity of
$J^{(1)}_{\lambda\bfk}(\tt)$ in $\tt$ we can simplify Eq.\
(\ref{eq:J1ans0}) to
\be
{\hat J}^{(1)}_\lambda(\bfPsi,\tt) = \sum_{\bfk \ne {\bf 0}}
\frac{ G^{(1)}_{\lambda\,\bfk}[\bfcalJ^{(0)}(\tt),\tt] }
{i \bfk \cdot \bfOmega^{(0)}(\tt)} e^{i \bfk \cdot \bfPsi},
\label{eq:J1ans1}
\ee
where any terms of the form $0/0$ that appear in the coefficients are
interpreted to be $0$.
This yields the first term in the result (\ref{sec2:eq:J1ans}) for
$\bfJ^{(1)}$ when we use the notation (\ref{sec2:eq:calIdef}).

Next, we simplify the first order equation (\ref{sec2:eq:1var1orderA})
using the zeroth order solutions (\ref{sec2:eq:J0ans}) and
(\ref{sec2:eq:q0ans}) and the $O(\sqrt{\varepsilon})$ solutions
(\ref{sec2:eq:J0.5ans}) and (\ref{sec2:eq:q0.5ans}), to obtain
\bea
\label{sec2:eq:1var1orderAexplicit}
&& \Omega^{(0)}_\beta(\tt) q^{(1)}_{\alpha,\Psi_\beta}(\bfPsi,\tt) =
g^{(1)}_\alpha[\bfPsi,\bfcalJ^{(0)}(\tt),\tt] - \Omega^{(1)}_\alpha(\tt)
 \nn \\ &&
\ \ \ +\omega_{\alpha,J_\lambda}[\bfcalJ^{(0)}(\tt),\tt] J^{(1)}_\lambda[\bfPsi,\tt]
 \nn \\ &&
\ \ \ +\frac{1}{2} \omega_{\alpha,J_\lambda J_\mu}[\bfcalJ^{(0)}(\tt),\tt]
{\cal J}^{(1/2)}_\lambda(\tt) {\cal J}^{(1/2)}_\mu(\tt).
\eea
Averaging with respect to $\bfPsi$ and using the decompositions
(\ref{sec2:eq:Jdecomposition}) and (\ref{sec2:eq:qdecomposition}) of
$\bfJ^{(1)}$ and $\bfq^{(1)}$ now gives the formula (\ref{sec2:eq:Omega1ans}) for
$\Omega^{(1)}(\tt)$.  Note however that the function
$\bfcalJ^{(1)}(\tt)$ in that formula has not yet been determined; it
will be necessary to go to two higher orders in $\sqrt{\varepsilon}$ to compute this function.

Finally, we subtract from Eq.\ (\ref{sec2:eq:1var1orderAexplicit}) its average
over $\bfPsi$ using the decompositions (\ref{sec2:eq:Jdecomposition}) and
(\ref{sec2:eq:qdecomposition}), and then solve the resulting
partial differential equation using the notation
(\ref{sec2:eq:calIdef})
and the convention described after Eq.\ (\ref{eq:J1ans1}).  This gives
\bea
\label{sec2:eq:hatq1ansA}
{\hat q}^{(1)}_\alpha(\bfPsi,\tt) &=&
\frac{\partial \omega_\alpha}{\partial J_\lambda}
[\bfcalJ^{(0)}(\tt),\tt] \,
{\cal I}_{\bfOmega^{(0)}(\tt)}{\hat J}^{(1)}_\lambda[\bfPsi,\tt] \nn \\
&& + {\cal I}_{\bfOmega^{(0)}(\tt)}{\hat
  g}^{(1)}_\alpha[\bfPsi,\bfcalJ^{(0)}(\tt),\tt].
\eea
Using the result for ${\hat J}^{(1)}_\beta$ given by the first term in Eq.\
(\ref{sec2:eq:J1ans}) now yields
the formula (\ref{sec2:eq:hatq1ans}) for ${\hat q}^{(1)}_\alpha(\bfPsi,\tt)$, and
the result (\ref{sec2:eq:q1ans}) for $q^{(1)}_\alpha$ then follows from the decomposition
(\ref{sec2:eq:qdecomposition}) together with the initial condition
(\ref{sec2:eq:ansatz5}).

\subsubsection{Order $O(\varepsilon^{3/2})$ analysis}

The $O(\varepsilon^{3/2})$ equation (\ref{sec2:eq:1var1.5orderB}) can be written more
explicitly as
\bea
\label{sec2:eq:1var1.5orderBexplicit}
&&\Omega^{(0)}_\beta(\tt) J^{(3/2)}_{\lambda,\Psi_\beta}(\bfPsi,\tt) =
- \Omega^{(1/2)}_\beta(\tt) J^{(1)}_{\lambda,\Psi_\beta}(\bfPsi,\tt)
-{\cal J}^{(1/2)}_{\lambda,\tt}(\tt) \nn \\
&&
\ \ \ + G^{(1)}_{\lambda,J_\mu}[\bfPsi,\bfcalJ^{(0)}(\tt),\tt] {\cal J}^{(1/2)}_\mu(\tt),
\eea
where we have simplified using the lower order solutions
(\ref{sec2:eq:J0ans}), (\ref{sec2:eq:q0ans}), (\ref{sec2:eq:J0.5ans})
and (\ref{sec2:eq:q0.5ans}).
We now take the average with
respect to $\bfPsi$ of this equation.  Two terms vanish since
they are total derivatives, and we obtain using the definition
(\ref{sec2:eq:Gkdef}) the differential equation (\ref{sec2:eq:calJ0.5ans}) for
$\bfcalJ^{(1/2)}(\tt)$.
The remaining non-zero Fourier components of Eq.\
(\ref{sec2:eq:1var1.5orderBexplicit})
can be used to solve for ${\hat \bfJ}^{(3/2)}$, which we will not need in what
follows.

Next, we simplify the $O(\varepsilon^{3/2})$ equation (\ref{sec2:eq:1var1.5orderA})
using the lower order solutions
(\ref{sec2:eq:J0ans}), (\ref{sec2:eq:q0ans}), (\ref{sec2:eq:J0.5ans})
and (\ref{sec2:eq:q0.5ans}) to obtain
\bea
\label{sec2:eq:1var1.5orderAexplicit}
&& \Omega^{(0)}_\beta(\tt) q^{(3/2)}_{\alpha,\Psi_\beta}(\bfPsi,\tt) =
g^{(1)}_{\alpha,J_\lambda}[\bfPsi,\bfcalJ^{(0)}(\tt),\tt] {\cal J}^{(1/2)}_\lambda(\tt)
 \nn \\ &&
\ \ \  - \Omega^{(3/2)}_\alpha(\tt) -\Omega^{(1/2)}_\beta(\tt) q^{(1)}_{\alpha,\Psi_\beta}(\bfPsi,\tt)
 \nn \\ &&
\ \ \ +\omega_{\alpha,J_\lambda}[\bfcalJ^{(0)}(\tt),\tt] J^{(3/2)}_\lambda[\bfPsi,\tt]
 \nn \\ &&
\ \ \ + \omega_{\alpha,J_\lambda J_\mu}[\bfcalJ^{(0)}(\tt),\tt]
J^{(1)}_\lambda[\bfPsi,\tt] {\cal J}^{(1/2)}_\mu(\tt)
 \nn \\ &&
\ \ \ +\frac{1}{2} \omega_{\alpha,J_\lambda J_\mu J_\sigma}[\bfcalJ^{(0)}(\tt),\tt]
{\cal J}^{(1/2)}_\lambda(\tt) {\cal J}^{(1/2)}_\mu(\tt) {\cal
  J}^{(1/2)}_\sigma(\tt). \nn \\ &&
\eea
The $\bfk=0$ component of this equation yields a formula for
$\bfOmega^{(3/2)}(\tt)$ in terms of $\bfcalJ^{(1/2)}(\tt)$ and
$\bfcalJ^{(3/2)}(\tt)$, and the Fourier components with $\bfk \ne
\bfzero$ yield a formula for ${\hat {\bf q}}^{(3/2)}$ which we shall
not need.

\subsubsection{Order $O(\varepsilon^2)$ analysis}

We simplify the second order equation (\ref{sec2:eq:1var2orderB})
using the lower order solutions
(\ref{sec2:eq:J0ans}), (\ref{sec2:eq:q0ans}), (\ref{sec2:eq:J0.5ans})
and (\ref{sec2:eq:q0.5ans}),
average
over $\bfPsi$, and simplify using the
decompositions (\ref{sec2:eq:Jdecomposition}) and
(\ref{sec2:eq:qdecomposition}) and the identities (\ref{sec2:eq:ident1}).
The result is
\bea
\frac{ d }{d\tt} {\cal J}^{(1)}_\lambda(\tt)
&=&   \frac{ \partial G^{(1)}_{\lambda\,\bfzero} }{\partial J_\mu}
[\bfcalJ^{(0)}(\tt),\tt] {\cal J}^{(1)}_\mu(\tt)
+ G^{(2)}_{\lambda\,\bfzero}[ \bfcalJ^{(0)}(\tt), \tt ]
\nn \\ &&
\frac{1}{2} \frac{ \partial^2 G^{(1)}_{\lambda\,\bfzero} }{\partial
  J_\mu \partial J_\sigma}
[\bfcalJ^{(0)}(\tt),\tt] {\cal J}^{(1/2)}_\mu(\tt) {\cal J}^{(1/2)}_\sigma(\tt)
\nn \\ &&
+ \left\langle {\hat q}^{(1)}_\alpha(\bfPsi,\tt) \,
\frac{ \partial {\hat G}^{(1)}_\lambda}{\partial \Psi_\alpha}
\left[ \bfPsi, {\cal J}^{(0)}(\tt),\tt\right] \right\rangle \nn \\
&& + \left\langle {\hat J}^{(1)}_\mu(\bfPsi,\tt) \,
\frac{ \partial {\hat G}^{(1)}_\lambda}{\partial J_\mu}
\left[ \bfPsi, {\cal J}^{(0)}(\tt),\tt\right] \right\rangle.
\label{sec2:eq:calJ1ansA}
\eea
Using the expressions (\ref{sec2:eq:hatq1ans}) and (\ref{sec2:eq:J1ans}) for ${\hat
q}^{(1)}_\alpha$ and ${\hat J}^{(1)}_\alpha$
now gives the differential equations (\ref{sec2:eq:calJ1ans}) for
$\bfcalJ^{(1)}$.\footnote{
We remark that a slight inconsistency arises in our solution ansatz
(\ref{sec2:eq:ansatzg})
at this order,
$O(\varepsilon^2)$.  Consider the $\bfk \ne 0$ Fourier components of the second
order equations (\ref{sec2:eq:1var2order}).  For resonant n-tuples
$\bfk$, the left hand sides of these two equations vanish by
definition, but the right hand sides are generically nonzero,
due to the effects of subleading resonances.
A similar inconsistency would arise in the $O(\varepsilon)$ equations (\ref{sec2:eq:1var1order}),
but for the fact that our no-resonance assumption
(\ref{eq:noresonance}) forces the right hand sides of those equations
to vanish for resonant n-tuples.
However, the no-resonance assumption (\ref{eq:noresonance}) is
insufficient to make the right hand sides of the $O(\varepsilon^2)$
equations (\ref{sec2:eq:1var2order}) vanish, because of the occurrence
of quadratic cross terms such as
$$
 g^{(1)}_{\alpha\,\bfk} \,
g^{(1)}_{\beta\,\bfk'} \,e^{i (\bfk + \bfk^\prime) \cdot \bfPsi}.
$$
It can be shown, by an analysis similar to that given in Ref.\
\cite{FH08b}, that the effect of these subleading resonances is
to (i) restrict the domain of validity of the expansion
(\ref{sec2:eq:ansatzg}) to exclude times $\tt$ at which subleading
resonances occur, and (ii) to add source terms to the differential
equation
for $\bfcalJ^{(3/2)}$
which encode the effect of passing through a subleading resonance.
These modifications do not affect any of the conclusions in the
present paper.
}

\section{NUMERICAL INTEGRATION OF AN ILLUSTRATIVE EXAMPLE}
\label{sec:numerics}

In this section we present a numerical integration of a particular example
of a dynamical system, in order to illustrate and validate the general theory of
Secs.\ \ref{sec:derivation_single} and \ref{sec2:manyvariables}.

\begin{figure}
\begin{center}
\epsfig{file=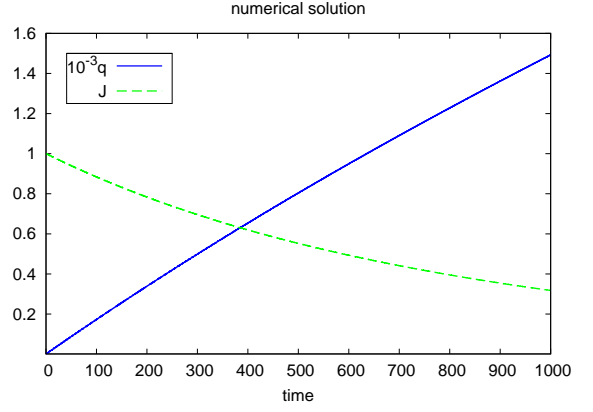,width=8.0cm}
\caption{The exact numerical solution of the system of equations
  (\ref{examplesystem}).  After a time $\sim 1/\varepsilon$, the
  action variable $J$ is $O(1)$, while the angle variable $q$ is
  $O(1/\varepsilon)$.}
\label{fig:exact}
\end{center}
\end{figure}

Consider the system of equations
\bes
\label{examplesystem}
\bea
{\dot q} &=& \omega(J) + \varepsilon g^{(1)}(q,J) \\
{\dot J} &=& \varepsilon G^{(1)}(q,J),
\eea
\ees
where
\begin{eqnarray}
\omega(J) &=& 1 + J - J^2/4, \nonumber \\
g^{(1)}(q,J) &=& \sin(q)/J, \nonumber \\
G^{(1)}(q,J) &=& - J - J^2/4 - J \cos(q) - J^2 \sin(q), \ \ \
\end{eqnarray}
together with the initial conditions $q(0) =1$, $J(0) =1$, and with
$\varepsilon = 10^{-3}$.  The exact solution of this system is shown
in Fig. \ref{fig:exact}.

\begin{figure}
\begin{center}
\epsfig{file=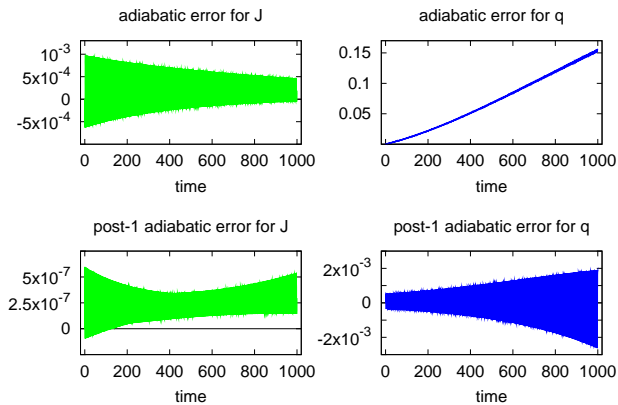,width=8.0cm}
\caption{{\it Upper panels:} The difference between the solution of the
  exact dynamical system (\ref{examplesystem}) and the adiabatic
  approximation given by Eqs.\ (\ref{adiabaticexample}) and
  (\ref{adiabaticexample1}). For the action variable
  $J$, this difference is $O(\varepsilon)$, while for the angle
  variable $q$, this difference is $O(1)$, as expected.
 {\it Lower panels:} The difference between
 the exact solution and the post-1-adiabatic
  approximation given by Eqs.\ (\ref{adiabaticexample}),
  (\ref{p1adiabaticexample})
   and
  (\ref{p1xexamplesoln}). Again the magnitudes of these errors are as
  expected: $O(\varepsilon^2)$ for $J$ and $O(\varepsilon)$ for $q$.}

\label{fig:approx}
\end{center}
\end{figure}

Consider now the adiabatic approximation to this system.
From Eqs.\ (\ref{psisdef}) -- (\ref{eq:Omega0ans})
the adiabatic approximation is given by the system
\bes
\label{adiabaticexample}
\bea
\frac{d \psi^{(0)}}{d \tt} &=& \omega({\cal J}^{(0)}),\\
\frac{ d {\cal J}^{(0)}}{d \tt} &=& - {\cal J}^{(0)} - {\cal J}^{(0)\,2}/4,
\eea
\ees
where $\tt = \varepsilon t$.
The adiabatic solution $(q_{\rm ad}, J_{\rm ad})$ is given in terms of
the functions $\psi^{(0)}(\tt)$ and ${\cal J}^{(0)}(\tt)$
by
\be
q_{\rm ad}(t,\varepsilon) = \varepsilon^{-1} \psi^{(0)}(\varepsilon
t), \ \ \ \ \ J_{\rm ad}(t,\varepsilon) = {\cal J}^{(0)}(\varepsilon
t).
\label{adiabaticexample1}
\ee
To this order, the initial conditions on $(q_{\rm ad}, J_{\rm ad})$ are the same
as those for $(q,J)$, which gives $\psi^{(0)}(0)=\varepsilon$
\footnote{Strictly speaking, our derivations assumed that
  $\psi^{(0)}(\tt)$ is independent of $\varepsilon$, and so it is
  inconsistent to use this initial condition for $\psi^{(0)}(0)$.
  Instead we should set $\psi^{(0)}(0) = 0$, and take account of the
  nonzero initial phase $q(0)$ at the next order, in the variable $\psi^{(1)}(0)$.  However,
  moving a constant from $\psi^{(1)}(\tt)$ to $\varepsilon^{-1}
  \psi^{(0)}(\tt)$ does not affect the solution,
 and so we are free to choose the initial data as done here.}
and
${\cal J}^{(0)}(0) =1$.
We expect to
find that after a time $t \sim 1/\varepsilon$, the errors are of order
$\sim 1$ for $q_{\rm ad}(t)$, and of order $\sim \varepsilon$
for $J_{\rm ad}(t)$.  This is confirmed by the two upper panels in Fig.\
\ref{fig:approx}, which show the differences $q - q_{\rm ad}$ and $J -
J_{\rm ad}$.

Consider next the post-1-adiabatic approximation.
From Eqs.\ (\ref{eq:calJ1ans}) and
(\ref{eq:Omega1ans}) this approximation is given
by the system of equations
\bes
\label{p1adiabaticexample}
\bea
\frac{ d \psi^{(1)}}{d \tt} &=&  \omega_{,J}({\cal
  J}^{(0)}) {\cal J}^{(1)}, \\
\frac{ d {\cal J}^{(1)}}{d \tt} &=& - (1 + {\cal J}^{(0)}/2) {\cal J}^{(1)}
+ \frac{ {\cal J}^{(0)} ( {\cal J}^{(0)} + 1)}{2 \omega(
  {\cal J}^{(0)}) },\ \ \ \
\eea
\ees
together with the adiabatic system (\ref{adiabaticexample}).
From Eqs.\ (\ref{eq:qfinalans}) and (\ref{eq:J1ans}) the
post-1-adiabatic solution $(q_{\rm p1a},J_{\rm p1a})$ is given by
\bes
\label{p1xexamplesoln}
\bea
q_{\rm p1a}(t,\varepsilon) &=& \varepsilon^{-1} \psi^{(0)}(\varepsilon
t) + \psi^{(1)}(\varepsilon t), \\
J_{\rm p1a}(t,\varepsilon) &=& {\cal J}^{(0)}(\varepsilon t) +
\varepsilon {\cal J}^{(1)}(\varepsilon t) \nn \\
&& + \varepsilon H[ {\cal J}^{(0)}(\varepsilon t), q_{\rm
  p1a}(t,\varepsilon) ],
\label{p1xexamplesoln2}
\eea
\ees
where the function $H$ is given by
\be
H({\cal J},q) = \frac{ {\cal J}^2 \cos q - {\cal J} \sin
  q}{\omega({\cal J})}.
\ee

Consider next the choice of initial conditions $\psi^{(0)}(0)$,
$\psi^{(1)}(0)$, ${\cal J}^{(0)}(0)$ and
${\cal J}^{(1)}(0)$ for the system of equations
(\ref{adiabaticexample}) and (\ref{p1adiabaticexample}).
From Eqs.\ (\ref{p1xexamplesoln}) these choices are constrained by, to
O$(\varepsilon^2)$,
\bes
\bea
q(0) &=& \varepsilon^{-1} \psi^{(0)}(0) + \psi^{(1)}(0), \\
J(0) &=& {\cal J}^{(0)}(0) +
\varepsilon {\cal J}^{(1)}(0) + \varepsilon H[ J(0), q(0) ].\ \ \
\eea
\ees
We solve these equations by taking $\psi^{(0)}(0) =0$, $\psi^{(1)}(0)
= q(0) = 1$, ${\cal J}^{(0)}(0) = J(0) = 1$, and ${\cal J}^{(1)}(0) = - H[
J(0),q(0)]$.
We expect to
find that after a time $t \sim 1/\varepsilon$, the errors are of order
$\sim \varepsilon$ for $q_{\rm p1a}(t)$, and of order $\sim \varepsilon^2$
for $J_{\rm p1a}(t)$.  This is confirmed by the two lower panels in Fig.\
\ref{fig:approx}, which show the differences $q - q_{\rm p1a}$ and $J -
J_{\rm p1a}$.

\section{Discussion}
\label{sec:discussion}

In Sec.\ \ref{sec:kerrapplication} above we derived the set of equations
(\ref{eq:eomsimplified}) describing the
radiation-reaction driven inspiral of a particle into a spinning black
hole, in terms of generalized action angle variables.
Although those equations contain some functions which are
currently unknown, it is possible to give a general analysis of the
dependence of the solutions on the mass ratio $\varepsilon = \mu/M$ as
$\varepsilon \to 0$, using two-timescale expansions.
That analysis was presented in Secs.\ \ref{sec:generalsystem} --
\ref{sec:numerics} above, for the
general class of equation systems (\ref{sec2:eq:eom1}) of which the Kerr inspiral
example (\ref{eq:eomsimplified}) is a special case.
In this final section we combine these various results and discuss the
implications for our understanding of inspirals into black holes.

\subsection{Consistency and uniqueness of approximation scheme}

Our analysis has demonstrated that the adiabatic approximation
method gives a simple and unique prescription for computing
successive approximations to the exact solution, order by order, which
is free of
ambiguities.  In this sense it is similar to the post-Newtonian
approximation method.\footnote{The analogy is closer when
the two-timescale method is extended to include the field equations and
wave generation as well as the inspiral motion \cite{FH08c}.}
This is shown explicitly in Sec.\ \ref{subsec:induction},
which shows that the adiabatic method can be extended to all orders
for the case of a single degree of freedom, and in Sec.\ \ref{sec:numerics}, which
shows how the method works in practice in a numerical example.  In
particular there is no ambiguity in the assignment of initial
conditions when computing adiabatic or post-1-adiabatic
approximations.

This conclusion appears to be at odds with a recent analysis of Pound
and Poisson (PP) \cite{Pound:2007ti}.  These authors conclude
that ``An adiabatic approximation to the exact differential
equations and initial conditions, designed
to capture the secular changes in
the orbital elements and to discard the
oscillations, would be very difficult to formulate
without prior knowledge of the exact
solution.''
The reason for the disagreement is in part a matter of terminology:
PP's definition of ``adiabatic approximation'' is
different to ours.\footnote{In a later version of their
paper they call it instead a ``secular approximation''.}  They take it
to mean an approximation which (i) discards all the pieces of the true
solutions that vary on the rapid timescale $\sim 1$, and retains the
pieces that vary on the slow timescale $\sim 1/\varepsilon$; and (ii)
is globally accurate to some specified order in $\varepsilon$ over an inspiral time --
throughout their paper they work to the first subleading order,
i.e. post-1-adiabatic order.
In our terminology, their approximation would consist of the adiabatic
approximation, plus the secular piece of the post-1-adiabatic approximation
[given by omitting the first term in Eq.\ (\ref{sec2:eq:J1ans})].

The difference in the terminology used here and in PP is not the only
reason for the different conclusions.
Our formalism shows that PP's ``adiabatic approximation'' is actually
straightforward to formulate,
and that prior knowledge of the exact
solution is not required.  The reason for the different conclusions is
as follows.
By ``exact solution'' PP in fact meant
any approximation which includes the rapidly oscillating pieces
at post-1-adiabatic order.  Their intended meaning
was that, since the secular and rapidly oscillating pieces
are coupled together at post-1-adiabatic order, any
approximation which completely neglects the oscillations cannot be
accurate to post-1-adiabatic order \cite{PPprivate}.  We agree with
this conclusion.

On the other hand, we disagree with the overall pessimism of PP's
conclusion, because we disagree with their premise.
Since the qualitative arguments that were originally presented for the
radiative approximation involved discarding oscillatory effects
\cite{Mino2003,Hughes:2005qb}, PP chose to examine general
approximation schemes that neglect oscillatory effects\footnote{In the
strong sense of neglecting the influence of the oscillatory pieces of
the solution on the secular pieces, as well as neglecting the
oscillatory pieces themselves.}
and correctly concluded that such schemes cannot be accurate
beyond the leading order.  However, our viewpoint is that there
is no need to restrict attention to schemes that neglect all
oscillatory effects.  The two timescale scheme presented here
yields leading order solutions which are not influenced by oscillatory
effects, and higher order solutions whose secular pieces are.
The development of a systematic approximation scheme that exploits the
disparity in orbital and radiation reaction timescales need not be
synonymous with neglecting all oscillatory effects.

\subsection{Effects of conservative and dissipative pieces of the self force}

As we have discussed in Secs.\ \ref{resultsdiscussion} and
\ref{sec:discuss} above, our analysis
shows rigorously that the dissipative piece of the self
force contributes to the leading order, adiabatic motion, while the
conservative piece does not, as argued in Refs.\ \cite{Mino2003,Hughes:2005qb}.
It is possible to understand this fundamental difference in a simple
way as follows.  We use units where the orbital timescale is $\sim 1$
and the inspiral timescale is $\sim 1/\varepsilon$.  Then the
total phase accumulated during the inspiral is $\sim 1/\varepsilon$,
and this accumulated phase is driven by the dissipative piece of the self force.

Consider now the effect of the conservative piece of the self force.
As a helpful thought experiment, imagine setting to zero the
dissipative piece of the first order self force.  What then is the effect of the
conservative first order self-force on the dynamics?  We believe that the
perturbed motion is likely to still be integrable; arguments for this
will be presented elsewhere \cite{FH08a,FH08b}.  However, even if the
perturbed motion is not integrable, the Kolmogorov-Arnold-Moser (KAM)
theorem \cite{Arnold} implies that the perturbed motion will
generically be
confined to a torus in phase space for sufficiently small
$\varepsilon$.  The effect of the conservative
self force is therefore roughly to give an $O(\varepsilon)$ distortion to this
torus, and to give $O(\varepsilon)$ corrections to the fundamental
frequencies.\footnote{This corresponds to adding to the frequency
 $\omega_\alpha$ in Eq.\ (\ref{sec5:eq:eom1a}) the average over $\bfq$ of the term
 $\varepsilon g^{(1)}_\alpha$.}
If one now adds the effects of dissipation, we see that
after the inspiral time $\sim 1/\varepsilon$, the corrections due to
the conservative force will give a fractional phase correction of
order $\sim \varepsilon$, corresponding to a total phase correction
$\sim 1$.  This correction therefore comes in at post-1-adiabatic order.

Another way of describing the difference is that the dissipative
self-force produces secular changes in the orbital elements, while the
conservative self-force does not at the leading order in
$\varepsilon$.  In Ref.\ \cite{Hughes:2005qb} this difference was
overstated: it was claimed that the conservative self-force does
not produce any secular effects.  However, once one goes beyond the
leading order, adiabatic approximation, there are in fact conservative secular
effects.  At post-1-adiabatic order these are  described by the first
term on the right hand side of Eq.\ (\ref{sec2:eq:Omega1ans}).  This
error was pointed out by Pound and
Poisson \cite{Pound:2005fs,Pound:2007ti}.

\subsection{The radiative approximation}
\label{sec:radiative}

So far in this paper we have treated the self force as fixed,
and have focused on how to compute successive approximations to the
inspiralling motion.  However, as explained in the introduction,
the first order self force is currently not yet known explicitly.
The time-averaged, dissipative\footnote{We use the terms radiative and
  dissipative interchangeably; both denote the time-odd piece of the
  self force, as defined by Eq.\ (\ref{eq:adissdef}) above.} piece is
known from work of Mino and others
\cite{Mino2003,Hughes:2005qb,scalar,2005PThPh.114..509S,2006PThPh.115..873S}.
The remaining, fluctuating piece of the dissipative first order self
force has not been computed but will be straightforward to
compute\footnote{For example, by evaluating $J_{\omega
    lmkn}$ from Eq.\ (8.21) of Ref.\ \cite{scalar} at $\omega = \omega_{mk'n'}$
  instead of $\omega = \omega_{mkn}$.}.  The conservative piece of the
first order self force will be much more difficult to compute, and is
the subject of much current research
\cite{Poisson:2003nc,Barack:2002mh,Gralla:2005et,Keidl:2006wk,Barack:2007tm}.

It is natural therefore to consider the {\it radiative approximation}
obtained by using only the currently available, radiative piece of the first
order self force, as suggested by Mino \cite{Mino2003}, and by integrating the
orbital equations exactly (eg numerically).
How well will this approximation perform?

From our analysis it follows that the motion as computed in this
approximation will agree with the true motion to adiabatic order, and
will differ at post-1-adiabatic order.  At post-1-adiabatic order, it will omit
effects due to the conservative first order force, and also effects
due to the dissipative second order self force.  It will include
post-1-adiabatic effects due to the fluctuating pieces of the first
order, dissipative self force, and so would be
expected to be more accurate than the adiabatic approximation.\footnote{It is of course possible
that, due to an accidental near-cancellation of different post-1-adiabatic terms,
the adiabatic approximation may be closer to the true solution than
the radiative approximation.}
EMRI waveforms computed using this approximation will likely be the
state of the art for quite some time.

Our conclusions about the radiative approximation appear to differ
from those of PP \cite{Pound:2007ti}, who argue that ``
The radiative approximation does not achieve the goals of
an adiabatic approximation''.  Here, however, the different
conclusions arise entirely from a difference in terminology, since PP
define ``adiabatic approximation'' to include slowly varying pieces of
the solution to at least post-1-adiabatic order.  The radiative
approximation does produce solutions that are accurate to adiabatic
order, as we have defined it.

We now discuss in more detail the errors that arise in the radiative
approximation. These errors occur at post-1-adiabatic order.
For discussing these errors, we will neglect post-2-adiabatic effects,
and so it is sufficient to use our post-1-adiabatic dynamical
equations (\ref{sec2:eq:calJ1ans})
and (\ref{sec2:eq:Omega1ans}).  These equations have the structure
\begin{equation}
\label{schematic}
{\cal D} \left[ \begin{array}{l}
         \psi^{(1)}_\alpha(\tt)    \\
         {\cal J}^{(1)}_\lambda(\tt) \\
                \end{array}
        \right] = {\cal S},
\end{equation}
where ${\cal D}$ is a linear differential operator and ${\cal S}$ is a
source term.  The appropriate initial conditions are
[see Sec.\ \ref{sec:initialconditions}]
\bes
\bea
\psi_\alpha^{(0)} =0,    && {\cal J}^{(0)}_\lambda(0) = J_{\lambda}(0),
\\
\psi_\alpha^{(1)} =q_\alpha(0),    && {\cal J}^{(1)}_\lambda(0) =
- H_{\lambda}[\bfq(0),\bfJ(0)],
\eea
\label{ics}\ees
where $\bfq(0)$ and $\bfJ(0)$ are the exact initial conditions and the
function $H_\lambda$ is given by, from Eq. (\ref{sec2:eq:J1ans}),
\be
H_\lambda(\bfq,\bfJ)  = {\cal I}_{\bfOmega^{(0)}(0)}{\hat
  G}^{(1)}_\lambda[\bfq,\bfJ,0].
\label{Hlambdadef}
\ee

In terms of these quantities,
the radiative approximation is equivalent to making the
replacements
\bes
\bea
g^{(1)}_\alpha(\bfq,\bfJ) &\to& g^{(1)}_{\alpha\,{\rm diss}}(\bfq,\bfJ),
\\
G^{(1)}_i(\bfq,\bfJ) &\to& G^{(1)}_{i\,{\rm diss}}(\bfq,\bfJ),
\label{replacement1}
\\
G^{(2)}_i(\bfq,\bfJ) &\to& 0.
\eea
\ees
These replacements have two effects: (i) they give rise to an error in
the source term ${\cal S}$ in Eq.\ (\ref{schematic}), and (ii) they
give rise to an error in the function $H_\lambda$ and hence in the
initial conditions (\ref{ics}).  There are thus two distinct types of
errors that occur in the radiative approximation.\footnote{These two
errors are both secular, varying on long timescales.  There is in
addition a rapidly oscillating error caused by the correction to the
first term in the expression (\ref{sec2:eq:J1ans}) for $J^{(1)}_\lambda$.}

The second type of error could in principle be removed by adjusting the initial
conditions appropriately.
For fixed initial conditions $\bfq(0)$ and $\bfJ(0)$, such an
adjustment would require knowledge of the conservative piece of the
self force, and so is not currently feasible.  However, in the context
of searches for gravitational wave signals, matched filtering searches
will automatically vary over a wide range of initial conditions.  Therefore the second
type of error will not be an impediment to detecting gravitational
wave signals.  It will, however, cause errors in parameter extraction.

This fact that the error in the radiative approximation can be reduced
by adjusting the initial conditions was discovered 
by Pound and Poisson \cite{Pound:2007th}, who
numerically integrated inspirals in Schwarzschild
using post-Newtonian self-force expressions.
Their ``time-averaged'' initial conditions, which they found to give the
highest accuracy, correspond to removing the second type of error
discussed above, that is, using the initial conditions (\ref{ics})
with the exact function $H_\lambda$ rather than the radiative
approximation to $H_\lambda$.

Finally, we note that given the radiative approximation to the self
force, one can compute waveforms using the radiative approximation as
described above, or compute waveforms in the adiabatic approximation by
solving equations (\ref{jks}), (\ref{sec2:eq:calJ0ans}) and (\ref{sec2:eq:Omega0ans})
using the replacement (\ref{replacement1}).  This second option would
be easier although somewhat less accurate.

\subsection{Utility of adiabatic approximation for  detection of
  gravitational wave signals}

The key motivation for accurate computations of waveforms from inspiral
events is of course their use for detecting and analyzing
gravitational wave signals.  How well will the adiabatic and
radiative approximations perform in practice?
In this section, we review the studies that have been made of
this question.  These studies are largely consistent with one another,
despite differences in emphasis and interpretation that can be found in the literature.
We restrict attention to inspirals in Schwarzschild, and to circular
or equatorial inspirals in Kerr; fully general orbits present
additional features that will be discussed elsewhere \cite{FH08a,FH08b}.

First, we note that in this paper we have focused on how the
post-1-adiabatic error in phase scales with the mass ratio $\varepsilon=\mu/M$.  However, one can
also ask
how the error scales with the post-Newtonian expansion parameter $v/c
\sim \sqrt{M/r}$.  From Eq.\ (A10) of Ref.\ \cite{scalar} it follows
that the post-1-adiabatic phase errors scale as
$$
\sim \left( \frac{\mu}{M} \right)^0 \, \left({v \over c} \right)^{-3};
$$
this scaling is consistent with the more recent analysis of Ref.\ \cite{Pound:2007th}.
This scaling does imply that the error gets large in the weak field
regime, as correctly argued in Ref.\ \cite{Pound:2007th}.
However, it does not necessarily imply large errors in the
relativistic regime $v/c \sim 1$ relevant to LISA observations.

The first, order of magnitude estimates of the effects of the conservative piece
of the self force were made by Burko in Refs.\
\cite{2001IJMPA..16.1471B,2003PhRvD..67h4001B}.
Refs.\ \cite{Hughes:2005qb,scalar} computed the post-1-adiabatic phase
error within the post-Newtonian approximation for circular orbits,
minimized over some of the template parameters, and evaluated at frequencies
relevant for LISA.  The results indicated a total phase error of order
one cycle, not enough to impede detection given that maximum coherent
integration times are computationally limited to $\sim 3$ weeks
\cite{Gair:2004iv}.  This result was extended to eccentric orbits with
eccentricities $\alt 0.4$ in Refs.\ \cite{Favata:2006,2006APS..APRS11006F}, with similar
results.  Similar computations were performed by Burko in Refs.\
\cite{2006CQGra..23.4281B,2006AIPC..873..269B}, although without
minimization over template parameters.

These analyses all focused on extreme mass ratio inspirals for LISA.
For intermediate mass ratio inspirals, potential sources for LIGO,
the post-1-adiabatic corrections were studied within the
post-Newtonian approximation in Refs.\ \cite{Brown2006,2005CQGra..22S1179A}.
Ref. \cite{Brown2006} computed fitting factors in addition to phase errors,
found that the associated loss of signal to noise ratio would be less than 10\%
in all but the most rapidly spinning cases, and concluded that it
would be ``worthwhile but not essential'' to go beyond adiabatic order
for detection templates.

The most definitive study to date of post-adiabatic errors for LISA in the
Schwarzschild case was performed by Pound and Poisson (PP1)
\cite{Pound:2007th}.  PP1 numerically integrated the geodesic equations with post-Newtonian
expressions for the self force, with and without conservative terms.
PP1 found large phase errors, $\delta \phi \agt 100$, in the weak
field regime.  However, the regime relevant to LISA observations is
$p \alt 30$ \cite{Finn:2000sy}\footnote{It is true that there will be some binaries visible
to LISA at higher values of $p$, that do not merge within the LISA
mission lifetime.  However post-Newtonian templates should be sufficient for
the detection of these systems.}, where $p$ is the dimensionless semilatus rectum parameter defined by PP1,
and PP1's results are focused mostly on values of $p$ larger than
this\footnote{The second panel of their Fig.\ 6 does show phase shifts
for smaller values of $p$, but these are all for a mass ratio of
$\varepsilon = 0.1$, too large to be a good model of LISA
observations; although the phase shift becomes independent of
$\varepsilon$ as $\varepsilon\to0$, their Fig. 6 shows that it can
vary by factors of up to $\sim 10$ as $\varepsilon$ varies between $0.1$ and
$0.001$.}.  It is therefore difficult to compare the results of PP1
with earlier estimates or to use them directly to make inferences
about signal detection with LISA.
PP1's results do show clearly that the errors increase rapidly with increasing eccentricity.

We have repeated PP1's calculations, reproducing the results
of their Fig.\ 6, and extended their calculations to
more relativistic systems at lower values of $p$.
More specifically, we performed the following computation:
(i) Select values of the mass parameters $M$ and $\mu$, and the initial eccentricity $e$;
(ii) Choose the initial value of semilatus rectum $p$ to correspond to
one year before the last stable orbit, which occurs on the separatrix
$p = 6 + 2 e$ \cite{PhysRevD.50.3816}; (iii) Choose the radiative
evolution and the exact evolution to line up at some matching time
$t_{\rm m}$ during the last year of inspiral; (iv) Start the radiative and
exact evolutions with slightly different initial conditions in order
that the secular pieces of the evolutions initially coincide -- this
is the ``time-averaged'' initial data prescription of PP1;  (v)
Compute the maximum of the absolute value of the phase error $\delta
\phi$ incurred during the last year; (vi) Minimize over the
matching time $t_{\rm m}$; and (vii) Repeat for different values of $M$,
$\mu$ and $e$.
As an example, for $M = 10^6 M_\odot$ and $\mu=10 M_\odot$, an
inspiral starting at $(p,e) = (10.77,0.300)$ ends up at $(6.31,0.153)$
after one year.  We match the two evolutions at 0.2427 years before plunge, with
the exact evolution starting at $(p,e) = (8.81933,0.210700)$ and the radiative evolution starting
at $(p,e) = (8.81928,0.210681)$.  The maximum phase error incurred
in the last year is then $0.91$ cycles.

The phase error incurred during an inspiral from some initial values
of $e$ and $p$ to the plunge is independent of the masses $M$ and
$\mu$ in the small mass ratio limit.  However the phase error incurred
during the last year of inspiral is not, since the initial value of
$p$ depends on the inspiral timescale $\sim M^2/\mu$.  The result is
that the phase error depends only on the combination of masses
$M^2/\mu$ to a good approximation.

\begin{figure}
\begin{center}
\epsfig{file=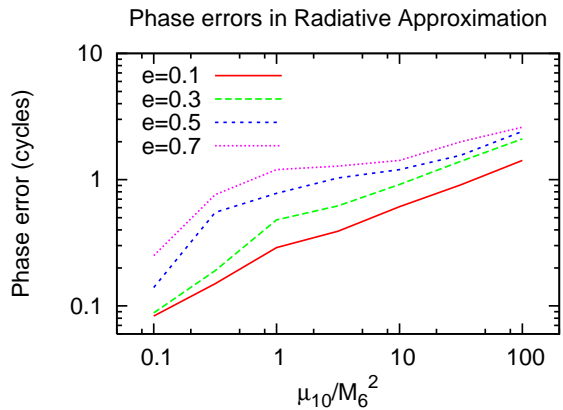,width=8.0cm}
\caption{
The maximum orbital phase error in cycles, $\delta N = \delta \phi/(2 \pi)$,
incurred in the radiative approximation during the last
year of inspiral, as a function of the mass $M_6$ of the central black hole in units
of $10^6 M_\odot$, the mass $\mu_{10}$ of the small object in units of $10 \, M_\odot$,
and the eccentricity $e$ of the system at the start of the final year
of inspiral.
The exact and radiative inspirals are chosen to line up at some time
$t_{\rm m} $ during the final year, and the value of $t_{\rm m}$ is
chosen to minimize the phase error.  The initial data at time
$t_{\rm m}$ for the
radiative evolution is slightly different to that used for the exact
evolution in order that the secular pieces of the two evolutions
initially coincide; this is the ``time-averaged'' initial
data prescription of Pound and Poisson \protect{\cite{Pound:2007th}}.
All evolutions are computed using the hybrid equations of motion of Kidder,
Will and Wiseman \protect{\cite{PhysRevD.47.3281}} in the osculating-element form given
by Pound and Poisson.}
\label{fig:phaseerrors}
\end{center}
\end{figure}

Our results are shown in Fig.\ \ref{fig:phaseerrors}.
This figure shows, firstly, that the computational
method of PP1 gives results for low eccentricity systems that
are roughly consistent with earlier, cruder, estimates, with total phase errors
of less than one cycle over most of the parameter space.  It also
shows that for large eccentricity systems the total phase error
can be as large as two or three cycles.

How much will the phase errors shown in Fig.\ \ref{fig:phaseerrors}
impede the use of the radiative approximation to detect signals?
There are two factors which will help.  First,
Fig.\ \ref{fig:phaseerrors} shows the maximum phase error during the
last year of inspiral, while for detection phase coherence is needed
only for periods of $\sim 3$ weeks \cite{Gair:2004iv}.
Second, the matched filtering search process will automatically select
parameter values to maximize the overlap between the template and true
signal, and parameter mismatches will therefore be likely to reduce
the effect of the phase error\footnote{We note that there
  are already two minimizations over parameters included in the phase
  errors shown in Fig.\ \protect{\ref{fig:phaseerrors}}: a
  minimization over $t_{\rm m}$ as discussed above, and the
  replacement $m_1 \to m_1 + m_2$ used by PP1 in the derivation of
  their self-force expressions in order to eliminate the leading order
  piece of the self-force.}.
On the other hand, for large eccentricities, the phase error $\delta
\phi(t)$ is typically a rapidly oscillating function, rather than a
smooth function, which may counteract the helpful effects of smaller
time windows or parameter mismatches.  Also we note that a sign flip
will occur in the integrand of an overlap integral once the gravitational wave phase
error $2 \delta \phi$ exceeds $\pi$, corresponding to the number of
cycles plotted in Fig.\ \ref{fig:phaseerrors} exceeding $1/4$.
This occurs in a large part of the parameter space.

Thus, there is a considerable amount of uncertainty
as to whether the radiative approximation will be sufficiently
accurate for signal detection.  A detailed study would require
computation of fitting
factors and optimizing over all template parameters, and modeling the
hierarchical detection algorithm discussed in Ref.\
\cite{Gair:2004iv}.  Such a study is beyond the scope of this paper.
Based on the results shown in Fig.\ \ref{fig:phaseerrors},
we agree with the conclusions of PP1 that the early estimates based on circular orbits
\cite{Hughes:2005qb,scalar} were too optimistic, and that it is not
clear that the radiative approximation is sufficiently accurate.
(Moreover parameter extraction will clearly require going beyond
the radiative approximation.)

For gravitational wave searches, it might therefore be advisable to
use {\it hybrid} waveforms, computed using the fully relativistic
dissipative piece of the self force, and using post-Newtonian
expressions for the conservative piece.  Although the post-Newtonian
expressions are not expected to be very accurate in the relativistic
regime, improved versions have been obtained recently based on
comparisons between post-Newtonian and fully numerical waveforms from
binary black hole mergers; see, for example, the effective one body
approximation of Refs.\
\cite{1999PhRvD..59h4006B,2001PhRvD..64l4013D,Buonanno:2002gi,Damour:2002vi,Damour:2007xr,Buonanno:2007pf}.
It seems likely that hybrid EMRI waveforms incorporating such improved
post-Newtonian expressions for the conservative self force
will be more accurate than radiative waveforms.  Hybrid waveforms may
be the best that can be done until the fully relativistic conservative
self-force is computed.

\section{Conclusions}

In this paper we have developed a systematic two-timescale
approximation method for computing the inspirals of particles into
spinning black holes.  Future papers in this series will deal with the
effects of transient resonances \cite{FH08a,FH08b}, and will give more
details of the two-timescale expansion of the Einstein equations
\cite{FH08c} that meshes consistently with the approximation method
for orbital motion discussed here.

\acknowledgments

We thank Steve Drasco, Marc Favata, John Friedman, Scott Hughes, Yasushi Mino,
Eric Poisson, Adam Pound and Eran Rosenthal for helpful conversations.  This
research was supported in part by NSF grant
PHY-0457200 and NASA grant NAGW-12906.  TH was supported in part by
the John and David Boochever Prize Fellowship in Theoretical Physics
at Cornell.

\appendix

\section{Explicit expressions for the coefficients in the action-angle
  equations of motion}
\label{appendix:frequencies}

From the formulae (\ref{sec4:eq:BLactions}) for the action variables
together with the definitions (\ref{sec4:eq:potentials}) of the
potentials $V_r$ and $V_\theta$ we can compute the partial derivatives
$\partial J_\alpha / \partial P_\beta$.  The non-trivial derivatives are
\bes
\bea
\frac{\partial J_r}{\partial H}&=&\frac{Y}{\pi},\\
\frac{\partial J_r}{\partial E}&=&\frac{W}{\pi},\\
\frac{\partial J_r}{\partial L_z}&=& - \frac{Z}{\pi},\\
\frac{\partial J_r}{\partial Q}&=&-\frac{X}{2\pi},\\
\frac{\partial J_\theta}{\partial H}&=& \frac{2 \sqrt{z_+} a^2}{\pi
\beta}\left[K(k)-E(k)\right], \\
\frac{\partial J_\theta}{\partial E}&=& \frac{2 \sqrt{z_+} Ea^2}{\pi
\beta}\left[K(k)-E(k)\right],\\
\frac{\partial J_\theta}{\partial L_z}&=&\frac{2L_z}{\pi \beta \sqrt{z_+}}
\left[K(k)-\Pi(\pi/2,z_-,k)\right],\\
\frac{\partial J_\theta}{\partial Q}&=& \frac{1}{\pi \beta
\sqrt{z_+}}K(k).
\eea
\label{sec4:eq:Jptransformation}\ees
Here the quantities $W$, $X$, $Y$ and $Z$ are the
radial integrals defined by Schmidt
\footnote{There is a typo in the definition of $W$ given in Eq.\ (44)
of Schmidt \protect{\cite{Schmidt}}.}
as {\cite{Schmidt}}
\bes
\bea
W&=&\int^{r_2}_{r_1}\frac{r^2 E (r^2+a^2) - 2 M r a (L_z - a E)}{\Delta
\sqrt{V_r}}dr, \ \ \ \ \ \\
X&=&\int^{r_2}_{r_1}\frac{dr}{\sqrt{V_r}},\\
Y&=&\int^{r_2}_{r_1}\frac{r^2}{\sqrt{V_r}} dr,\\
Z&=&\int^{r_2}_{r_1}\frac{r\left[L_zr-2 M (L_z-aE)\right]}{\Delta
\sqrt{V_r}}dr,\ \ \
\label{sec4:eq:radialintegr}
\eea
\ees
where $r_1$ and $r_2$ are the turning points of the radial motion,
i.e. the two largest roots of $V_r(r)=0$.  In these equations $K(k)$
is the complete
elliptic integral of the
first kind, $E(k)$ is the complete elliptic integral of the second
kind, and $\Pi(\phi,n,k)$ is the Legendre elliptic integral of the
third kind \cite{NR}:
\begin{align}
K(k) &= \int_0^{\pi/2} \frac{d\theta}{\sqrt{1 - k^2 \sin^2 \theta}},&
\\
E(k) &= \int_{0}^{\pi/2} d\theta \sqrt{1 - k^2 \sin^2\theta},&
\\
\Pi(\phi,n,k) &=  \int_0^{\phi} \frac{d\theta}{(1 - n \sin^2 \theta)\sqrt{1 - k^2 \sin^2 \theta}}.&
\label{sec4:eq:complelliptic}
\end{align}
Also we have defined $\beta^2=a^2(\mu^2-E^2)$ and $k =
\sqrt{z_-/z_+}$, where $z = \cos^2\theta$ \footnote{Here we follow
  Drasco and Hughes \cite{Drasco:2005kz} rather than Schmidt who defines $z=\cos\theta$.}
and $z_-$ and $z_+$ are the
two roots of $V_\theta(z)=0$ with $0 < z_- < 1 < z_+$.

Combining the derivatives (\ref{sec4:eq:Jptransformation}) with the chain rule in the form
\be
\frac{\partial P_\alpha}{ \partial J_\beta} \frac{\partial
  J_\beta }{\partial P_\gamma}  = \delta^\alpha_\gamma
\ee
yields the following expression for the frequencies
(\ref{sec22:eq:omegadef0}) as functions of $P_\alpha$:
\bes
\bea
  \Omega_t &=&
  \frac{K(k)W +
        a^{2}z_{+}E\left[K(k)-E(k)\right]X
        }{K(k)Y +
          a^{2}z_{+}\left[K(k)-E(k)\right]X},\ \ \  \\
  \Omega_r &=&
  \frac{\pi K(k)}{K(k)Y + a^{2}z_{+}\left[K(k)-E(k)\right]
                  X}, \ \ \ \\
  \Omega_\theta &=&
  \frac{\pi\beta \sqrt{z_+}X/2}{K(k)Y +
    a^{2}z_{+}\left[K(k)-E(k)\right]X}, \ \ \ \\
  \Omega_\phi &=&
  \frac{K(k)Z + L_{z}[\Pi(\pi/2,z_{-},k)-K(k)]X
        }{K(k)Y + a^{2}z_{+}[K(k)-E(k)]X}. \ \ \
\eea
\ees

\section{Comparison with treatment of Kevorkian and Cole}
\label{app:Kevorkian}

As explained in Sec.\ \ref{sec:generalsystem} above, our two-timescale analysis of the
general system of equations (\ref{sec2:eq:eom1}) follows closely that of the
textbook \cite{Kevorkian} by Kevorkian and Cole (KC), which is a standard reference
on asymptotic methods.
In this appendix we explain the minor ways in which our treatment of Secs.\
\ref{sec:derivation_single} and \ref{sec2:manyvariables} extends and corrects that of KC.
Section 4.4 of KC covers the one variable case.  We simplify this
treatment by using action angle variables, and also extend it by
showing that the method works to all orders in $\varepsilon$.
Our general system of equations (\ref{sec2:eq:eom1}) is studied by KC in their
section 4.5.  We generalize this analysis by including the
half-integer powers of $\varepsilon$, which are required for the
treatment of resonances.  A minor correction is that
their solution (4.5.54a) is not generally valid, since it requires
$\Omega_i$ and $\tau_i$ to be collinear, which will not always be the
case. However it is easy to repair this error by replacing the
expression with one constructed using Fourier methods, cf.\ Eq.\
(\ref{eq:J1ans1}) above.
Finally, our treatment of resonances \cite{FH08a,FH08b} will closely
follow KC's section 5.4, except that our analysis will apply to the
general system (\ref{sec2:eq:eom1}), generalizing KC's treatment of
special cases.


\end{document}